\documentclass[11pt,english]{article}
\usepackage[T1]{fontenc}
\usepackage[latin9]{inputenc}
\usepackage[a4paper]{geometry}
\usepackage{color}
\usepackage{mathrsfs}
\usepackage{mathtools}
\usepackage{amsmath}
\usepackage{amsthm}
\usepackage{amssymb}
\usepackage{graphicx}
\usepackage{setspace}
\usepackage[authoryear]{natbib}
\makeatletter
  \theoremstyle{plain}
  \newtheorem{assumption}{\protect\assumptionname}
  \theoremstyle{plain}
  \newtheorem{lem}{\protect\lemmaname}
\theoremstyle{plain}
\newtheorem{thm}{\protect\theoremname}

\@ifundefined{date}{}{\date{}}
\bibpunct{(}{)}{,}{a}{,}{,}

\usepackage{babel}

\usepackage[multiple]{footmisc}

\usepackage{chngcntr}
\usepackage{apptools}
\AtAppendix{\counterwithin{lemma}{section}}
\newtheorem{lemma}{Lemma}

\makeatother

\usepackage{babel}
  \providecommand{\assumptionname}{Assumption}
  \providecommand{\lemmaname}{Lemma}
\providecommand{\theoremname}{Theorem}

\begin{document}

\title{\vspace{20pt}
Testing for observation-dependent regime switching\\
in mixture autoregressive models\thanks{The authors thank the Academy of Finland for financial support. Contact
addresses: Mika Meitz, Department of Political and Economic Studies,
University of Helsinki, P. O. Box 17, FI\textendash 00014 University
of Helsinki, Finland; e-mail: mika.meitz@helsinki.fi. Pentti Saikkonen,
Department of Mathematics and Statistics, University of Helsinki,
P. O. Box 68, FI\textendash 00014 University of Helsinki, Finland;
e-mail: pentti.saikkonen@helsinki.fi.}\vspace{20pt}
}

\author{Mika Meitz\\\small{University of Helsinki} \and Pentti Saikkonen\\\small{University of Helsinki}\vspace{20pt}
}

\date{October 2017}
\maketitle
\begin{abstract}
\noindent Testing for regime switching when the regime switching
probabilities are specified either as constants (`mixture models')
or are governed by a finite-state Markov chain (`Markov switching
models') are long-standing problems that have also attracted recent
interest. This paper considers testing for regime switching when the
regime switching probabilities are time-varying and depend on observed
data (`observation-dependent regime switching'). Specifically, we
consider the likelihood ratio test for observation-dependent regime
switching in mixture autoregressive models. The testing problem is
highly nonstandard, involving unidentified nuisance parameters under
the null, parameters on the boundary, singular information matrices,
and higher-order approximations of the log-likelihood. We derive the
asymptotic null distribution of the likelihood ratio test statistic
in a general mixture autoregressive setting using high-level conditions
that allow for various forms of dependence of the regime switching
probabilities on past observations, and we illustrate the theory using
two particular mixture autoregressive models. The likelihood ratio
test has a nonstandard asymptotic distribution that can easily be
simulated, and Monte Carlo studies show the test to have satisfactory
finite sample size and power properties.

\bigskip{}
\bigskip{}
\bigskip{}


\bigskip{}

\end{abstract}
\vfill{}
\pagebreak{}

\section{Introduction}

Different regime switching models are in widespread use in economics,
finance, and other fields. When the regime switching probabilities
are constants, these models are often referred to as `mixture models',
and when these probabilities depend on past regimes and are governed
by a finite-state Markov chain, the term (time homogeneous) `Markov
switching models' is typically used. In this paper, we are interested
in the case where the regime switching probabilities depend on observed
data, a case we refer to as `observation-dependent regime switching'.
Models of this kind can be viewed as special cases of time inhomogeneous
Markov switching models (in which regime switching probabilities depend
on both past regimes and observed data). Overviews of regime switching
models can be found, for example, in \citet{fruhwirth2006finite}
and \citet{hamilton2016macroeconomic}. Of critical interest in all
these models is whether the use of several regimes is warranted or
if a single-regime model would suffice. Testing for regime switching
in all these models is plagued by several irregular features such
as unidentified parameters and parameters on the boundary and is consequently
notoriously difficult. 

Tests for Markov switching have been considered by several authors
in the econometrics literature. \citet{hansen1992likelihood} and
\citet{garcia1998asymptotic} both considered sup-type likelihood
ratio (LR) tests in Markov switching models but they did not present
complete solutions. Hansen derived a bound for the distribution of
the LR statistic, leading to a conservative procedure, while Garcia
did not treat all the non-standard features of the problem in detail.
\citet{cho2007testing} analyzed the use of a LR statistic for a mixture
model to test for Markov-switching type regime switching. They found
their test based on a mixture model to have power against Markov switching
alternatives even though it ignores the temporal dependence of the
Markov chain. \citet*{carrasco2014optimal} took a different approach
and proposed an information matrix type test that they showed to be
asymptotically optimal against Markov switching alternatives. Very
recently, both \citet{Qu2017Likelihood} and \citet{Kasahara2017Testing}
have studied the LR statistic for regime switching in Markov switching
models. 

Regarding testing for mixture type regime switching, the existing
literature is extensive, and several early references can be found,
for instance, in \citet[Sec. 6.5.1]{mclachlan2000finite}. Most papers
in this literature consider the case of independent observations without
regressors. Notable exceptions allowing for regressors (but not dependent
data) and having set-ups closer to the present paper are \citet{zhu2004hypothesis,zhu2006asymptotics}
and \citet{kasahara2015testing} who consider (among other things)
LR tests for regime switching. Further comparison to these works will
be provided in later sections. 

In contrast to testing for Markov switching or mixture type regime
switching, there exists almost no literature on testing for observation-dependent
regime switching. The only two exceptions we are aware of are the
unpublished PhD thesis of \citet{jeffries1998logistic} and the recent
paper of \citet{shen2015inference}. Jeffries's thesis, which appears
to have gone largely unnoticed, analyses the LR test in a specific
(first-order) mixture autoregressive model; we will discuss his work
further in later sections. \citet{shen2015inference} consider the
case of independent observations with regressors and observation-dependent
regime switching, and propose an `expectation maximization test' for
regime switching.

In this paper we consider testing for observation-dependent regime
switching in a time series context. Specifically, we analyze the asymptotic
distribution of the LR test statistic for testing a linear autoregressive
model against a two-regime mixture autoregressive model with observation-dependent
regime switching. Mixture autoregressive models have been discussed
for instance in \citet{wong2000mixture,wong2001logistic}, \citet*{dueker2007contemporaneous},
\citet*{dueker2011multivariate}, and \citet*{kalliovirta2015gaussian,kalliovirta2016gaussian};
further discussion of this previous work will be provided in Section
2. Motivation for allowing the regime switching probabilities to depend
on observed data stems, for instance, from the desire to associate
changes in regime to observable economic variables.\footnote{See, for instance, \citet{hamilton2016macroeconomic}, whose \emph{Handbook
of Macroeconomics} chapter begins ``Many economic time series exhibit
dramatic breaks associated with events such as economic recessions,
financial panics, and currency crises. Such changes in regime may
arise from tipping points or other nonlinear dynamics and are core
to some of the most important questions in macroeconomics.'' }\footnote{More general models in which the regime switching probabilities are
allowed to depend on both past regimes and observable variables have
also been considered, see, e.g., \citet*{diebold1994regime}, \citet{filardo1994business},
and \citet*{kim2008estimation}.} Following \citet{kasahara2015testing} it would also be possible
to consider the more general testing problem that in a model with
more than two regimes the number of regimes can be reduced. However,
as even the case of two regimes is quite complex in our set-up, we
leave this extension to future research.

We consider mixture autoregressive (MAR) models in a rather general
setting employing high-level conditions that allow for various forms
of observation-dependent regime switching. As specific examples, we
treat the so-called logistic MAR (LMAR) model of \citet{wong2001logistic}
and (a version of the) Gaussian MAR (GMAR) model of \citet{kalliovirta2015gaussian}
in detail. The technical challenges we face in analyzing the LR test
statistic are similar to those when testing for Markov switching and
mixture type regime switching. First, there are nuisance parameters
that are unidentified under the null hypothesis. This is the classical
\citet{davies1977hypothesis,davies1987hypothesis} type problem. Second,
under the null hypothesis, there are parameters on the boundary of
the permissible parameter space. Such problems (also allowing for
unidentified nuisance parameters under the null) are discussed in
\citet{andrews1999estimation,andrews2001testing}. Third, the Fisher
information matrix is (potentially) singular, preventing the use of
conventional second-order expansions of the log-likelihood to analyze
the LR test statistic. Such problems are discussed by \citet*{rotnitzky2000likelihood},
and suitable reparameterizations and higher-order expansions are needed
to analyze the LR statistic. A particular challenge in the present
paper is to deal with these three problems simultaneously. Similar
problems were faced by \citet{kasahara2015testing}, and inspired
by their work we consider a suitably reparameterized model, write
a higher-order expansion of the log-likelihood function as a quadratic
function of the new parameters, and then derive the asymptotic distribution
of the LR test statistic by slightly extending and adapting the arguments
of \citet{andrews1999estimation,andrews2001testing} and \citet{zhu2006asymptotics}
(who partially generalize results of Andrews). Our two examples demonstrate
that, compared to the mixture type regime switching considered by
\citet{kasahara2015testing}, observation-dependent regime switching
can either simplify or complicate the analysis of the LR test statistic. 

We contribute to the literature in several ways. (1) To the best of
our knowledge, apart from the unpublished PhD thesis of \citet{jeffries1998logistic},
we are the first to study testing for observation-dependent regime
switching using the LR test statistic and among the rather few to
allow for dependent observations. (2) We provide a general framework
to cover various forms of observation-dependent regime switching,
making our results potentially applicable to several models not explicitly
discussed in the present paper. (3) From a methodological perspective,
we slightly extend and adapt certain arguments of \citet{andrews1999estimation,andrews2001testing}
and \citet{zhu2006asymptotics}, which may be of independent interest. 

The rest of the paper is organized as follows. Section 2 reviews mixture
autoregressive models. Section 3 analyzes the LR test statistic for
testing a linear autoregressive model against a two-regime mixture
autoregressive model. Simulation-based critical values and a Monte
Carlo study are discussed in Section 4, and Section 5 concludes. Appendices
A\textendash C contain technical details and proofs. Supplementary
Appendices D\textendash E, available from the authors upon request,
contain further technical details omitted from the paper.

Finally, a few notational conventions are given. All vectors will
be treated as column vectors and, for the sake of uncluttered notation,
we shall write $x=(x_{1},\ldots,x_{n})$ for the (column) vector $x$
where the components $x_{i}$ may be either scalars or vectors (or
both). For any vector or matrix $x$, the Euclidean norm is denoted
by $\left\Vert x\right\Vert $. We let $X_{T\alpha}=o_{p\alpha}(1)$
and $X_{T\alpha}=O{}_{p\alpha}(1)$ stand for $\sup_{\alpha\in A}\left\Vert X_{T\alpha}\right\Vert =o_{p}(1)$
and $\sup_{\alpha\in A}\left\Vert X_{T\alpha}\right\Vert =O{}_{p}(1)$,
respectively, and $\lambda_{\min}(\cdot)$ and $\lambda_{\max}(\cdot)$
for the smallest and largest eigenvalue of the indicated matrix. 

\section{Mixture autoregressive models}

\subsection{General formulation}

Let $y_{t}$ ($t=1,2,\ldots$) be a real-valued time series of interest,
and let $\mathcal{F}_{t-1}=\sigma\left(y_{s},\textrm{ }s<t\right)$
denote the $\sigma$\textendash algebra generated by past $y_{t}$'s.
We use $P_{t-1}\left(\cdot\right)$ to signify the conditional probability
of the indicated event given $\mathcal{F}_{t-1}$. In the general
two component mixture autoregressive model we consider the $y_{t}$'s
are generated by 
\begin{equation}
y_{t}=s_{t}\biggl(\tilde{\phi}_{0}+\sum_{i=1}^{p}\tilde{\phi}_{i}y_{t-i}+\tilde{\sigma}_{1}\varepsilon_{t}\biggr)+(1-s_{t})\biggl(\tilde{\varphi}_{0}+\sum_{i=1}^{p}\tilde{\varphi}_{i}y_{t-i}+\tilde{\sigma}_{2}\varepsilon_{t}\biggr),\label{Model 1}
\end{equation}
where the parameters $\tilde{\sigma}_{1}$ and $\tilde{\sigma}_{2}$
are positive, and conditions required for the autoregressive parameters
$\tilde{\phi}_{i}$ and $\tilde{\varphi}_{i}$ ($i=1,\ldots,p$) will
be discussed later. Furthermore, $\varepsilon_{t}$ and $s_{t}$ are
(unobserved) stochastic processes which satisfy the following conditions:
(a) $\varepsilon_{t}$ is a sequence of independent standard normal
random variables such that $\varepsilon_{t}$ is independent of $\{y_{t-j},\ j>0\}$,
(b) $s_{t}$ is a sequence of Bernoulli random variables such that,
for each $t$, $P_{t-1}(s_{t}=1)=\alpha_{t}$ with $\alpha_{t}$ a
function of $\boldsymbol{y}_{t-1}=(y_{t-1},\ldots,y_{t-p})$, and
(c) conditional on $\mathcal{F}_{t-1}$, $s_{t}$ and $\varepsilon_{t}$
are independent. 

The conditional probabilities $\alpha_{t}$ and $1-\alpha_{t}$ ($=P_{t-1}(s_{t}=0)$)
are referred to as mixing weights. They can be thought of as (conditional)
probabilities that determine which one of the two autoregressive components
of the mixture generates the next observation $y_{t}$. In condition
(b) it is assumed that the mixing weight $\alpha_{t}$ (and hence
also the conditional distribution of $y_{t}$ given its past) only
depends on $p$ lags of $y_{t}$; allowing for more than $p$ lags
in the mixing weight would be possible at the cost of more complicated
notation.

We assume that of the original parameters $\tilde{\phi}=(\tilde{\phi}_{0},\tilde{\phi}_{1},\ldots,\tilde{\phi}_{p},\tilde{\sigma}_{1}^{2})$
and $\tilde{\varphi}=(\tilde{\varphi}_{0},\tilde{\varphi}_{1},\ldots,\tilde{\varphi}_{p},\tilde{\sigma}_{2}^{2})$
in the two regimes, $q_{1}$ parameters are a priori assumed the same
in both regimes and the remaining $q_{2}$ parameters are potentially
different in the two regimes (with $q_{1}+q_{2}=p+2$). For instance,
one may assume that $\tilde{\phi}_{0}$ and $\tilde{\varphi}_{0}$
are equal, or alternatively that $\tilde{\sigma}_{1}^{2}$ and $\tilde{\sigma}_{2}^{2}$
are equal. If such an assumption is plausible, taking it into account
when devising a test for regime switching will be advantageous (it
will lead to a test with better power). To this end, let $\beta$
be a $q_{1}\times1$ vector of common parameters, and let $\phi$
and $\varphi$ be $q_{2}\times1$ vectors of (potentially) different
parameters. Then, for some known $(p+2)$\textendash dimensional permutation
matrix $P$, $(\beta,\phi)=P\tilde{\phi}$ and $(\beta,\varphi)=P\tilde{\varphi}$.
For simplicity, we assume that $\beta$ and $\phi$ are variation-free,
requiring the autoregressive coefficients $\tilde{\phi}_{1},\ldots,\tilde{\phi}_{p}$
to be contained in either $\beta$ or $\phi$ (the same variation-freeness
is assumed of $\beta$ and $\varphi$). If there are no common coefficients
in the two regimes, the parameter $\beta$ can be dropped and $\phi=\tilde{\phi}$
and $\varphi=\tilde{\varphi}$. 

As for the mixing weight $\alpha_{t}$, in addition to past $y_{t}$'s
it depends on unknown parameters which may include components of the
parameter vector $(\beta,\phi,\varphi)$ and an additional parameter
$\alpha$ (scalar or vector). When this dependence needs to be emphasized
we use the notation $\alpha_{t}(\alpha,\beta,\phi,\varphi)$.

Using equation (\ref{Model 1}) and the conditions following it, the
conditional density function of $y_{t}$ given its past, $f(\cdot\mid\mathcal{F}_{t-1})$,
is obtained as 
\begin{equation}
f(y_{t}\mid\mathcal{F}_{t-1})=\alpha_{t}f_{t}(\beta,\phi)+(1-\alpha_{t})f_{t}(\beta,\varphi),\label{Model 2}
\end{equation}
where the notation $f_{t}(\beta,\phi)$ signifies the density function
of a (univariate) normal distribution with mean $\tilde{\phi}_{0}+\tilde{\phi}_{1}y_{t-1}+\cdots+\tilde{\phi}_{p}y_{t-p}$
and variance $\tilde{\sigma}_{1}^{2}$ evaluated at $y_{t}$, that
is,

\begin{equation}
f_{t}(\beta,\phi)=\frac{1}{\tilde{\sigma}_{1}}{\scriptstyle \mathscr{\mathfrak{N}}}\biggl(\frac{y_{t}-(\tilde{\phi}_{0}+\tilde{\phi}_{1}y_{t-1}+\cdots+\tilde{\phi}_{p}y_{t-p})}{\tilde{\sigma}_{1}}\biggr),\label{eq:f_t}
\end{equation}
with $\mathscr{{\scriptstyle \mathfrak{N}}}(u)=(2\bar{\pi})^{-1/2}\exp(-u^{2}/2)$
the density function of a standard normal random variable and $\bar{\pi}=3.14\ldots$
the number pi. The notation $f_{t}(\beta,\varphi)$ is defined similarly
by using the parameters $\tilde{\varphi}_{i}$ $\left(i=0,\ldots,p\right)$
and $\tilde{\sigma}_{2}^{2}$ instead of $\tilde{\phi}_{i}$ $\left(i=0,\ldots,p\right)$
and $\tilde{\sigma}_{1}^{2}$. Thus, the distribution of $y_{t}$
given its past is specified as a mixture of two normal densities with
time varying mixing weights $\alpha_{t}$ and $1-\alpha_{t}$.

Different mixture autoregressive models are obtained by different
specifications of the mixing weights (or in our case the single mixing
weight $\alpha_{t}$). In some of the proposed models more than two
mixture components are allowed but for reasons to be discussed below
we will not consider these extensions. If the mixing weights are assumed
constant over time the general mixture autoregressive model introduced
above reduces to (a two component version) of the MAR\ model studied
by \citet{wong2000mixture}. In the LMAR\ model of \citet{wong2001logistic},
a logistic transformation of the two mixing weights is assumed to
be a linear function of past observed variables. Related two-regime
mixture models with time-varying mixing weights have also been considered
by \citet{gourieroux2006stochastic}, \citet{dueker2007contemporaneous}
and \citet*{bec2008acr} whereas \citet{lanne2003modeling} and \citet{kalliovirta2015gaussian}
have considered mixture autoregressive models in which multiple regimes
are allowed. 

A common problem with the application of mixture autoregressive models
is determining the value of the (usually unknown) number of component
models or regimes. As discussed in the Introduction, several irregular
features make this problem difficult and these difficulties are encountered
even when the observations are a random sample from a mixture of (two)
normal distributions. To our knowledge the only solution presented
for mixture autoregressive models is provided for a simple first order
case with no intercept terms in the unpublished PhD thesis of \citet{jeffries1998logistic}.
As discussed in the recent papers by \citet{kasahara2012testing,kasahara2015testing}
and the references therein, some of the difficulties involved stem
from properties of the normal distribution. 

The difficulties referred to above also partly explain the complexity
of our testing problem and why we only consider test procedures that
can be used to test the null hypothesis that a two component mixture
autoregressive model reduces to a conventional linear autoregressive
model. Following the ideas in \citet{zhu2006asymptotics} and \citet{kasahara2012testing,kasahara2015testing},
we derive a LR test in the general set-up described above and apply
it to two particular cases, the LMAR model of \citet{wong2001logistic}
and the GMAR model of \citet{kalliovirta2015gaussian}. Next, we shall
discuss these two models in more detail.

\subsection{Two particular examples }

\paragraph{LMAR Example.}

\noindent The LMAR model of \citet{wong2001logistic} is defined by
specifying the mixing weight $\alpha_{t}$ as 
\[
\alpha_{t}^{L}=\alpha_{t}^{L}(\alpha)=\frac{\exp(\alpha_{0}+\alpha_{1}y_{t-1}+\cdots+\alpha_{r}y_{t-m})}{1+\exp(\alpha_{0}+\alpha_{1}y_{t-1}+\cdots+\alpha_{r}y_{t-m})},
\]
where the vector $\alpha=(\alpha_{0},\alpha_{1},\ldots,\alpha_{m})$
contains $m+1$ unknown parameters and the order $m$ ($1\leq m\leq p$)
is assumed known.

\paragraph{GMAR Example.}

\noindent In the GMAR model of \citet{kalliovirta2015gaussian} the
mixing weight is defined as 
\begin{equation}
\alpha_{t}^{G}=\alpha_{t}^{G}(\alpha,\tilde{\phi},\tilde{\varphi})=\frac{\alpha\mathsf{n}_{p}(\boldsymbol{y}_{t-1};\tilde{\phi})}{\alpha\mathsf{n}_{p}(\boldsymbol{y}_{t-1};\tilde{\phi})+(1-\alpha)\mathsf{n}_{p}(\boldsymbol{y}_{t-1};\tilde{\varphi})},\label{Mixing Weights}
\end{equation}
where $\alpha\in(0,1)$ is an unknown parameter and $\mathsf{n}_{p}(\boldsymbol{y}_{t-1};\cdot)$
denotes the density function of a particular $p$\textendash dimensional
($p\geq1$) normal distribution defined as follows. 

First, define the auxiliary Gaussian AR($p$) processes (cf. equation
(\ref{Model 1})) 
\[
\nu{}_{1,t}=\tilde{\phi}_{0}+\sum_{i=1}^{p}\tilde{\phi}_{i}\nu_{1,t-i}+\tilde{\sigma}_{1}\varepsilon_{t}\ \ \ \textrm{and}\ \ \ \nu_{2,t}=\tilde{\varphi}_{0}+\sum_{i=1}^{p}\tilde{\varphi}_{i}\nu_{2,t-i}+\tilde{\sigma}_{2}\varepsilon_{t},
\]
where the autoregressive coefficients are assumed to satisfy
\begin{equation}
\tilde{\phi}(z):=1-\sum_{i=1}^{p}\tilde{\phi}_{i}z^{i}\neq0\,\,\text{for}\,\,\left\vert z\right\vert \leq1\,\,\,\,\textrm{and}\,\,\,\,\tilde{\varphi}(z):=1-\sum_{i=1}^{p}\tilde{\varphi}_{i}z^{i}\neq0\,\,\text{for}\,\,\left\vert z\right\vert \leq1.\label{Stat. condition}
\end{equation}
This condition implies that the processes $\nu{}_{1,t}$ and $\nu{}_{2,t}$
are stationary and that each of the two component models in (\ref{Model 1})
satisfies the usual stationarity condition of the conventional linear
AR($p$) model. Now set $\mathbf{\boldsymbol{\nu}}_{m,t}=\left(\nu_{m,t},\ldots,\nu_{m,t-p+1}\right)$
and $\mathbf{1}_{p}=\left(1,\ldots,1\right)$ ($p\times1$), and let
$\mu_{m}\mathbf{1}_{p}$ and $\mathbf{\Gamma}_{m,p}$ signify the
mean vector and covariance matrix of $\mathbf{\boldsymbol{\nu}}_{m,t}$
($m=1,2$).\footnote{We have $\mu_{1}=\tilde{\phi}_{0}/\tilde{\phi}\left(1\right)$ and
$\mu_{2}=\tilde{\varphi}_{0}/\tilde{\varphi}\left(1\right)$, whereas
each of $\mathbf{\Gamma}_{m,p}$, $(m=1,2)$, has the familiar form
of being a $p\times p$ symmetric Toeplitz matrix with $\gamma_{m,0}=Cov[\nu_{m,t},\nu_{m,t}]$
along the main diagonal, and $\gamma_{m,i}=Cov[\nu_{m,t},\nu{}_{m,t-i}]$,
$i=1,\ldots,p-1$, on the diagonals above and below the main diagonal.
Similarly to $\mu_{1}$ and $\mu_{2}$ the elements of the covariance
matrices $\mathbf{\Gamma}_{1,p}$ and $\mathbf{\Gamma}_{2,p}$ are
treated as functions of the parameters $\tilde{\phi}$ and $\tilde{\varphi}$,
respectively (for details of this dependence, see \citet[eqn. (2.1.39)]{lutkepohl2005new}).} The random vector $\mathbf{\boldsymbol{\nu}}_{1,t}$ follows the
$p$\textendash dimensional multivariate normal distribution with
density 
\begin{equation}
\mathsf{n}_{p}(\mathbf{\boldsymbol{\nu}}_{1,t};\tilde{\phi})=\left(2\bar{\pi}\right)^{-p/2}\det(\mathbf{\Gamma}_{1,p}\mathbf{)}^{-1/2}\exp\left\{ -\tfrac{1}{2}\left(\mathbf{\boldsymbol{\nu}}_{1,t}-\mu_{1}\mathbf{1}_{p}\right)^{\prime}\mathbf{\Gamma}_{1,p}^{-1}\left(\mathbf{\boldsymbol{\nu}}_{1,t}-\mu_{1}\mathbf{1}_{p}\right)\right\} ,\label{Normal p}
\end{equation}
and the density of $\mathbf{\boldsymbol{\nu}}_{2,t}$, denoted by
$\mathsf{n}_{p}(\mathbf{\boldsymbol{\nu}}_{2,t};\tilde{\varphi})$,
is defined similarly. Equation (\ref{Model 1}) and conditions (\ref{Mixing Weights})\textendash (\ref{Normal p})
define the (two component) GMAR model (condition (\ref{Stat. condition})
is part of the definition of the model because it is used to define
the mixing weights).

\section{Test procedure}

We now consider a test procedure of the null hypothesis that a two
component mixture autoregressive model reduces to a conventional linear
autoregressive model.

\subsection{The null hypothesis and the LR test statistic}

We denote the conditional density function corresponding to the unrestricted
model as (see (\ref{Model 2})) 
\[
f_{2,t}(\alpha,\beta,\phi,\varphi):=f_{2}(y_{t}\mid\boldsymbol{y}_{t-1};\alpha,\beta,\phi,\varphi):=\alpha_{t}(\alpha,\beta,\phi,\varphi)f_{t}(\beta,\phi)+\left(1-\alpha_{t}(\alpha,\beta,\phi,\varphi)\right)f_{t}(\beta,\varphi),
\]
where we now make the dependence of the mixing weight on the parameters
explicit. With this notation the log-likelihood function of the model
based on a sample $(y_{-p+1},\ldots,y_{T})$ (and conditional on the
initial values $(y_{-p+1},\ldots,y_{0})$) is $L_{T}(\alpha,\beta,\phi,\varphi)=\sum_{t=1}^{T}l_{t}(\alpha,\beta,\phi,\varphi)$
where 
\[
l_{t}(\alpha,\beta,\phi,\varphi)=\log[f_{2,t}(\alpha,\beta,\phi,\varphi)]=\log\left[\alpha_{t}(\alpha,\beta,\phi,\varphi)f_{t}(\beta,\phi)+\left(1-\alpha_{t}(\alpha,\beta,\phi,\varphi)\right)f_{t}(\beta,\varphi)\right].
\]
The following assumption provides conditions on the data generation
process, the parameter space of $\left(\alpha,\beta,\phi,\varphi\right)$,
and the mixing weight $\alpha_{t}\left(\alpha,\beta,\phi,\varphi\right)$. 
\begin{assumption}
\label{assu:DGPParSpaceMixWeight}~\vspace{-5pt}

\begin{description}
\item [{(i)}] The $y_{t}$'s are generated by a stationary linear Gaussian
AR($p$) model with (the true but unknown) parameter value $\tilde{\phi}^{\ast}$
an interior point of $\tilde{\Phi}$, a compact subset of $\{\tilde{\phi}=(\tilde{\phi}_{0},\tilde{\phi}_{1},\ldots,\tilde{\phi}_{p},\tilde{\sigma}^{2})\in\mathbb{R}^{p+2}:\tilde{\phi}_{0}\in\mathbb{R};\,1-\sum_{i=1}^{p}\tilde{\phi}_{i}z^{i}\neq0\text{ \ for }\left\vert z\right\vert \leq1;\,\tilde{\sigma}^{2}\in(0,\infty)\}$.
\item [{(ii)}] The parameter space of $\left(\alpha,\beta,\phi,\varphi\right)$
is $A\times B\times\Phi\times\Phi$, where $A$ is a compact subset
of $\mathbb{R}^{a}$ and $B$ and $\Phi$ are those compact subsets
of $\mathbb{R}^{q_{1}}$ and $\mathbb{R}^{q_{2}}$, respectively,
that satisfy $(\beta,\phi)\in B\times\Phi$ if and only if $P^{-1}(\beta,\phi)\in\tilde{\Phi}$
(here $P$ is as in the third paragraph of Section 2.1).
\item [{(iii)}] For all $t$ and all $\left(\alpha,\beta,\phi,\varphi\right)\in A\times B\times\Phi\times\Phi$,
the mixing weight $\alpha_{t}\left(\alpha,\beta,\phi,\varphi\right)$,
is $\sigma(\boldsymbol{y}{}_{t-1})$\textendash measurable (with $\sigma(\boldsymbol{y}{}_{t-1})$
denoting the $\sigma$\textendash algebra generated by $\boldsymbol{y}{}_{t-1}$)
and satisfies $\alpha_{t}\left(\alpha,\beta,\phi,\varphi\right)\in(0,1)$. 
\end{description}
\end{assumption}
As our interest is to study the asymptotic null distribution of the
LR test statistic, Assumption 1(i) requires the data to be generated
by a stationary linear Gaussian AR($p$) model. Assuming a compact
parameter space in Assumptions 1(i) and (ii) is a standard requirement
which facilitates proofs. Assumption 1(ii) accommodates to the main
cases of interest, namely $\beta=\tilde{\phi}_{0}$, $\beta=(\tilde{\phi}_{1},\ldots,\tilde{\phi}_{p})$,
$\beta=\tilde{\sigma}^{2}$, or any combination of these.\footnote{Note that assuming the autoregressive parameters $\phi$ and $\varphi$
to have a common parameter space is made for simplicity and could
be relaxed; for an example where such a relaxation would be needed,
see the ACR model of \citet{bec2008acr}.} 

Assumption 1(iii) implies that our two-component mixture autoregressive
model reduces to a linear autoregression only when $\phi=\varphi$,
regardless of the values of $\alpha\in A$ and $\beta\in B$. The
null hypothesis to be tested is therefore $\phi=\varphi$ and the
alternative is $\phi\neq\varphi$ or, more precisely, 
\[
H_{0}:(\phi,\varphi)\in(\Phi\times\Phi)^{0},\;\alpha\in A,\,\beta\in B\qquad\textrm{vs.}\qquad H_{1}:(\phi,\varphi)\in(\Phi\times\Phi)\setminus(\Phi\times\Phi)^{0},\;\alpha\in A,\,\beta\in B,
\]
where 
\[
(\Phi\times\Phi)^{0}=\{(\phi,\varphi)\in\Phi\times\Phi:\phi=\varphi\}.
\]
Note that under the null hypothesis the parameter $\alpha$ vanishes
from the likelihood function and is therefore unidentified. 

Let $f_{t}^{0}(\tilde{\phi}):=f^{0}(y_{t}\mid\boldsymbol{y}_{t-1};\tilde{\phi})$
and $L_{T}^{0}(\tilde{\phi})$ denote the conditional density and
log-likelihood corresponding to the restricted model, that is, 
\[
f_{t}^{0}(\tilde{\phi})=f_{2,t}(\alpha,\beta,\phi,\phi)=f_{t}(\tilde{\phi})\,\,\,\,\,\text{and}\,\,\,\,\,L_{T}^{0}(\tilde{\phi})=\sum_{t=1}^{T}l_{t}^{0}(\tilde{\phi})\,\,\,\,\,\text{with}\,\,\,\,\,l_{t}^{0}(\tilde{\phi})=\log[f_{t}(\tilde{\phi})]
\]
(here the superscript $0$ refers to the model restricted by the null
hypothesis). Note that these quantities are obtained from a linear
Gaussian AR($p$) model. As $f_{2,t}(\alpha,\beta^{*},\phi^{\ast},\phi^{\ast})=f_{t}(\tilde{\phi}^{\ast})$
for any $\alpha\in A$, in the unrestricted model the parameter vector
$(\alpha,\beta^{*},\phi^{\ast},\phi^{\ast})$ corresponds to the true
model for any $\alpha\in A$.

As already indicated, Assumption 1(iii) implies that the restriction
$\phi=\varphi$ is the only possibility to formulate the null hypothesis.
However, this is not necessarily the case if (against Assumption 1(iii))
the mixing weight $\alpha_{t}\left(\alpha,\beta,\phi,\varphi\right)$
were allowed to take the boundary values zero and one. Of our two
examples this would be possible for the GMAR model of \citet{kalliovirta2015gaussian}
but not for the LMAR model of \citet{wong2001logistic}. In the GMAR
model $\alpha_{t}\left(\alpha,\beta,\phi,\varphi\right)$ takes the
boundary values zero and one when the parameter $\alpha$ takes these
values (see Section 2.2). In both cases a linear autoregression results
and either the parameter $\phi$ or $\varphi$ is unidentified (see
(\ref{Model 2})) (the MAR model of \citet{wong2000mixture} provides
a similar example). It would be possible to obtain tests for the GMAR
model by using the null hypotheses which specifies $\alpha=0$ or
$\alpha=1$. However, as in \citet{kasahara2012testing} (see also
\citet{kasahara2015testing}), this approach would require rather
restrictive assumptions and would also lead to very complicated derivations.\footnote{See, for instance, the remarks following Proposition 5 in \citet{kasahara2012testing}
or property (a) on p. 1633 of \citet{kasahara2015testing}.} Therefore, we will not consider this option.

As the parameter $\alpha$ is unidentified under the null hypothesis,
the appropriate likelihood ratio type test statistic is 
\[
LR_{T}=\sup_{\alpha\in A}LR_{T}\left(\alpha\right),
\]
where, for each fixed $\alpha\in A$, 
\[
LR_{T}\left(\alpha\right)=2[\sup_{(\beta,\phi,\varphi)\in B\times\Phi\times\Phi}L_{T}(\alpha,\beta,\phi,\varphi)-\sup_{\tilde{\phi}\in\tilde{\Phi}}L_{T}^{0}(\tilde{\phi})].
\]
To obtain an operational test statistic let, for each fixed $\alpha\in A$,
$(\hat{\beta}_{T\alpha},\hat{\phi}_{T\alpha},\hat{\varphi}_{T\alpha})$
denote an (approximate) unrestricted maximum likelihood (ML) estimator
of the parameter vector $\left(\beta,\phi,\varphi\right)$. We make
the following assumption.
\begin{assumption}
\label{assu:unrestricted_est}The unrestricted ML estimator satisfies
the following conditions:~

\begin{description}
\item [{(i)}] $L_{T}(\alpha,\hat{\beta}_{T\alpha},\hat{\phi}_{T\alpha},\hat{\varphi}_{T\alpha})=\sup\nolimits _{(\beta,\phi,\varphi)\in B\times\Phi\times\Phi}L_{T}(\alpha,\beta,\phi,\varphi)+o_{p\alpha}(1)$,
\item [{(ii)}] $(\hat{\beta}_{T\alpha},\hat{\phi}_{T\alpha},\hat{\varphi}_{T\alpha})=(\beta^{*},\phi^{\ast},\phi^{\ast})+o_{p\alpha}(1)$.
\end{description}
\end{assumption}
Assumption 2(i) means that $(\hat{\beta}_{T\alpha},\hat{\phi}_{T\alpha},\hat{\varphi}_{T\alpha})$
is assumed to maximize the likelihood function only asymptotically.
This assumption is technical and made for ease of exposition (see
\citet{andrews1999estimation} and \citet{zhu2006asymptotics} for
similar assumptions in related problems). Assumption 2(ii) is a high
level condition on (uniform) consistency of the ML estimator and is
analogous to Assumption 1 of \citet{andrews2001testing}. It has to
be verified on a case by case basis (this is exemplified below for
the LMAR model and GMAR model). 

As for the term $\sup_{\tilde{\phi}\in\tilde{\Phi}}L_{T}^{0}(\tilde{\phi})$
in the LR test statistic, note that $L_{T}^{0}(\tilde{\phi})$ is
the (conditional) log-likelihood function of a linear Gaussian AR($p$)
model. Let $\hat{\tilde{\phi}}_{T}$ denote an (approximate) maximum
likelihood estimator of the parameters of a linear Gaussian AR($p$)
model, that is, $\hat{\tilde{\phi}}_{T}$ satisfies\footnote{Note that the parameter space for $\tilde{\phi}$ is the compact set
$\tilde{\Phi}$ and not the entire stationarity region of a (causal)
linear AR($p$) model. Asymptotically, also the OLS estimator can
be used. }
\[
L_{T}^{0}(\hat{\tilde{\phi}}_{T})=\sup_{\tilde{\phi}\in\tilde{\Phi}}L_{T}^{0}(\tilde{\phi})+o_{p}(1)\,\,\,\textrm{and}\,\,\,\hat{\tilde{\phi}}_{T}=\tilde{\phi}^{\ast}+o_{p}(1).
\]
Noting that $L_{T}(\alpha,\beta^{*},\phi^{\ast},\phi^{\ast})=L_{T}^{0}(\tilde{\phi}^{\ast})$
for any $\alpha$ now allows us to write $LR_{T}\left(\alpha\right)$
as
\begin{equation}
LR_{T}\left(\alpha\right)=2[L_{T}(\alpha,\hat{\beta}_{T\alpha},\hat{\phi}_{T\alpha},\hat{\varphi}_{T\alpha})-L_{T}(\alpha,\beta^{*},\phi^{\ast},\phi^{\ast})]-2[L_{T}^{0}(\hat{\tilde{\phi}}_{T})-L_{T}^{0}(\tilde{\phi}^{\ast})]+o_{p\alpha}(1).\label{eq:LR_1}
\end{equation}
The analysis of the second term on the right hand side is standard
while dealing with the first term is more demanding requiring a substantial
amount of preparation.

\subsubsection{Examples (continued)}

\paragraph{LMAR Example. }

In the LMAR example, we assume there are no common parameters in the
two regimes so that the parameter $\beta$ is omitted, $\phi=\tilde{\phi}$,
$\varphi=\tilde{\varphi}$, $q_{1}=0$, and $q_{2}=p+2$. To satisfy
conditions (ii) and (iii) in Assumption 1, $A$ can be any compact
subset of $\{(\alpha_{0},\alpha_{1},\ldots,\alpha_{m})\in\mathbb{R}^{m+1}:(\alpha_{1},\ldots,\alpha_{m})\neq(0,\ldots,0)\}$
where $1\leq m\leq p$. This ensures that the mixing weight $\alpha_{t}^{L}$
is not equal to a constant. For the verification of Assumption 2(ii),
see Appendix B.

When $m=0$, the mixing weight $\alpha_{t}^{L}$ (and hence $1-\alpha_{t}^{L}$)
is constant and the LMAR model reduces to the MAR model of \citet{wong2000mixture}.
In this special case our testing problem requires different and more
complicated analyses than in the `real' LMAR case where $m\geq1$
and $(\alpha_{1},\ldots,\alpha_{r})\neq(0,\ldots,0)$ (we shall discuss
this point more later). Therefore, the conditions $m\geq1$ and $(\alpha_{1},\ldots,\alpha_{m})\neq(0,\ldots,0)$
will be assumed in the sequel. A similar restriction is made by \citet{jeffries1998logistic}
in his (first-order) logistic mixture autoregressive model to facilitate
the derivation of the LR test (see the hypotheses at the end of p.
95 and the following discussion, as well as the end of p. 110). 

\paragraph{GMAR Example. }

\noindent The GMAR model exemplifies the setting with common coefficients
by assuming that the intercept terms in the two regimes are the same
(note that this still allows for different means in the two regimes).
As will be discussed in more detail in Section 3.3.1, this assumption
is partly due to the fact that otherwise the derivation of the LR
test would become extremely complicated. Hence, in this example $\beta=\tilde{\phi}_{0}\,(=\tilde{\varphi}_{0})$,
$\phi=(\tilde{\phi}_{1},\ldots,\tilde{\phi}_{p},\tilde{\sigma}_{1}^{2})$,
$\varphi=(\tilde{\varphi}_{1},\ldots,\tilde{\varphi}_{p},\tilde{\sigma}_{2}^{2})$,
$q_{1}=1$, and $q_{2}=p+1$. To satisfy Assumptions 1(ii) and (iii)
the parameter space $A$ of $\alpha$ can be any compact and convex
subset of $(0,1)$ (this also rules out the possibility that $\alpha=0$
or $\alpha=1$ discussed after Assumption 1). For the verification
of Assumption 2(ii), see Appendix C.

It may be worth noting that there are cases where the mixing weight
$\alpha_{t}^{G}$ is time invariant and equals $\alpha$. If this
happens the GMAR model reduces to the MAR model of \citet{wong2000mixture}.\footnote{An example is when $p=1$, $\tilde{\phi}_{0}=\tilde{\varphi}_{0}=0$,
and $\tilde{\sigma}_{1}^{2}/(1-\tilde{\phi}_{1}^{2})=\tilde{\sigma}_{2}^{2}/(1-\tilde{\varphi}_{1}^{2})$
where the last equality can hold even if $(\tilde{\phi},\tilde{\sigma}_{1}^{2})$
is different from $(\tilde{\varphi},\tilde{\sigma}_{2}^{2})$.} However, unlike in the case of the LMAR model this fact does not
complicate the derivation of our test. The reason seems to be that
in the GMAR model the reduction occurs only for certain values of
the parameters $\tilde{\phi}$ and $\tilde{\varphi}$ whereas in the
LMAR model it occurs for all values of $\tilde{\phi}$ and $\tilde{\varphi}$.

\subsection{Reparameterized model}

In standard testing problems the derivation of the asymptotic distribution
of the LR test would rely on a quadratic expansion of the log-likelihood
function $L_{T}(\alpha,\beta,\phi,\varphi)=\sum_{t=1}^{T}l_{t}(\alpha,\beta,\phi,\varphi)$;
when the parameter $\alpha$ is not identified under the null hypothesis,
the relevant derivatives in this expansion would be with respect to
$(\beta,\phi,\varphi)$ for fixed values of $\alpha\in A$. In problems
with a singular information matrix it turns out to be convenient to
follow \citet{rotnitzky2000likelihood} and \citet{kasahara2012testing,kasahara2015testing}
and employ an appropriately reparameterized model. 

The employed reparameterization is model specific and aims to have
two conveniences. First, it transforms the null hypothesis $\phi=\varphi$
into a point null hypothesis where some components of the parameter
vector are restricted to zero and the rest are left unrestricted.
Second, and more importantly, it simplifies derivations in cases where
the conventional quadratic expansion of the log-likelihood function
breaks down because, under the null hypothesis, the scores of the
parameters $(\beta,\phi,\varphi)$ are linearly dependent and, consequently,
the (Fisher) information matrix is singular. As will be seen later,
this is the case for the GMAR model of \citet{kalliovirta2015gaussian}
but not for the LMAR model of \citet{wong2001logistic}. 

General requirements for the reparameterization are described in the
following assumption. Only the parameters restricted by the null hypothesis,
$\phi$ and $\varphi$, are reparameterized. The examples in this
and the following subsection illustrate how the reparameterization
could be chosen. 

\pagebreak{}
\begin{assumption}
\label{assu:pi-repam}~\vspace{-5pt}

\begin{description}
\item [{(i)}] For every $\alpha\in A$, the mapping $(\pi,\varpi)=\boldsymbol{\pi}_{\alpha}(\phi,\varphi)$
from $\Phi\times\Phi$ to $\Pi_{\alpha}$ is one-to-one with $\boldsymbol{\pi}_{\alpha}$
and $\boldsymbol{\pi}_{\alpha}^{-1}$ continuous.
\item [{(ii)}] For every $\alpha\in A$, $\boldsymbol{\pi}_{\alpha}((\Phi\times\Phi)^{0})=\Phi\times\{0\}$
and $\boldsymbol{\pi}_{\alpha}(\phi^{\ast},\phi^{\ast})=(\pi^{\ast},0):=(\phi^{\ast},0)$.
\item [{(iii)}] $(\hat{\beta}_{T\alpha},\hat{\pi}_{T\alpha},\hat{\varpi}_{T\alpha})=(\beta^{*},\pi^{\ast},0)+o_{p\alpha}(1)$,
where $(\hat{\pi}_{T\alpha},\hat{\varpi}_{T\alpha})\coloneqq\boldsymbol{\pi}_{\alpha}(\hat{\phi}_{T\alpha},\hat{\varphi}_{T\alpha})$.
\end{description}
\end{assumption}
We sometimes refer to the reparameterization described in Assumption
\ref{assu:pi-repam} as the `$\pi$-parameterization' and the original
reparameterization as the `$\phi$-parameterization'. Note that the
transformed parameters $\pi$ and $\varpi$ generally depend on $\alpha$
but, for brevity, we suppress this dependence from the notation. The
parameter space of $(\pi,\varpi)$ also depends on $\alpha$ and is
given, for any $\alpha\in A$, by
\[
\Pi_{\alpha}=\{(\pi,\varpi)\in\mathbf{\mathbb{R}}^{2q_{2}}:(\pi,\varpi)=\boldsymbol{\pi}_{\alpha}(\phi,\varphi)\;\;\textrm{for some}\;\;(\phi,\varphi)\in\Phi\times\Phi\}.
\]
By Assumption \ref{assu:pi-repam}(ii), the null hypothesis $\phi=\varphi$
can be equivalently written as $\varpi=0$ or, more precisely, as
\[
H_{0}:\pi\in\Phi,\;\varpi=0,\;\alpha\in A,\,\beta\in B\qquad\textrm{vs.}\qquad H_{1}:(\pi,\varpi)\in\Pi_{\alpha}\setminus(\Phi\times\{0\}),\;\alpha\in A,\,\beta\in B.
\]
Note that under $H_{0}$, the parameters $\beta$ and $\pi$ are identified,
but $\alpha$ is not. As for Assumption \ref{assu:pi-repam}(iii),
it is a high level condition similar to Assumption \ref{assu:unrestricted_est}(ii)
from which it can be derived with appropriate additional assumptions.
A simple Lipschitz condition similar to \citet[Assumption SE-1(b)]{andrews1992generic},
given in Lemma \ref{lem:LipCondForUnifCons} in Appendix A, is one
possibility.

To develop further notation, partition $\boldsymbol{\pi}_{\alpha}^{-1}(\pi,\varpi)$
into two $q_{2}$\textendash dimensional components as $\boldsymbol{\pi}_{\alpha}^{-1}(\pi,\varpi)=(\boldsymbol{\pi}_{\alpha,1}^{-1}(\pi,\varpi),\boldsymbol{\pi}_{\alpha,2}^{-1}(\pi,\varpi))$,
and define
\[
\hspace{-10pt}f_{2,t}^{\pi}(\alpha,\beta,\pi,\varpi):=f_{2}(y_{t}\mid\boldsymbol{y}_{t-1};\alpha,\beta,\phi,\varphi)=\alpha_{t}^{\pi}(\alpha,\beta,\pi,\varpi)f_{t}(\beta,\boldsymbol{\pi}_{\alpha,1}^{-1}(\pi,\varpi))+(1-\alpha_{t}^{\pi}(\alpha,\beta,\pi,\varpi))f_{t}(\beta,\boldsymbol{\pi}_{\alpha,2}^{-1}(\pi,\varpi)),
\]
where $\alpha_{t}^{\pi}(\alpha,\beta,\pi,\varpi)=\alpha_{t}(\alpha,\beta,\boldsymbol{\pi}_{\alpha,1}^{-1}(\pi,\varpi),\boldsymbol{\pi}_{\alpha,2}^{-1}(\pi,\varpi))$
and the function $f_{t}(\cdot)$ is as in (\ref{Model 2}). The log-likelihood
function of the reparameterized model can now be expressed as 
\begin{equation}
L_{T}^{\pi}(\alpha,\beta,\pi,\varpi)=\sum_{t=1}^{T}l_{t}^{\pi}(\alpha,\beta,\pi,\varpi),\label{eq:LogLikPi}
\end{equation}
where $l_{t}^{\pi}(\alpha,\beta,\pi,\varpi)=\log[f_{2,t}^{\pi}(\alpha,\beta,\pi,\varpi)]$,
and in the $\pi$\textendash parameterization equation (\ref{eq:LR_1})
reads as
\begin{equation}
LR_{T}\left(\alpha\right)=2[L_{T}^{\pi}(\alpha,\hat{\beta}_{T\alpha},\hat{\pi}_{T\alpha},\hat{\varpi}_{T\alpha})-L_{T}^{\pi}(\alpha,\beta^{*},\phi^{\ast},0)]-2[L_{T}^{0}(\hat{\tilde{\phi}}_{T})-L_{T}^{0}(\tilde{\phi}^{\ast})]+o_{p\alpha}(1).\label{eq:LR_pi-repam}
\end{equation}

\subsubsection{Examples (continued)}

\paragraph{LMAR Example. }

The reparameterization we employ in the LMAR model is 
\[
(\pi,\varpi)=\boldsymbol{\pi}_{\alpha}(\phi,\varphi)=(\phi,\phi-\varphi)\;\;\textrm{so that}\;\;(\phi,\varphi)=\boldsymbol{\pi}_{\alpha}^{-1}(\pi,\varpi)=(\pi,\pi-\varpi).
\]
Note that in this case the reparameterization (via $\boldsymbol{\pi}_{\alpha}(\cdot,\cdot)$)
does not depend on $\alpha$, and the same is true for the parameter
space of $(\pi,\varpi)$. Verification of Assumption \ref{assu:pi-repam}
is straightforward using Lemma \ref{lem:LipCondForUnifCons} (for
details, see Appendix B). In the LMAR case, the only benefit of the
reparameterization is to transform the null hypothesis into a point
null hypothesis.

\paragraph{GMAR Example. }

\noindent In the GMAR model our reparameterization is obtained by
setting, for any fixed $\alpha\in A$, 
\[
(\pi,\varpi)=\boldsymbol{\pi}_{\alpha}(\phi,\varphi)=(\alpha\phi+(1-\alpha)\varphi,\phi-\varphi)\;\;\textrm{so that}\;\;(\phi,\varphi)=\boldsymbol{\pi}_{\alpha}^{-1}(\pi,\varpi)=(\pi+(1-\alpha)\varpi,\pi-\alpha\varpi).
\]
Verification of Assumption \ref{assu:pi-repam} is again straightforward
using Lemma \ref{lem:LipCondForUnifCons} (for details, see Appendix
C). In the GMAR model, simplifying the null hypothesis is not the
only benefit of the reparameterization, as will be discussed next. 

As discussed before Assumption \ref{assu:pi-repam}, the relevant
derivatives when expanding $L_{T}(\alpha,\beta,\phi,\varphi)$ are
with respect to $(\beta,\phi,\varphi)$ and, in the GMAR case, these
derivatives are linearly dependent under the null hypothesis. To see
this and how the reparameterization affects this feature, note first
that straightforward differentiation yields
\[
\nabla_{(\beta,\phi,\varphi)}l_{t}(\alpha,\beta,\phi,\phi)=\left(\alpha\frac{\nabla_{\beta}f_{t}(\beta,\phi)}{f_{t}(\beta,\phi)},\alpha\frac{\nabla_{\phi}f_{t}(\beta,\phi)}{f_{t}(\beta,\phi)},(1-\alpha)\frac{\nabla_{\phi}f_{t}(\beta,\phi)}{f_{t}(\beta,\phi)}\right),
\]
where the null hypothesis $\phi=\varphi$ is imposed and $\nabla$
denotes differentiation with respect to the indicated parameters.
As $(f_{t}(\beta,\phi))^{-1}\nabla_{(\beta,\phi)}f_{t}(\beta,\phi)$
is the score vector obtained from a linear Gaussian AR($p$) model,
it is clear that the covariance matrix of the $(2p+3)$\textendash dimensional
vector $\nabla_{(\beta,\phi,\varphi)}l_{t}(\alpha,\beta,\phi,\phi)$,
and hence the (Fisher) information matrix, is singular with rank $p+2$.
In contrast to the above, in the $\pi$\textendash parameterization
the score vector is given by (see Supplementary Appendix C)
\[
\nabla_{(\beta,\pi,\varpi)}l_{t}^{\pi}(\alpha,\beta,\pi,0)=\left(\frac{\nabla_{\beta}f_{t}(\beta,\pi)}{f_{t}(\beta,\pi)},\frac{\nabla_{\pi}f_{t}(\beta,\pi)}{f_{t}(\beta,\pi)},\mathbf{0}_{p+1}\right)
\]
when the null hypothesis $\varpi=0$ is imposed. Now the score of
$\varpi$ is identically zero so that the reparameterization simplifies
linear dependencies of the scores which turns out to be very useful
in subsequent asymptotic analyses.

\subsection{Quadratic expansion of the (reparameterized) log-likelihood function}

As alluded to above, in standard testing problems the asymptotic analysis
of a LR test statistic is based on a second order Taylor expansion
of the (average) log-likelihood function around the true parameter
value. An essential assumption here is positive definiteness of the
(limiting) information matrix but, as illustrated in the previous
section, this assumption does not necessarily hold in our testing
problem due to linear dependencies among the derivatives of the log-likelihood
function. As in \citet{rotnitzky2000likelihood}, \citet{zhu2006asymptotics},
and \citet{kasahara2012testing,kasahara2015testing}, we therefore
consider a quadratic expansion of the log-likelihood function that
is not based on a second order Taylor expansion but (possibly) on
a higher order Taylor expansion. The need for higher-order derivatives
is illustrated by the GMAR example: as the score of $\varpi$ is identically
zero, the second derivative (which turns out to be linearly independent
of the score of $(\beta,\pi)$) now provides the first (nontrivial)
local approximation for $\varpi$. 

The following assumption ensures that the (reparameterized) log-likelihood
function (\ref{eq:LogLikPi}) is (at least) twice continuously differentiable.
\begin{assumption}
\label{assu:cont_differentiability}For some integer $k\geq2$, and
for every fixed $\alpha\in A$, the functions $\alpha_{t}(\alpha,\beta,\phi,\varphi)$
and $\boldsymbol{\pi}_{\alpha}^{-1}(\pi,\varpi)$ are $k$ times continuously
differentiable (with respect to $(\beta,\phi,\varphi)$ and $(\pi,\varpi)$
in the interior of $B\times\Phi\times\Phi$ and $\Pi_{\alpha}$, respectively).
\end{assumption}
In our general framework the reparameterized log-likelihood function
is assumed to have, for each $\alpha\in A$, a quadratic expansion
in a transformed parameter vector $\boldsymbol{\theta}(\alpha,\beta,\pi,\varpi)$
around $(\beta^{*},\pi^{\ast},0)$ given by
\begin{multline}
L_{T}^{\pi}(\alpha,\beta,\pi,\varpi)-L_{T}^{\pi}(\alpha,\beta^{*},\pi^{*},0)\\
=(T^{-1/2}S_{T\alpha})'[T^{1/2}\boldsymbol{\theta}(\alpha,\beta,\pi,\varpi)]-\frac{1}{2}[T^{1/2}\boldsymbol{\theta}(\alpha,\beta,\pi,\varpi)]^{\prime}\mathcal{I}_{\alpha}[T^{1/2}\boldsymbol{\theta}(\alpha,\beta,\pi,\varpi)]+R_{T}(\alpha,\beta,\pi,\varpi).\label{eq:Quadr-Exp_1}
\end{multline}
To illustrate this expansion, suppose the information matrix is positive
definite so that the quantities on the right hand side are (typically)
based on a second order Taylor expansion with $S_{T\alpha}$ and $\mathcal{I}_{\alpha}$
functions of $(\alpha,\beta^{*},\pi^{*},0)$. As already mentioned,
this is the case for the LMAR model where (the following will be justified
shortly) the parameter $\boldsymbol{\theta}(\alpha,\beta,\pi,\varpi)$
is independent of $\alpha$ and given by $(\pi-\pi^{\ast},\varpi)$
and, for each $\alpha\in A$, $S_{T\alpha}$ is the score vector,
$\mathcal{I}_{\alpha}$ is the (positive definite) Fisher information
matrix, and $R_{T}(\alpha,\beta,\pi,\varpi)$ is a remainder term.
As the notation indicates, these three terms depend on $\alpha$,
and in general they may involve partial derivatives of the log-likelihood
function of order higher than two (this is the case for the GMAR model,
as will be demonstrated shortly). Then it may also get more complicated
to find the reparameterization of the previous subsection and the
transformed parameter vector $\boldsymbol{\theta}(\alpha,\beta,\pi,\varpi)$,
as the examples of Kasahara and Shimotsu (2015, 2017) and the discussion
on the GMAR model below show; one possibility is to consider the iterative
procedure discussed by \citet[Sections 4.4 and 4.5]{rotnitzky2000likelihood}
(for a recent illuminating illustration of this approach, see \citet{hallin2014skew}). 

Our next assumption provides further details on expansion (\ref{eq:Quadr-Exp_1}).
We use $\Rightarrow$ to signify weak convergence of a sequence of
stochastic processes on a function space. In the assumption below,
the weak convergence of interest is that of the process $S_{T\alpha}$
(indexed by $\alpha\in A$) to a process $S_{\alpha}$. The two function
spaces relevant in this paper are $\mathcal{B}(A,\mathbb{R}^{k})$
and $\mathcal{C}(A,\mathbb{R}^{k})$, the former is the space of all
$\mathbb{R}^{k}$\textendash valued bounded functions defined on (the
compact set) $A$ equipped with the uniform metric ($d(x,y)=\sup_{a\in A}\left\Vert x(a)-y(a)\right\Vert $),
and the latter is the same but with the continuity of the functions
(with respect to $\alpha\in A$) also assumed.
\begin{assumption}
\label{assu:quadexp}For each $\alpha\in A$, the log-likelihood function
$L_{T}^{\pi}(\alpha,\beta,\pi,\varpi)$ has a quadratic expansion
given in (\ref{eq:Quadr-Exp_1}), where

\begin{description}
\item [{(i)}] for each $\alpha\in A$, $\boldsymbol{\theta}(\alpha,\beta,\pi,\varpi)$
is a mapping from $B\times\Pi_{\alpha}$ to\\
$\Theta_{\alpha}=\{\boldsymbol{\theta}\in\mathbb{R}^{r}:\boldsymbol{\theta}=\boldsymbol{\theta}(\alpha,\beta,\pi,\varpi)\text{ for some }(\beta,\pi,\varpi)\in B\times\Pi_{\alpha}\}$
such that (a) $\boldsymbol{\theta}(\alpha,\beta^{*},\pi^{*},0)=0$
and (b) for all $\epsilon>0$, $\inf_{\alpha\in A}\inf$$_{(\beta,\pi,\varpi)\in B\times\Pi_{\alpha}:\left\Vert (\beta,\pi,\varpi)-(\beta^{*},\pi^{\ast},0)\right\Vert \geq\epsilon}\left\Vert \boldsymbol{\theta}(\alpha,\beta,\pi,\varpi)\right\Vert \geq\delta_{\epsilon}$
for some $\delta_{\epsilon}>0$.
\item [{(ii)}] $S_{T\alpha}=\sum_{t=1}^{T}s_{t\alpha}$ is a sequence of
$\mathbb{R}^{r}$\textendash valued $\mathcal{F}_{T}$\textendash measurable
stochastic processes indexed by $\alpha\in A$; $S_{T\alpha}$ does
not depend on $(\beta,\pi,\varpi)$; $S_{T\alpha}$ has sample paths
that are continuous as functions of $\alpha$; the process $T^{-1/2}S_{T\alpha}$
obeys $T^{-1/2}S_{T\bullet}\Rightarrow S_{\bullet}$ for some mean
zero $\mathbb{R}^{r}$-valued Gaussian process $\{S_{\alpha}:\alpha\in A\}$
that satisfies $E[S_{\alpha}S_{\alpha}']=E[s_{t\alpha}s_{t\alpha}']=\mathcal{I}_{\alpha}$
for all $\alpha\in A$ and has continuous sample paths (as functions
of $\alpha$) with probability 1.
\item [{(iii)}] $\mathcal{I}_{\alpha}$ is, for each $\alpha\in A$, a
non-random symmetric $r\times r$ matrix (independent of $(\beta,\pi,\varpi)$);
$\mathcal{I}_{\alpha}$ is continuous as a function of $\alpha$ and
such that $0<\inf_{\alpha\in A}\lambda_{\min}(\mathcal{I}_{\alpha})$,
$\sup{}_{\alpha\in A}\lambda_{\max}(\mathcal{I}_{\alpha})<\infty$\emph{.}
\item [{(iv)}] $R_{T}(\alpha,\beta,\pi,\varpi)$ is a remainder term such
that 
\[
\sup_{(\beta,\pi,\varpi)\in B\times\Pi_{\alpha}:\left\Vert (\beta,\pi,\varpi)-(\beta^{*},\pi^{\ast},0)\right\Vert \leq\gamma_{T}}\frac{\lvert R_{T}(\alpha,\beta,\pi,\varpi)\rvert}{(1+\lVert T^{1/2}\boldsymbol{\theta}(\alpha,\beta,\pi,\varpi)\rVert)^{2}}=o_{p\alpha}(1)
\]
for all sequences of (non-random) positive scalars $\{\gamma_{T},\ T\geq1\}$
for which $\gamma_{T}\rightarrow0$ as $T\rightarrow\infty$.
\end{description}
\end{assumption}
Assumption \ref{assu:quadexp}(i) describes the transformed parameter
$\boldsymbol{\theta}(\alpha,\beta,\pi,\varpi)$, with part (b) being
an identification condition. Assumption \ref{assu:quadexp}(ii) is
the main ingredient needed to derive the limiting distribution of
our LR test whereas \ref{assu:quadexp}(iv) ensures that the remainder
term $R_{T}(\alpha,\beta,\pi,\varpi)$ has no effect on the final
result. Assumption \ref{assu:quadexp}(iii) imposes rather standard
conditions on the counterpart of the information matrix. 

As in \citet{andrews1999estimation,andrews2001testing}, \citet{zhu2006asymptotics},
and \citet{kasahara2012testing,kasahara2015testing}, for further
developments it will be convenient to write the expansion (\ref{eq:Quadr-Exp_1})
in an alternative form as
\begin{multline}
L_{T}^{\pi}(\alpha,\beta,\pi,\varpi)-L_{T}^{\pi}(\alpha,\beta^{*},\pi^{*},0)\\
=\frac{1}{2}Z_{T\alpha}^{\prime}\mathcal{I}_{\alpha}Z_{T\alpha}-\frac{1}{2}\bigl[T^{1/2}\boldsymbol{\theta}(\alpha,\beta,\pi,\varpi)-Z_{T\alpha}\bigr]'\mathcal{I}_{\alpha}\bigl[T^{1/2}\boldsymbol{\theta}(\alpha,\beta,\pi,\varpi)-Z_{T\alpha}\bigr]+R_{T}(\alpha,\beta,\pi,\varpi),\label{eq:Quadr-Exp_2}
\end{multline}
where $Z_{T\alpha}=\mathcal{I}_{\alpha}^{-1}T^{-1/2}S_{T\alpha}$.
Assumptions 5(ii) and (iii) imply the following facts (that will be
justified in the proof of Lemma \ref{lem:AsInsigOfRemainder} in Appendix
A): $Z_{T\alpha}$ is $\mathcal{F}_{T}$\textendash measurable, independent
of $(\beta,\pi,\varpi)$, continuous as a function of $\alpha$ with
probability 1, and $Z_{T\bullet}\Rightarrow Z_{\bullet}$ where the
mean zero $\mathbb{R}^{r}$\textendash valued Gaussian process $Z_{\alpha}=\mathcal{I}_{\alpha}^{-1}S_{\alpha}$
satisfies $E[Z_{\alpha}Z_{\alpha}']=\mathcal{I}_{\alpha}^{-1}$ for
all $\alpha\in A$ and has continuous sample paths (as functions of
$\alpha$) with probability 1.

\subsubsection{Examples (continued)}

\paragraph{LMAR Example.}

For the LMAR model, expansion (\ref{eq:Quadr-Exp_1}) (with the unnecessary
$\beta$ being dropped everywhere) is obtained from a standard second-order
Taylor expansion. Specifically, for an arbitrary fixed $\alpha\in A$,
a standard second-order Taylor expansion of $L_{T}^{\pi}(\alpha,\pi,\varpi)=\sum_{t=1}^{T}l_{t}^{\pi}(\alpha,\pi,\varpi)$
around $\left(\pi^{\ast},0\right)$ with respect to the parameters
$(\pi,\varpi)$ yields
\begin{align}
L_{T}^{\pi}(\alpha,\pi,\varpi)-L_{T}^{\pi}(\alpha,\pi^{\ast},0) & =(\pi-\pi^{\ast},\varpi)^{\prime}\nabla_{(\pi,\varpi)}L_{T}^{\pi}(\alpha,\pi^{\ast},0)\nonumber \\
 & \qquad+\frac{1}{2}(\pi-\pi^{\ast},\varpi)^{\prime}\nabla_{(\pi,\varpi)(\pi,\varpi)'}^{2}L_{T}^{\pi}(\alpha,\dot{\pi},\dot{\varpi})(\pi-\pi^{\ast},\varpi),\label{LMAR_TaylorExp}
\end{align}
where $(\dot{\pi},\dot{\varpi})$ denotes a point between $(\pi,\varpi)$
and $(\pi^{\ast},0)$, $\nabla_{(\pi,\varpi)}L_{T}^{\pi}(\alpha,\pi^{\ast},0)=\sum_{t=1}^{T}\nabla_{(\pi,\varpi)}l_{t}^{\pi}(\alpha,\pi^{\ast},0)$
and $\nabla_{(\pi,\varpi)(\pi,\varpi)'}^{2}L_{T}^{\pi}(\alpha,\pi,\varpi)=\sum_{t=1}^{T}\nabla_{(\pi,\varpi)(\pi,\varpi)'}^{2}l_{t}^{\pi}(\alpha,\pi,\varpi)$
(explicit expressions for the required derivatives are provided in
Appendix B), and $\nabla$ and $\nabla^{2}$ denote first and second
order differentiation with respect to the indicated parameters. Set
$\boldsymbol{\theta}(\alpha,\pi,\varpi)=(\pi-\pi^{\ast},\varpi)=(\theta,\vartheta)$
and note that the parameter space $\Theta_{\alpha}=\Theta$ is independent
of $\alpha$ and contains the origin, corresponding to the true model,
as an interior point. Then define the vector $S_{T\alpha}$ and the
matrix $\mathcal{I}_{\alpha}$ as\footnote{In what follows, $\nabla f_{t}(\cdot)$ denotes differentiation of
$f_{t}(\cdot)$ in (\ref{eq:f_t}) with respect to $\tilde{\phi}=(\tilde{\phi}_{0},\tilde{\phi}_{1},\ldots,\tilde{\phi}_{p},\tilde{\sigma}_{1}^{2})$.} 
\begin{align}
S_{T\alpha} & =\nabla_{(\pi,\varpi)}L_{T}^{\pi}(\alpha,\pi^{\ast},0)=\sum_{t=1}^{T}\left(\frac{\nabla f_{t}(\pi^{\ast})}{f_{t}(\pi^{\ast})},-(1-\alpha_{1,t}^{L}(\alpha))\frac{\nabla f_{t}(\pi^{\ast})}{f_{t}(\pi^{\ast})}\right),\label{eq:LMAR_S_Ta}\\
\mathcal{I}_{\alpha} & =\begin{bmatrix}E\left[\frac{\nabla f_{t}(\pi^{\ast})}{f_{t}(\pi^{\ast})}\frac{\nabla'f_{t}(\pi^{\ast})}{f_{t}(\pi^{\ast})}\right] & -E\left[(1-\alpha_{1,t}^{L}(\alpha))\frac{\nabla f_{t}(\pi^{\ast})}{f_{t}(\pi^{\ast})}\frac{\nabla'f_{t}(\pi^{\ast})}{f_{t}(\pi^{\ast})}\right]\\
-E\left[(1-\alpha_{1,t}^{L}(\alpha))\frac{\nabla f_{t}(\pi^{\ast})}{f_{t}(\pi^{\ast})}\frac{\nabla'f_{t}(\pi^{\ast})}{f_{t}(\pi^{\ast})}\right] & E\left[(1-\alpha_{1,t}^{L}(\alpha))^{2}\frac{\nabla f_{t}(\pi^{\ast})}{f_{t}(\pi^{\ast})}\frac{\nabla'f_{t}(\pi^{\ast})}{f_{t}(\pi^{\ast})}\right]
\end{bmatrix}.\nonumber 
\end{align}
Adding and subtracting terms and reorganizing, expansion (\ref{LMAR_TaylorExp})
can be written as
\begin{align}
L_{T}^{\pi}(\alpha,\pi,\varpi)-L_{T}^{\pi}(\alpha,\pi^{*},0) & =(T^{-1/2}S_{T\alpha})'[T^{1/2}\boldsymbol{\theta}(\alpha,\pi,\varpi)]\nonumber \\
 & \qquad-\frac{1}{2}[T^{1/2}\boldsymbol{\theta}(\alpha,\pi,\varpi)]^{\prime}\mathcal{I}_{\alpha}[T^{1/2}\boldsymbol{\theta}(\alpha,\pi,\varpi)]+R_{T}(\alpha,\pi,\varpi),\label{LMAR_QuadraticExp}
\end{align}
with the remainder term
\begin{equation}
R_{T}(\alpha,\pi,\varpi)=\frac{1}{2}[T^{1/2}\boldsymbol{\theta}(\alpha,\pi,\varpi)]^{\prime}[T^{-1}\nabla_{(\pi,\varpi)(\pi,\varpi)'}^{2}L_{T}^{\pi}(\alpha,\dot{\pi},\dot{\varpi})-(-\mathcal{I}_{\alpha})][T^{1/2}\boldsymbol{\theta}(\alpha,\pi,\varpi)].\label{LMAR_RemainderTerm}
\end{equation}
These equations yield the expansion (\ref{eq:Quadr-Exp_1}) in the
case of the LMAR model. For details of verifying Assumptions \ref{assu:cont_differentiability}
and \ref{assu:quadexp}, we refer to Appendix B.

As mentioned in the LMAR example of Section 3.1.1, the treatment of
the special case where the mixing weight $\alpha_{t}^{L}$ is constant
is more complicated than that of the `real' LMAR case. Indeed, replacing
the mixing weight $\alpha_{t}^{L}$ by a constant in the preceding
expression of the score vector $S_{T\alpha}$ immediately shows that
the second-order Taylor expansion (\ref{LMAR_QuadraticExp}) breaks
down because, contrary to Assumption 5(iii), the components of $S_{T\alpha}$
are not linearly independent and, consequently, the Fisher information
matrix $\mathcal{I}_{\alpha}$ is singular. Thus, a higher order Taylor
expansion is needed to analyze the LR test statistic. 

To give an idea of how one could proceed, we first note that the partial
derivatives of the log-likelihood function behave in the same way
as their counterparts in \citet{kasahara2015testing} where mixtures
of normal regression models (with constant mixing weights) are considered
(see particularly the discussion following their Proposition 1). This
is due to the fact that in the special case of constant mixing weights
the LMAR model is obtained from the model considered in \citet{kasahara2015testing}
by replacing the exogenous regressors therein by lagged values of
$y_{t}$. Thus, the arguments employed in that paper could be used
to obtain the asymptotic distribution of the LR test statistic. Instead
of a conventional second-order Taylor expansion this would require
a more complicated reparameterization and an expansion based on partial
derivatives of the log-likelihood function up to order eight. As most
of the details appear very similar to those in \citet{kasahara2015testing}
we have preferred not to pursue this matter in this paper. 

The preceding discussion means that, in the case of the LMAR model,
time varying mixing weights are beneficial when the purpose is to
derive a LR test for the adequacy of a single-regime model. A similar
observation was made already by \citet[p. 80]{jeffries1998logistic}.
However, this does not happen in all mixture autoregressive models
with time varying mixing weights, as the following discussion on the
GMAR model demonstrates.

\paragraph{GMAR Example. }

As alluded to earlier, in the case of the GMAR model the expansion
(\ref{eq:Quadr-Exp_1}) cannot be based on a second order Taylor expansion
of the log-likelihood function. A higher order expansion is required,
and similarly to \citet{kasahara2012testing} the appropriate order
turns out to be the fourth one with the elements of $\nabla_{\beta}l_{t}^{\pi}(\alpha,\beta^{*},\pi^{*},0)$
and $\nabla_{\pi}l_{t}^{\pi}(\alpha,\beta^{*},\pi^{*},0)$ and the
distinctive elements of $\nabla_{\varpi\varpi^{\prime}}l_{t}^{\pi}(\alpha,\beta^{*},\pi^{*},0)$
(suitably normalized) used to define the vector $S_{T\alpha}$. In
Appendix C we present, for an arbitrary fixed $\alpha\in A$, the
explicit form of a standard fourth-order Taylor expansion of $L_{T}^{\pi}(\alpha,\beta,\pi,\varpi)=\sum_{t=1}^{T}l_{t}^{\pi}(\alpha,\beta,\pi,\varpi)$
around $(\beta^{*},\pi^{*},0)$ with respect to the parameters $(\beta,\pi,\varpi)$.
Therein we also demonstrate that this fourth-order Taylor expansion
can be written as a quadratic expansion of the form (\ref{eq:Quadr-Exp_1})
(or (\ref{eq:Quadr-Exp_2})) with the different quantities appearing
therein defined as follows. 

Define the vector $\boldsymbol{\theta}(\alpha,\beta,\pi,\varpi)$
in (\ref{eq:Quadr-Exp_1}) as 
\[
\boldsymbol{\theta}(\alpha,\beta,\pi,\varpi)=\begin{bmatrix}\theta(\alpha,\beta,\pi,\varpi)\\
\vartheta(\alpha,\beta,\pi,\varpi)
\end{bmatrix}=\begin{bmatrix}\beta-\beta^{\ast}\\
\pi-\pi^{\ast}\\
\alpha(1-\alpha)v(\varpi)
\end{bmatrix},
\]
where $\theta(\alpha,\beta,\pi,\varpi)$ is $(q_{1}+q_{2})\times1$
and $\vartheta(\alpha,\beta,\pi,\varpi)$ is $q_{\vartheta}\times1$
with $q_{\vartheta}=q_{2}(q_{2}+1)/2$ (where $q_{1}=1$ and $q_{2}=p+1$)
and where the vector $v(\varpi)$ contains the unique elements of
$\varpi\varpi^{\prime}$, that is, 
\[
v(\varpi)=(\varpi_{1}^{2},\ldots,\varpi_{q_{2}}^{2},\varpi_{1}\varpi_{2},\ldots,\varpi_{1}\varpi_{q_{2}},\varpi_{2}\varpi_{3},\ldots,\varpi_{q_{2}-1}\varpi_{q_{2}})
\]
(note that $v(\varpi)$ is just a re-ordering of $vech(\varpi\varpi^{\prime})$).
The parameter space 
\[
\Theta_{\alpha}=\{\boldsymbol{\theta}=(\theta,\vartheta)\in\mathbb{R}^{q_{1}+q_{2}+q_{\vartheta}}:\theta=(\beta-\beta^{\ast},\pi-\pi^{\ast}),\vartheta=\alpha(1-\alpha)v(\varpi)\text{ for some }(\beta,\pi,\varpi)\in B\times\Pi_{\alpha}\}
\]
now depends on $\alpha$ and has the origin, corresponding to the
true model, as a boundary point (due to the particular shape of the
range of $\vartheta=\alpha(1-\alpha)v(\varpi)$); both of these features
will complicate the subsequent analysis. 

Next set 
\begin{equation}
S_{T}\,(=S_{T\alpha})=\sum_{t=1}^{T}\tilde{\nabla}_{\boldsymbol{\theta}}l_{t}^{\pi\ast}\ \ \ \ \ \text{where }\,\,\,\,\,\tilde{\nabla}_{\boldsymbol{\theta}}l_{t}^{\pi\ast}=(\tilde{\nabla}_{\theta}l_{t}^{\pi\ast},\tilde{\nabla}_{\vartheta}l_{t}^{\pi\ast})\ \ \ \ \ ((q_{1}+q_{2}+q_{\vartheta})\times1)\label{eq:GMAR_S_T}
\end{equation}
with the component vectors $\tilde{\nabla}_{\theta}l_{t}^{\pi\ast}$
and $\tilde{\nabla}_{\vartheta}l_{t}^{\pi\ast}$ given by 
\begin{align}
\tilde{\nabla}_{\theta}l_{t}^{\pi\ast} & =(\nabla_{\beta}l_{t}^{\pi}(\alpha,\beta^{*},\pi^{\ast},0),\nabla_{\pi}l_{t}^{\pi}(\alpha,\beta^{*},\pi^{\ast},0)),\nonumber \\
\tilde{\nabla}_{\vartheta}l_{t}^{\pi\ast} & =\bigl(c_{11}\nabla_{\varpi_{1}\varpi_{1}}^{2}l_{t}^{\pi}(\alpha,\beta^{*},\pi^{\ast},0),\ldots,c_{q_{2}q_{2}}\nabla_{\varpi_{q_{2}}\varpi_{q_{2}}}^{2}l_{t}^{\pi}(\alpha,\beta^{*},\pi^{\ast},0),\nonumber \\
 & \ \ \ \ \ c_{12}\nabla_{\varpi_{1}\varpi_{2}}^{2}l_{t}^{\pi}(\alpha,\beta^{*},\pi^{\ast},0),\ldots,c_{q_{2}-1,q_{2}}\nabla_{\varpi_{q_{2}-1}\varpi_{q_{2}}}^{2}l_{t}^{\pi}(\alpha,\beta^{*},\pi^{\ast},0)\bigr)/(\alpha(1-\alpha)),\label{eq:GMAR_S_T_component3}
\end{align}
where $c_{ij}=1/2$ if $i=j$ and $c_{ij}=1$ if $i\neq j$. Explicit
expressions for $\tilde{\nabla}_{\theta}l_{t}^{\pi\ast}$ and $\tilde{\nabla}_{\vartheta}l_{t}^{\pi\ast}$
can be found in Appendix C, and from them it can be seen that $S_{T}$
depends on $(\beta^{*},\pi^{\ast})$ only and not on $(\alpha,\beta,\pi,\varpi)$.
The same is true for the matrix $\mathcal{I}\,(=\mathcal{I_{\alpha}})$
($(q_{1}+q_{2}+q_{\vartheta})\times(q_{1}+q_{2}+q_{\vartheta})$)
whose expression is also given in Appendix C. Finally, an explicit
expression of the remainder term $R_{T}(\alpha,\beta,\pi,\varpi)$
is given in Appendix C. For the verification of Assumptions \ref{assu:cont_differentiability}
and \ref{assu:quadexp}, see Appendix C. 

In the GMAR example we have assumed that the intercept terms $\tilde{\phi}_{0}$
and $\tilde{\varphi}_{0}$ in the two regimes are the same. We are
now in a position to describe the difficulties that allowing for $\tilde{\phi}_{0}\neq\tilde{\varphi}_{0}$
(and, hence, dropping $\beta$) would entail. In this case, the additional
parameter $\tilde{\varphi}_{0}$ would correspond to $\varpi_{1}$,
the first component of $\varpi$. As in Section 3.2.1, it would again
be the case that $\nabla_{\varpi_{1}}l_{t}^{\pi}(\alpha,\pi^{*},0)=0$,
leading us to consider second derivatives. But now, due to the properties
of the Gaussian distribution, it would be the case that $\nabla_{\varpi_{1}\varpi_{1}}^{2}l_{t}^{\pi}(\alpha,\pi^{*},0)$
is linearly dependent with the components of $\nabla_{\pi}l_{t}^{\pi}(\alpha,\pi^{*},0)$,
making it unsuitable to serve as a component of $S_{T}$. A reparameterization
more complicated than that used in Section 3.2.1 would be needed,
with the aim of obtaining $\nabla_{\varpi_{1}\varpi_{1}}^{2}l_{t}^{\pi}(\alpha,\pi^{*},0)=0$
and, instead of $\nabla_{\varpi_{1}\varpi_{1}}^{2}l_{t}^{\pi}(\alpha,\pi^{*},0)$,
using $\nabla_{\varpi_{1}\varpi_{1}\varpi_{1}}^{3}l_{t}^{\pi}(\alpha,\pi^{*},0)$
or perhaps $\nabla_{\varpi_{1}\varpi_{1}\varpi_{1}\varpi_{1}}^{4}l_{t}^{\pi}(\alpha,\pi^{*},0)$
as the counterpart of the score of the parameter $\varpi_{1}$. It
turns out that (restricting the discussion to the case $p=1$ only)
the third derivative is suitable when $\alpha\neq1/2$ and $\tilde{\phi}_{1}\neq-1/2$,
but that fourth (or higher) order derivatives are needed when $\alpha=1/2$
or $\tilde{\phi}_{1}=-1/2$. Similar difficulties (involving situations
comparable to the cases $\alpha=1/2$ vs. $\alpha\neq1/2$, but apparently
not ones involving also a counterpart of $\tilde{\phi}_{1}$) were
faced by \citet[Sec. 2.3.3]{cho2007testing} and \citet[Sec. 6.2]{Kasahara2017Testing},
whose analyses suggest that expanding the log-likelihood at least
to the eighth order is required. As the required analysis gets excessively
complicated, we have chosen to leave it for future research.

\subsection{Asymptotic analysis of the quadratic expansion}

We continue by analyzing the expansion (\ref{eq:Quadr-Exp_2}) evaluated
at $(\hat{\beta}_{T\alpha},\hat{\pi}_{T\alpha},\hat{\varpi}_{T\alpha})$.
Previously, a similar analysis is provided by \citet{andrews2001testing}
but his approach is not directly applicable in our setting. The reason
for this is that in the quadratic expansion in (\ref{eq:Quadr-Exp_2})
the dependence of the parameter $\boldsymbol{\theta}(\alpha,\beta,\pi,\varpi)$
and its parameter space $\Theta_{\alpha}$ on the nuisance parameter
$\alpha$ is not compatible with the formulation of \citet[eqn (3.3)]{andrews2001testing}.
The results of \citet{zhu2006asymptotics} probably cover our case,
but instead of trying to verify the assumptions employed by these
authors we prove the needed results by adapting the arguments used
in \citet{andrews1999estimation,andrews2001testing} and \citet{zhu2006asymptotics}
to our setting. We proceed in several steps.

\paragraph{Asymptotic insignificance of the remainder term.}

We first establish that the remainder term $R_{T}(\alpha,\beta,\pi,\varpi)$,
when evaluated at $(\hat{\beta}_{T\alpha},\hat{\pi}_{T\alpha},\hat{\varpi}_{T\alpha})$,
has no effect on the asymptotic distribution of the quadratic expansion.
A crucial ingredient in showing this is showing that the transformed
parameter vector $\boldsymbol{\theta}(\alpha,\hat{\beta}_{T\alpha},\hat{\pi}_{T\alpha},\hat{\varpi}_{T\alpha})$
is root-$T$ consistent in the sense that $\lVert T^{1/2}\boldsymbol{\theta}(\alpha,\hat{\beta}_{T\alpha},\hat{\pi}_{T\alpha},\hat{\varpi}_{T\alpha})\rVert=O_{p\alpha}(1)$.
This, together with part (iv) of Assumption \ref{assu:quadexp} allows
us to obtain the result $R_{T}(\alpha,\hat{\beta}_{T\alpha},\hat{\pi}_{T\alpha},\hat{\varpi}_{T\alpha})=o_{p\alpha}(1)$.
We collect these results in the following lemma.
\begin{lem}
\label{lem:AsInsigOfRemainder}If Assumptions 1\textendash 5 hold\,\footnote{Here and in what follows, a subset of the listed assumptions would
sometimes suffice for the stated results.}, then (i) $\lVert T^{1/2}\boldsymbol{\theta}(\alpha,\hat{\beta}_{T\alpha},\hat{\pi}_{T\alpha},\hat{\varpi}_{T\alpha})\rVert=O_{p\alpha}(1)$,
\\
(ii) $R_{T}(\alpha,\hat{\beta}_{T\alpha},\hat{\pi}_{T\alpha},\hat{\varpi}_{T\alpha})=o_{p\alpha}(1)$,
and (iii) 
\begin{align}
 & L_{T}^{\pi}(\alpha,\hat{\beta}_{T\alpha},\hat{\pi}_{T\alpha},\hat{\varpi}_{T\alpha})-L_{T}^{\pi}(\alpha,\beta^{*},\pi^{*},0)\nonumber \\
 & =\frac{1}{2}Z_{T\alpha}^{\prime}\mathcal{I}_{\alpha}Z_{T\alpha}-\frac{1}{2}\bigl[T^{1/2}\boldsymbol{\theta}(\alpha,\hat{\beta}_{T\alpha},\hat{\pi}_{T\alpha},\hat{\varpi}_{T\alpha})-Z_{T\alpha}\bigr]'\mathcal{I}_{\alpha}\bigl[T^{1/2}\boldsymbol{\theta}(\alpha,\hat{\beta}_{T\alpha},\hat{\pi}_{T\alpha},\hat{\varpi}_{T\alpha})-Z_{T\alpha}\bigr]+o_{p\alpha}(1).\label{eq:Lemma 2 kaava}
\end{align}
\end{lem}
Note that assertion (iii) of Lemma \ref{lem:AsInsigOfRemainder} is
analogous to \citet[Theorem 2b]{andrews1999estimation}. 

\paragraph{Maximization of the likelihood vs. minimization of a related quadratic
form.}

The first two terms on the right hand side of (\ref{eq:Quadr-Exp_2})
provide an approximation to the (reparameterized and centered) log-likelihood
function $L_{T}^{\pi}(\alpha,\beta,\pi,\varpi)-L_{T}^{\pi}(\alpha,\beta^{*},\pi^{*},0)$
evaluated at the (approximate unrestricted reparameterized) ML estimator
$(\hat{\beta}_{T\alpha},\hat{\pi}_{T\alpha},\hat{\varpi}_{T\alpha})$
in equation (\ref{eq:Lemma 2 kaava}). For later developments it would
be convenient if the ML estimator on the right hand side of (\ref{eq:Lemma 2 kaava})
could be replaced by an (approximate) minimizer of the quadratic form
$[T^{1/2}\boldsymbol{\theta}(\alpha,\beta,\pi,\varpi)-Z_{T\alpha}]'\mathcal{I}_{\alpha}[T^{1/2}\boldsymbol{\theta}(\alpha,\beta,\pi,\varpi)-Z_{T\alpha}]$.
In order to justify this replacement we first note that, by definition,
(cf. \citet[eqn. (3.6)]{andrews1999estimation}) 
\[
\inf_{(\beta,\pi,\varpi)\in B\times\Pi_{\alpha}}\bigl\{\bigl[T^{1/2}\boldsymbol{\theta}(\alpha,\beta,\pi,\varpi)-Z_{T\alpha}\bigr]'\mathcal{I}_{\alpha}\bigl[T^{1/2}\boldsymbol{\theta}(\alpha,\beta,\pi,\varpi)-Z_{T\alpha}\bigr]\bigr\}=\inf_{\boldsymbol{\lambda}\in\Theta_{\alpha,T}}\left\{ (\boldsymbol{\lambda}-Z_{T\alpha})^{\prime}\mathcal{I}_{\alpha}(\boldsymbol{\lambda}-Z_{T\alpha})\right\} 
\]
where, for each $T$,
\[
\Theta_{\alpha,T}=\{\boldsymbol{\lambda}\in\mathbb{R}^{r}:\boldsymbol{\lambda}=T^{1/2}\boldsymbol{\theta}\text{ for some }\boldsymbol{\theta}\in\Theta_{\alpha}\}
\]
and $\Theta_{\alpha}$ is as defined in Assumption \ref{assu:quadexp}(i).
Next, for each $\alpha\in A$, let $\boldsymbol{\hat{\lambda}}_{T\alpha q}=T^{1/2}\boldsymbol{\theta}(\alpha,\hat{\beta}_{T\alpha q},\hat{\pi}_{T\alpha q},\hat{\varpi}_{T\alpha q})$
(with the additional `$q$' in the subscripts referring to quadratic
form) denote an approximate minimizer of $(\boldsymbol{\lambda}-Z_{T\alpha})^{\prime}\mathcal{I}_{\alpha}(\boldsymbol{\lambda}-Z_{T\alpha})$
over $\Theta_{\alpha,T}$, that is, (cf. \citet[eqn. (12)]{zhu2006asymptotics})
\begin{equation}
(\boldsymbol{\hat{\lambda}}_{T\alpha q}-Z_{T\alpha})^{\prime}\mathcal{I}_{\alpha}(\boldsymbol{\hat{\lambda}}_{T\alpha q}-Z_{T\alpha})=\inf_{\boldsymbol{\lambda}\in\Theta_{\alpha,T}}\left\{ (\boldsymbol{\lambda}-Z_{T\alpha})^{\prime}\mathcal{I}_{\alpha}(\boldsymbol{\lambda}-Z_{T\alpha})\right\} +o_{p\alpha}(1).\label{eq:QuadFormMinimizer}
\end{equation}
Now we can state the following lemma justifying the discussed replacement
of $(\hat{\beta}_{T\alpha},\hat{\pi}_{T\alpha},\hat{\varpi}_{T\alpha})$
with $\boldsymbol{\hat{\lambda}}_{T\alpha q}$. 
\begin{lem}
\label{lem:MaxLikVsMinQuadForm}If Assumptions 1\textendash 5 hold,
then
\[
L_{T}^{\pi}(\alpha,\hat{\beta}_{T\alpha},\hat{\pi}_{T\alpha},\hat{\varpi}_{T\alpha})-L_{T}^{\pi}(\alpha,\beta^{*},\pi^{*},0)=\frac{1}{2}Z_{T\alpha}^{\prime}\mathcal{I}_{\alpha}Z_{T\alpha}-\frac{1}{2}(\boldsymbol{\hat{\lambda}}_{T\alpha q}-Z_{T\alpha})^{\prime}\mathcal{I}_{\alpha}(\boldsymbol{\hat{\lambda}}_{T\alpha q}-Z_{T\alpha})+o_{p\alpha}(1).
\]
\end{lem}

\paragraph{Approximating the parameter space with a cone.}

In the previous subsection the quadratic form $(\boldsymbol{\lambda}-Z_{T\alpha})^{\prime}\mathcal{I}_{\alpha}(\boldsymbol{\lambda}-Z_{T\alpha})$
was minimized over the set $\Theta_{\alpha,T}$ which can be complicated
and hence difficult to use. Therefore we next show that the quadratic
form $(\boldsymbol{\lambda}-Z_{T\alpha})^{\prime}\mathcal{I}_{\alpha}(\boldsymbol{\lambda}-Z_{T\alpha})$
can instead be minimized over a simpler set, and to this end we first
introduce some terminology.

We say that a collection of sets $\{\Gamma_{\alpha},\ \alpha\in A\}$
(where for each $\alpha\in A$, $\Gamma_{\alpha}\subset\mathbb{R}^{r}$)
is `locally (at the origin) uniformly equal' to a set $\Lambda\subset\mathbb{R}^{r}$
if there exists a $\delta>0$ such that $\Gamma_{\alpha}\cap(-\delta,\delta)^{r}=\Lambda\cap(-\delta,\delta)^{r}$
for all $\alpha\in A$. Note that `$\{\Gamma_{\alpha},\ \alpha\in A\}$
is locally uniformly equal to $\Lambda$' implies that (i) `for all
$\alpha\in A$, $\Gamma_{\alpha}$ is locally equal to $\Lambda$
in the sense of \citet[p. 1359]{andrews1999estimation}', but the
reverse does not hold; and also that (ii) $\{\Gamma_{\alpha},\ \alpha\in A\}$
is uniformly approximated by the set $\Lambda$ in the sense of \citet[Defn. 3]{zhu2006asymptotics}.
Finally, we say that a set $\Lambda\subset\mathbb{R}^{r}$ is a `cone'
if $\lambda\in\Lambda$ implies that $a\lambda\in\Lambda$ for all
positive real scalars $a$.

Based on the preceding discussion we state the following assumption.
\begin{assumption}
\label{assu:cone}The collection of sets $\{\Theta_{\alpha},\ \alpha\in A\}$
is locally uniformly equal to a cone $\Lambda\;(\subset\mathbb{R}^{r})$. 
\end{assumption}
Note that by Assumption \ref{assu:quadexp}(i)(a), $0\in\Theta_{\alpha}$
for all $\alpha\in A$, so that the cone $\Lambda$ in Assumption
\ref{assu:cone} necessarily contains $0\;(\in\mathbb{R}^{r})$. The
cone $\Lambda$ also does not depend on $\alpha.$ Now we can establish
the following result. 
\begin{lem}
\label{lem:Cone}If Assumptions 1\textendash 6 hold, then 
\[
\inf_{\boldsymbol{\lambda}\in\Theta_{\alpha,T}}\left\{ (\boldsymbol{\lambda}-Z_{T\alpha})^{\prime}\mathcal{I}_{\alpha}(\boldsymbol{\lambda}-Z_{T\alpha})\right\} =\inf_{\boldsymbol{\lambda}\in\Lambda}\left\{ (\boldsymbol{\lambda}-Z_{T\alpha})^{\prime}\mathcal{I}_{\alpha}(\boldsymbol{\lambda}-Z_{T\alpha})\right\} +o_{p\alpha}(1).
\]
\end{lem}

\paragraph{Describing the limiting random variable.}

From Lemmas \ref{lem:MaxLikVsMinQuadForm} and \ref{lem:Cone} and
the definition of $\boldsymbol{\hat{\lambda}}_{T\alpha q}$ we can
now conclude that 
\begin{equation}
2[L_{T}^{\pi}(\alpha,\hat{\beta}_{T\alpha},\hat{\pi}_{T\alpha},\hat{\varpi}_{T\alpha})-L_{T}^{\pi}(\alpha,\beta^{*},\pi^{*},0)]=Z_{T\alpha}^{\prime}\mathcal{I}_{\alpha}Z_{T\alpha}-\inf_{\boldsymbol{\lambda}\in\Lambda}\left\{ (\boldsymbol{\lambda}-Z_{T\alpha})^{\prime}\mathcal{I}_{\alpha}(\boldsymbol{\lambda}-Z_{T\alpha})\right\} +o_{p\alpha}(1).\label{eq:BeforeLemmaWeakConv}
\end{equation}
The assumed weak convergence of $S_{T\alpha}$ (and hence that of
$Z_{\alpha}=\mathcal{I}_{\alpha}^{-1}S_{\alpha}$) allows us to derive
the weak limit of this random process described in the following lemma.
\begin{lem}
\label{lem:WeakConv}If Assumptions 1\textendash 6 hold, then 
\[
2[L_{T}^{\pi}(\bullet,\hat{\beta}_{T\bullet},\hat{\pi}_{T\bullet},\hat{\varpi}_{T\bullet})-L_{T}^{\pi}(\bullet,\beta^{*},\pi^{\ast},0)]\Rightarrow Z_{\bullet}^{\prime}\mathcal{I}_{\bullet}Z_{\bullet}-\inf_{\boldsymbol{\lambda}\in\Lambda}\left\{ (\boldsymbol{\lambda}-Z_{\bullet})^{\prime}\mathcal{I}_{\bullet}(\boldsymbol{\lambda}-Z_{\bullet})\right\} .
\]
\end{lem}
The limiting random process in Lemma \ref{lem:WeakConv} can be written
in a somewhat simpler form (cf. \citet[Thm 4]{andrews1999estimation}
and \citet[Thm 2]{andrews2001testing}). The motivation for this comes
from the fact that in our applications $\boldsymbol{\theta}(\alpha,\beta,\pi,\varpi)$
can be decomposed into two parts as $\boldsymbol{\theta}(\alpha,\beta,\pi,\varpi)=(\theta(\alpha,\beta,\pi,\varpi),\vartheta(\alpha,\beta,\pi,\varpi))$
with $\theta(\alpha,\beta,\pi,\varpi)\in\mathbb{R}^{q_{\theta}}$
and $\vartheta(\alpha,\beta,\pi,\varpi)\in\mathbb{R}^{q_{\vartheta}}$
(with $q_{\theta}=q_{1}+q_{2}$ and $r=q_{\theta}+q_{\vartheta}$)
such that (i) the values of $\theta(\alpha,\beta,\pi,\varpi)$ are
not restricted by the null hypothesis and do not lie on the boundary
of the parameter space and (ii) the values of $\vartheta(\alpha,\beta,\pi,\varpi)$
are restricted by the null hypothesis and potentially lie on the boundary
of the parameter space. Specifically, we assume the following.
\begin{assumption}
\label{assu:cone2}The cone $\Lambda$ of Assumption \ref{assu:cone}
satisfies $\Lambda=\mathbb{R}^{q_{\theta}}\times\Lambda_{\vartheta}$
with $\Lambda_{\vartheta}$ a cone in $\mathbb{R}^{q_{\vartheta}}$.
\end{assumption}
Partition $S_{\alpha}$, $Z_{\alpha}$, $\boldsymbol{\lambda}$, and
$\mathcal{I}_{\alpha}$ conformably with the partition $\boldsymbol{\theta}(\alpha,\beta,\pi,\varpi)=(\theta(\alpha,\beta,\pi,\varpi),\vartheta(\alpha,\beta,\pi,\varpi))$
as
\[
S_{\alpha}=\left[\begin{array}{c}
S_{\theta\alpha}\\
S_{\vartheta\alpha}
\end{array}\right],\,\,\,Z_{\alpha}=\left[\begin{array}{c}
Z_{\theta\alpha}\\
Z_{\vartheta\alpha}
\end{array}\right],\,\,\,\boldsymbol{\lambda}=\begin{bmatrix}\boldsymbol{\lambda}_{\theta}\\
\boldsymbol{\lambda}_{\vartheta}
\end{bmatrix},\,\,\,\mathcal{I}_{\alpha}=\left[\begin{array}{cc}
\mathcal{I}_{\theta\theta\alpha} & \mathcal{I}_{\theta\vartheta\alpha}\\
\mathcal{I}_{\vartheta\theta\alpha} & \mathcal{I}_{\vartheta\vartheta\alpha}
\end{array}\right]
\]
and let $(\mathcal{I}_{\alpha}^{-1})_{\vartheta\vartheta}$ denote
the ($q_{\vartheta}\times q_{\vartheta}$) bottom right block of $\mathcal{I}_{\alpha}^{-1}$.
Assumption \ref{assu:cone2} together with properties of partitioned
matrices yields the following result.
\begin{lem}
\label{lem:WeakLimitSubvec}If Assumptions 1\textendash 7 hold, then
\begin{align*}
 & Z_{\alpha}'\mathcal{I}_{\alpha}Z_{\alpha}-\inf_{\boldsymbol{\lambda}\in\Lambda}\left\{ (\boldsymbol{\lambda}-Z_{\alpha})^{\prime}\mathcal{I}_{\alpha}(\boldsymbol{\lambda}-Z_{\alpha})\right\} \\
 & =Z_{\vartheta\alpha}'(\mathcal{I}_{\alpha}^{-1})_{\vartheta\vartheta}^{-1}Z_{\vartheta\alpha}-\inf_{\boldsymbol{\lambda}_{\vartheta}\in\Lambda_{\vartheta}}\left\{ (\boldsymbol{\lambda}_{\vartheta}-Z_{\vartheta\alpha})'(\mathcal{I}_{\alpha}^{-1})_{\vartheta\vartheta}^{-1}(\boldsymbol{\lambda}_{\vartheta}-Z_{\vartheta\alpha})\right\} +S_{\theta\alpha}'\mathcal{I}_{\theta\theta\alpha}^{-1}S_{\theta\alpha}.
\end{align*}
\end{lem}
Explicit expressions for $(\mathcal{I}_{\alpha}^{-1})_{\vartheta\vartheta}$
and $Z_{\vartheta\alpha}$ in terms of $S_{\alpha}$ and $\mathcal{I}_{\alpha}$
are given in the proof of this lemma in Supplementary Appendix D.

\subsection{The LR test statistic}

\subsubsection{Derivation of the test statistic}

The previous subsection described the asymptotic behavior of $2[L_{T}^{\pi}(\alpha,\hat{\beta}_{T\alpha},\hat{\pi}_{T\alpha},\hat{\varpi}_{T\alpha})-L_{T}^{\pi}(\alpha,\beta^{*},\pi^{*},0)]$,
the first term in the expression of $LR_{T}(\alpha)$ in (\ref{eq:LR_pi-repam}).
Now consider the second term, namely $2[L_{T}^{0}(\hat{\tilde{\phi}}_{T})-L_{T}^{0}(\tilde{\phi}^{\ast})]$,
corresponding to the model restricted by the null hypothesis. Recall
that $L_{T}^{0}(\tilde{\phi})=\sum_{t=1}^{T}l_{t}^{0}(\tilde{\phi})$
with $l_{t}^{0}(\tilde{\phi})=\log[f_{t}(\tilde{\phi})]$ so that
$\nabla_{\tilde{\phi}}l_{t}^{0}(\tilde{\phi}^{\ast})=(f_{t}(\tilde{\phi}^{*}))^{-1}\nabla f_{t}(\tilde{\phi}^{*})$
with $\tilde{\phi}^{*}$ an interior point of $\tilde{\Phi}$. Denote
the score vector and limiting information matrix by 
\[
S_{T}^{0}=\sum_{t=1}^{T}\frac{\nabla f_{t}(\tilde{\phi}^{\ast})}{f_{t}(\tilde{\phi}^{\ast})},\,\,\,\,\,\mathcal{I}^{0}=E\biggl[\frac{\nabla f_{t}(\tilde{\phi}^{\ast})}{f_{t}(\tilde{\phi}^{\ast})}\frac{\nabla'f_{t}(\tilde{\phi}^{\ast})}{f_{t}(\tilde{\phi}^{\ast})}\biggr],
\]
respectively. For the following assumption, partition the process
$S_{T\alpha}$ of Assumption \ref{assu:quadexp} as $S_{T\alpha}=(S_{T\theta\alpha},S_{T\vartheta\alpha})$
(with $S_{T\theta\alpha}$ $q_{\theta}$\textendash dimensional and
$S_{T\vartheta\alpha}$ $q_{\vartheta}$\textendash dimensional).
The following simplifying assumption, which holds in our examples
(see the expressions of $S_{T\alpha}$ in (\ref{eq:LMAR_S_Ta}) and
(\ref{eq:GMAR_S_T})), allows us to obtain a neat expression for the
likelihood ratio test statistic in Theorem 1 below. 
\begin{assumption}
\label{assu:ScoreEqualsLinearScore}$S_{T\theta\alpha}=S_{T}^{0}$.
\end{assumption}
Together with the earlier assumptions, Assumption \ref{assu:ScoreEqualsLinearScore}
implies that $T^{-1/2}S_{T\theta\alpha}=T^{-1/2}S_{T}^{0}\stackrel{d}{\to}S^{0}$,
a $q_{\theta}$\textendash dimensional Gaussian random vector with
mean zero and covariance matrix $\mathcal{I}_{\theta\theta\alpha}=E[S^{0}S^{0\prime}]=\mathcal{I}^{0}$.
Standard likelihood theory now implies the following result.
\begin{lem}
\label{lem:WeakConvLinear}If Assumptions 1\textendash 8 hold, then
$2[L_{T}^{0}(\hat{\tilde{\phi}}_{T})-L_{T}^{0}(\tilde{\phi}^{\ast})]\overset{d}{\rightarrow}S^{0\prime}(\mathcal{I}^{0})^{-1}S^{0}$,
and the convergence is joint with that in Lemma \ref{lem:WeakConv}. 
\end{lem}
The preceding results, in particular Lemmas \ref{lem:WeakConv}, \ref{lem:WeakLimitSubvec},
and \ref{lem:WeakConvLinear}, now yield the distribution of the LR
test statistic in the following theorem. 
\begin{thm}
If Assumptions 1\textendash 8 hold, then
\begin{description}
\item [{(i)}] $LR_{T}\left(\bullet\right)\Rightarrow Z_{\vartheta\bullet}'(\mathcal{I}_{\bullet}^{-1})_{\vartheta\vartheta}^{-1}Z_{\vartheta\bullet}-\inf_{\boldsymbol{\lambda}_{\vartheta}\in\Lambda_{\vartheta}}\{(\boldsymbol{\lambda}_{\vartheta}-Z_{\vartheta\bullet})'(\mathcal{I}_{\bullet}^{-1})_{\vartheta\vartheta}^{-1}(\boldsymbol{\lambda}_{\vartheta}-Z_{\vartheta\bullet})\}$,
and
\item [{(ii)}] $LR_{T}=\sup_{\alpha\in A}LR_{T}\left(\alpha\right)\overset{d}{\rightarrow}\sup_{\alpha\in A}\bigl\{ Z_{\vartheta\alpha}'(\mathcal{I}_{\alpha}^{-1})_{\vartheta\vartheta}^{-1}Z_{\vartheta\alpha}-\inf_{\boldsymbol{\lambda}_{\vartheta}\in\Lambda_{\vartheta}}\{(\boldsymbol{\lambda}_{\vartheta}-Z_{\vartheta\alpha})'(\mathcal{I}_{\alpha}^{-1})_{\vartheta\vartheta}^{-1}(\boldsymbol{\lambda}_{\vartheta}-Z_{\vartheta\alpha})\}\bigr\}$.
\end{description}
\end{thm}
This completes the derivation of the LR test statistic. The asymptotic
distribution is similar to that in \citet[Thm 4]{andrews2001testing}.
As we next discuss, this distribution simplifies in both the LMAR
and the GMAR examples.

\subsubsection{Examples (continued) }

\paragraph{LMAR Example. }

As was noted in Section 3.3.1, the LMAR case is rather standard in
the sense that a conventional second-order Taylor expansion with a
nonsingular information matrix and with no parameters on the boundary
was sufficient to study the LR test. The only nonstandard feature
in this case is the presence of unidentified parameters under the
null hypothesis. Validity of Assumptions \ref{assu:cone}\textendash \ref{assu:ScoreEqualsLinearScore}
is easy to check (see Appendix B) with the cone $\Lambda$ of Assumption
\ref{assu:cone} equal to $\mathbb{R}^{r}$. Thus the infimum in the
distribution of the $LR_{T}$ statistic in Theorem 1(ii) equals zero
and the result therein simplifies to\footnote{This result could also be obtained from \citet[Sec 2.2, 2.4]{andrews1995admissibility}
as their assumptions 1\textendash 5 appear to be satisfied in the
LMAR case.}
\[
LR_{T}=\sup_{\alpha\in A}LR_{T}\left(\alpha\right)\overset{d}{\rightarrow}\sup_{\alpha\in A}\left\{ Z_{\vartheta\alpha}'(\mathcal{I}_{\alpha}^{-1})_{\vartheta\vartheta}^{-1}Z_{\vartheta\alpha}\right\} .
\]
For every fixed $\alpha\in A$, the quantity $Z_{\vartheta\alpha}'(\mathcal{I}_{\alpha}^{-1})_{\vartheta\vartheta}^{-1}Z_{\vartheta\alpha}$
is a chi-squared random variable, so that the limiting distribution
is a supremum of a chi-squared process similarly as in, for example,
\citet{davies1987hypothesis}, \citet[Thm 1]{hansen1996inference},
and \citet[eqn. (5.7)]{andrews2001testing}.

\paragraph{GMAR Example. }

\noindent In Section 3.3.1 it was seen that in the GMAR example $Z_{\alpha}$
and $\mathcal{I}_{\alpha}$ do not depend on $\alpha$. As the cone
$\Lambda$ of Assumption \ref{assu:cone} does not depend on $\alpha$
either, the weak limit of $LR_{T}\left(\alpha\right)$ does not depend
on $\alpha$. Therefore the result of Theorem 1 (validity of Assumptions
\ref{assu:cone}\textendash \ref{assu:ScoreEqualsLinearScore} is
checked in Appendix C) simplifies to 
\[
LR_{T}=\sup_{\alpha\in A}LR_{T}\left(\alpha\right)\overset{d}{\rightarrow}Z_{\vartheta}'(\mathcal{I}^{-1})_{\vartheta\vartheta}^{-1}Z_{\vartheta}-\inf_{\boldsymbol{\lambda}_{\vartheta}\in\Lambda_{\vartheta}}\left\{ (\boldsymbol{\lambda}_{\vartheta}-Z_{\vartheta})'(\mathcal{I}^{-1})_{\vartheta\vartheta}^{-1}(\boldsymbol{\lambda}_{\vartheta}-Z_{\vartheta})\right\} ,
\]
where the unnecessary $\alpha$ has been dropped from the notation.
Here $Z_{\vartheta}$ follows an $q_{\vartheta}$-variate Gaussian
distribution with covariance matrix $(\mathcal{I}^{-1})_{\vartheta\vartheta}$,
and the limiting distribution, which is sometimes referred to as the
chi-bar-squared distribution, is similar to the one in \citet[Proposition 3c,d]{kasahara2012testing}.
Note that the cone $\Lambda_{\vartheta}=v(\mathbb{\mathbb{R}}^{q_{2}})$
(see Appendix C) is not convex (in contrast to (at least most of)
the examples in \citet{andrews2001testing}, but similarly to \citet[Proposition 3c,d]{kasahara2012testing})
and the dimension of this cone, $q_{\vartheta}=q_{2}(q_{2}+1)/2$,
may not be small either ($q_{\vartheta}=3,6,10,\ldots$ for $q_{2}=2,3,4,\ldots$). 

\section{Simulation-based critical values and a Monte Carlo study}

\subsection{Simulating the asymptotic null distribution}

Similarly to \citet{hansen1996inference} and \citet{andrews2001testing},
the asymptotic null distribution of the LR statistic in Theorem 1
is typically application-specific and cannot be tabulated. Following
these papers, we use simulation methods to obtain critical values
of the asymptotic null distribution. The following procedure is based
on \citet{hansen1996inference} and is analogous to the one used by
\citet[Sec 2.1]{zhu2004hypothesis} in a related mixture setting.\footnote{An alternative to this procedure is to use bootstrap. However, the
validity of bootstrap in the presence of parameters on the boundary
and singular information matrices is not clear (see, e.g., \citet{andrews2000inconsistency}).
Another reason for preferring the proposed simulation method is that
repeated estimation of the mixture model under the alternative may
be computationally rather demanding.}

Let $A_{G}$ be some finite grid of $\alpha$ values in $A$. For
each fixed $\alpha\in A_{G}$, let $\hat{s}_{t\alpha}$ signify an
empirical counterpart of $s_{t\alpha}$ (see Assumption 5) where the
unknown parameter $\tilde{\phi}^{*}$ (or $(\beta^{*},\pi^{*})$)
is replaced by its consistent estimator under the null, $\hat{\tilde{\phi}}_{T}$.
(The specific forms of $\hat{s}_{t\alpha}$ in the LMAR and GMAR examples
are provided in Appendices B and C, respectively.) Set $\hat{\mathcal{I}}_{T\alpha}=T^{-1}\sum_{t=1}^{T}\hat{s}_{t\alpha}\hat{s}_{t\alpha}'$.
Now, for each $j=1,\ldots,J$ (where $J$ denotes the number of repetitions),
do the following.
\begin{description}
\item [{(i)}] Generate a sequence $\{v_{tj}\}_{t=1}^{T}$ of $T$ i.i.d.
$N(0,1)$ random variables. 
\item [{(ii)}] For each $\alpha\in A_{G}$, set $\hat{S}_{T\alpha}^{j}=\sum_{t=1}^{T}\hat{s}_{t\alpha}v_{tj}$,
$\hat{Z}_{T\alpha}^{j}=\hat{\mathcal{I}}_{T\alpha}^{-1}T^{-1/2}\hat{S}_{T\alpha}^{j}$,
and (using similar partitioning notation as before) 
\[
\widehat{LR}_{T}^{j}(\alpha)=\hat{Z}_{T\vartheta\alpha}^{j\prime}(\hat{\mathcal{I}}_{T\alpha}^{-1})_{\vartheta\vartheta}^{-1}\hat{Z}_{T\vartheta\alpha}^{j}-\inf_{\boldsymbol{\lambda}_{\vartheta}\in\Lambda_{\vartheta}}\bigl\{(\boldsymbol{\lambda}_{\vartheta}-\hat{Z}_{T\vartheta\alpha}^{j})'(\hat{\mathcal{I}}_{T\alpha}^{-1})_{\vartheta\vartheta}^{-1}(\boldsymbol{\lambda}_{\vartheta}-\hat{Z}_{T\vartheta\alpha}^{j})\bigr\};
\]
here the minimization of the quadratic form over the cone $\Lambda_{\vartheta}$
has to be performed numerically. 
\item [{(iii)}] Set $\widehat{LR}_{T,A_{G}}^{j}=\max_{\alpha\in A_{G}}\widehat{LR}_{T}^{j}(\alpha)$. 
\end{description}
This yields a sample $\{\widehat{LR}_{T,A_{G}}^{1},\ldots,\widehat{LR}_{T,A_{G}}^{J}\}$
of $J$ realizations. An approximate $p$\textendash value corresponding
to an observed LR test statistic $LR_{T}$ is computed as $J^{-1}\sum_{j=1}^{J}\mathbf{1}(\widehat{LR}_{T,A_{G}}^{j}>LR_{T})$
(here $\mathbf{1}(\cdot)$ denotes the indicator function). The precision
of this approximation can be controlled by choosing $J$ large enough,
see \citet{hansen1996inference} (in the illustration below we use
$J=1000$).

\subsection{A small Monte Carlo study}

We now study the finite sample properties of the LR test statistics
and the simulation-based critical values. The results are presented
in Table 1. We consider two LR test statistics, one based on an estimated
LMAR model, and another based on an estimated GMAR model (as in our
two examples in the preceding sections). In all simulations, we use
an autoregressive order $p=1$, $J=1000$ repetitions (see the previous
subsection), and three different sample sizes: $T=250$, $500$, and
$1000$. 

The top part of Table 1 presents results for size simulations. Data
is generated from an AR(1) model (for a range of different parameter
values shown in Table 1) and AR(1), LMAR(1), and GMAR(1) models are
estimated (LMAR with $m=1$; GMAR with the restriction $\tilde{\phi}_{0}=\tilde{\varphi}_{0}$;
in estimation of the mixture models we use a genetic algorithm as
singularity of the information matrix may render gradient based methods
unreliable). Two LR test statistics are calculated based on the estimated
LMAR and GMAR models, respectively, and labelled `LMAR $LR_{T}$'
and `GMAR $LR_{T}$'. Simulation-based $p$\textendash values are
computed based on the asymptotic distributions in Section 3.5.2 and
using the simulation procedure in Section 4.1. Using nominal levels
$10\%$, $5\%$, and $1\%$, a reject/not-reject decision is recorded.
This exercise is repeated $1000$ times, and the six rightmost columns
in Table 1 present the empirical rejection frequencies (for the LMAR
$LR_{T}$ and GMAR $LR_{T}$ tests and the three nominal levels used).

As can be seen from the results in Table 1 (top part), the LMAR $LR_{T}$
test's size is satisfactory overall, typically being somewhat oversized
for sample sizes $T=250$ and $500$, and somewhat conservative for
the largest sample size ($T=1000$). The parameter values used in
simulation do not seem to have a large effect on the size. The GMAR
$LR_{T}$ test, on the other hand, appears to be moderately oversized
across all sample sizes and parameter values used. 

The lower part of Table 1 presents results for power simulations.
Data is generated either from a GMAR model or from an LMAR model (for
a range of different parameter values shown in Table 1), and empirical
rejection frequencies are calculated as above. Both the LMAR $LR_{T}$
test and the GMAR $LR_{T}$ test appear to have good overall power.
As expected, when the two regimes differ more from each other, the
tests have higher power, and the same happens when sample size is
increased. Besides having good power against the `right' alternatives,
the tests also turn out to have decent power against `wrong' alternatives:
When data is generated from the GMAR (resp., LMAR) model, the LMAR
$LR_{T}$ (resp., GMAR $LR_{T}$) test rejects reasonably often (the
GMAR $LR_{T}$ test in particular seems capable of picking up LMAR
type regime switching). Naturally, the power of the tests may be inflated
due to the tests being oversized. 

As a computational remark we note that the LMAR $LR_{T}$ and GMAR
$LR_{T}$ tests and their $p$\textendash values are reasonably straightforward
to compute in a matter of seconds using a standard, modern desktop
computer (for one particular model, one particular sample size, and
$J=1000$ repetitions). The GMAR $LR_{T}$ test is computationally
more demanding than the LMAR $LR_{T}$ test as it involves the minimization
of a quadratic form over a cone which is not needed in the LMAR case
(see Sections 3.5.2 and 4.1); this is also one potential reason for
the less precise size of the GMAR $LR_{T}$ test. 

\begin{table}[p]
\includegraphics[scale=0.88]{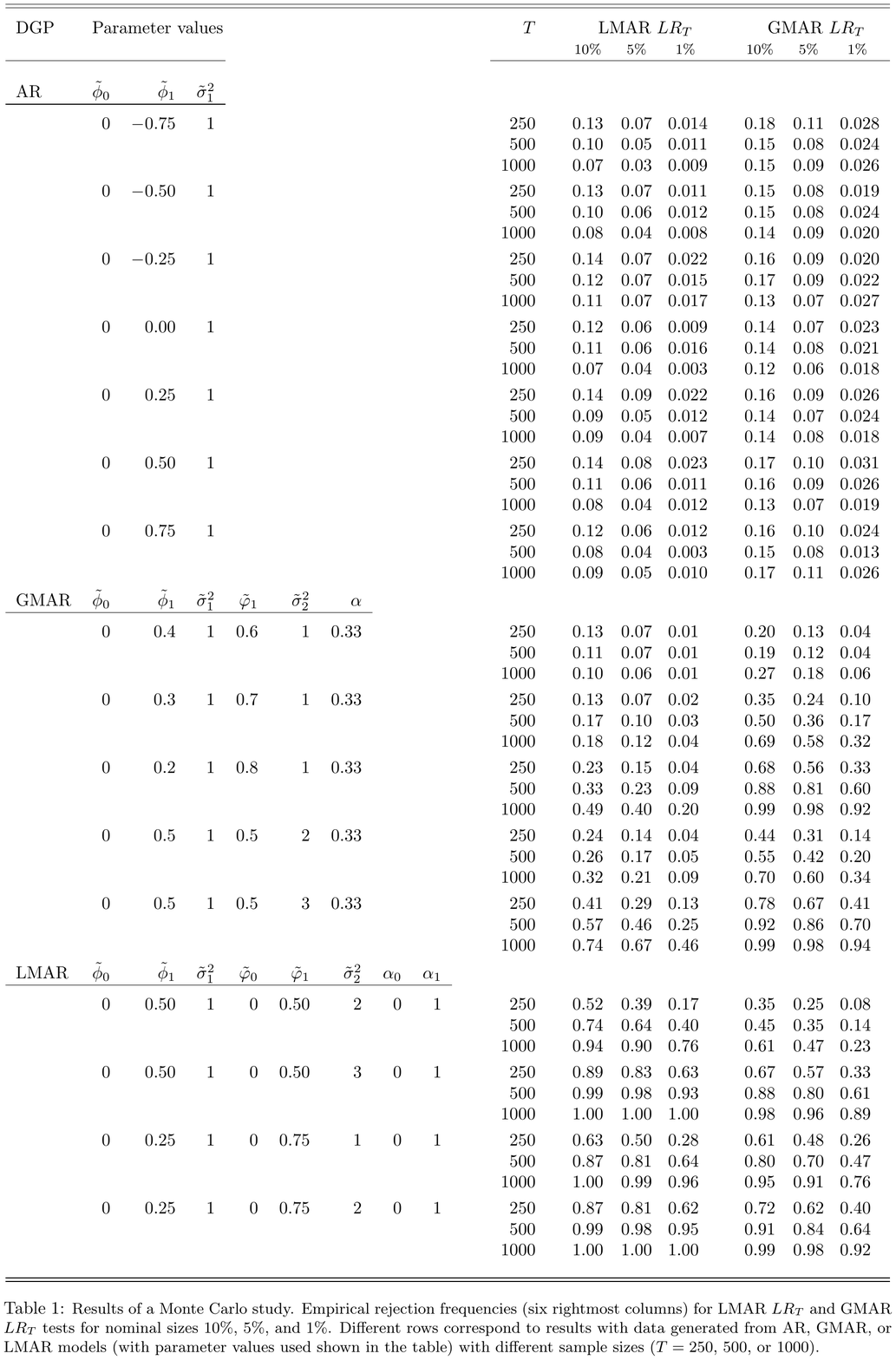}
\end{table}

\section{Conclusions}

This paper has studied the asymptotic distribution of the LR test
statistic for testing a linear autoregressive model against a two-regime
mixture autoregressive model. A distinguishing feature of the paper
is that the regime switching probabilities are observation-dependent.
Technical challenges resulting from unidentified parameters under
the null, parameters on the boundary, and singularity of the information
matrix were dealt with by considering an appropriately reparameterized
model and higher-order expansions of the log-likelihood function.
The resulting asymptotic distribution of the LR test statistic is
non-standard and application-specific. Critical values can be obtained
by a straightforward simulation procedure, and a Monte Carlo study
indicated the proposed tests to have satisfactory size and power properties.

The general theory of the paper was illustrated using two concrete
examples, the LMAR model of \citet{wong2001logistic} and (a version
of the) GMAR model of \citet{kalliovirta2015gaussian}. Considering
other mixture AR models, as well as the general GMAR model, is left
for future research. This paper was concerned with testing linearity
against a two-regime model, and considering tests of $M\geq2$ regimes
versus $M+1$ regimes, similarly as in \citet{kasahara2015testing}
in a related setting, forms another interesting research topic. 

\pagebreak{}

\appendix

\section*{Appendix}

\section{Details for the general results}

\begin{lemma}\label{lem:LipCondForUnifCons}When Assumptions \ref{assu:unrestricted_est}(ii)
and \ref{assu:pi-repam}(i,ii) hold, a sufficient condition for Assumption
\ref{assu:pi-repam}(iii) is that
\[
\left\Vert \boldsymbol{\pi}_{\alpha}(\phi,\varphi)-\boldsymbol{\pi}_{\alpha}(\phi^{\ast},\phi^{\ast})\right\Vert \leq Ch\left(\left\Vert (\phi,\varphi)-(\phi^{\ast},\phi^{\ast})\right\Vert _{*}\right)\,\,\,\textrm{for all}\,\,\,(\phi,\varphi)\in\Phi\times\Phi,
\]
where $C$ is a finite positive constant, $h:[0,\infty)\rightarrow[0,\infty)$
is a strictly increasing function such that $h\left(x\right)\downarrow0$
as $x\downarrow0$, and $\left\Vert \cdot\right\Vert _{*}$ is any
vector norm on $\mathbb{R}^{2q_{2}}$. \end{lemma}

\smallskip{}
\begin{proof}[\textbf{\emph{Proof of Lemma \ref{lem:LipCondForUnifCons}}}]
 First note that $\sup_{\alpha\in A}\lVert(\hat{\beta}_{T\alpha},\hat{\pi}_{T\alpha},\hat{\varpi}_{T\alpha})-(\beta^{*},\pi^{\ast},0)\rVert\leq\sup_{\alpha\in A}\lVert\hat{\beta}_{T\alpha}-\beta^{*}\rVert+\sup_{\alpha\in A}\lVert(\hat{\pi}_{T\alpha},\hat{\varpi}_{T\alpha})-(\pi^{\ast},0)\rVert$,
where the former term on the majorant side is $o_{p}(1)$ by Assumption
2(ii). Second, the latter term equals $\sup_{\alpha\in A}\lVert\boldsymbol{\pi}_{\alpha}(\hat{\phi}_{T\alpha},\hat{\varphi}_{T\alpha})-\boldsymbol{\pi}_{\alpha}(\phi^{\ast},\phi^{\ast})\rVert$
which, due to the assumptions made in the lemma, can be bounded by
$C\sup\nolimits _{\alpha\in A}h\bigl(\lVert(\hat{\phi}_{T\alpha},\hat{\varphi}_{T\alpha})-(\phi^{\ast},\phi^{\ast})\rVert_{*}\bigr)\leq Ch\bigl(\sup\nolimits _{\alpha\in A}\lVert(\hat{\phi}_{T\alpha},\hat{\varphi}_{T\alpha})-(\phi^{\ast},\phi^{\ast})\rVert_{*}\bigr)$.
By Assumption 2(ii) and the fact that all vector norms on $\mathbb{R}^{2q_{2}}$
are equivalent, the majorant side is $o_{p}(1)$.
\end{proof}
\smallskip{}
\begin{proof}[\textbf{\emph{Proof of Lemma \ref{lem:AsInsigOfRemainder}}}]
 Set $\boldsymbol{\theta}_{T\alpha}=\mathcal{I}_{\alpha}^{1/2}T^{1/2}\boldsymbol{\theta}(\alpha,\hat{\beta}_{T\alpha},\hat{\pi}_{T\alpha},\hat{\varpi}_{T\alpha})$,
and rewrite equation (\ref{eq:Quadr-Exp_1}) evaluated at $(\alpha,\hat{\beta}_{T\alpha},\hat{\pi}_{T\alpha},\hat{\varpi}_{T\alpha})$
as
\begin{equation}
o_{p\alpha}(1)\leq L_{T}^{\pi}(\alpha,\hat{\beta}_{T\alpha},\hat{\pi}_{T\alpha},\hat{\varpi}_{T\alpha})-L_{T}^{\pi}(\alpha,\beta^{*},\pi^{\ast},0)=(\mathcal{I}_{\alpha}^{1/2}Z_{T\alpha})^{\prime}\boldsymbol{\theta}_{T\alpha}-\frac{1}{2}\left\Vert \boldsymbol{\theta}_{T\alpha}\right\Vert ^{2}+R_{T}(\alpha,\hat{\beta}_{T\alpha},\hat{\pi}_{T\alpha},\hat{\varpi}_{T\alpha})\label{eq: Lemma 2_1}
\end{equation}
(the lower bound is due to Assumption \ref{assu:unrestricted_est}(i)
and the definitions of $(\hat{\beta}_{T\alpha},\hat{\pi}_{T\alpha},\hat{\varpi}_{T\alpha})$
and $L_{T}^{\pi}(\alpha,\beta,\pi,\varpi)$). As to the first term
on the right hand side, note that
\begin{equation}
\sup_{\alpha\in A}\lVert\mathcal{I}_{\alpha}^{1/2}Z_{T\alpha}\rVert=\sup_{\alpha\in A}\left(Z_{T\alpha}^{\prime}\mathcal{I}_{\alpha}Z_{T\alpha}\right)^{1/2}\leq\sup_{\alpha\in A}\left(\lambda_{\max}\left(\mathcal{I}_{\alpha}\right)\right)^{1/2}\sup_{\alpha\in A}\left\Vert Z_{T\alpha}\right\Vert \leq C\sup_{\alpha\in A}\left\Vert Z_{T\alpha}\right\Vert =O_{p}\left(1\right),\label{eq:Lemma 2_1_2}
\end{equation}
where the latter inequality holds with some finite $C$ in view of
Assumption \ref{assu:quadexp}(iii), and the last equality will be
justified below. Thus
\begin{equation}
\mathcal{I}_{\alpha}^{1/2}Z_{T\alpha}=O_{p\alpha}(1),\label{eq:Lemma 2 subresult}
\end{equation}
a result which also implies $\lVert(\mathcal{I}_{\alpha}^{1/2}Z_{T\alpha})^{\prime}\boldsymbol{\theta}_{T\alpha}\rVert\leq\lVert\boldsymbol{\theta}_{T\alpha}\rVert\lVert\mathcal{I}_{\alpha}^{1/2}Z_{T\alpha}\rVert=\lVert\boldsymbol{\theta}_{T\alpha}\rVert O_{p\alpha}(1)$. 

Next consider the third term on the right hand side of (\ref{eq: Lemma 2_1}),
where the assumption $(\hat{\beta}_{T\alpha},\hat{\pi}_{T\alpha},\hat{\varpi}_{T\alpha})=(\beta^{*},\pi^{\ast},0)+o_{p\alpha}(1)$
(see Assumption \ref{assu:pi-repam}(iii)) allows us to choose a sequence
$\{\gamma_{T},\ T\geq1\}$ of (non-random) positive scalars converging
to zero slowly enough to ensure that $P(\sup_{\alpha\in A}\lVert(\hat{\beta}_{T\alpha},\hat{\pi}_{T\alpha},\hat{\varpi}_{T\alpha})-(\beta^{*},\pi^{\ast},0)\rVert\leq\gamma_{T})\rightarrow1$,
and with this sequence $\{\gamma_{T},\ T\geq1\}$, Assumption \ref{assu:quadexp}(iv)
implies that
\begin{equation}
\lvert R_{T}(\alpha,\hat{\beta}_{T\alpha},\hat{\pi}_{T\alpha},\hat{\varpi}_{T\alpha})\rvert=(1+\lVert T^{1/2}\boldsymbol{\theta}(\alpha,\hat{\beta}_{T\alpha},\hat{\pi}_{T\alpha},\hat{\varpi}_{T\alpha})\rVert)^{2}o_{p\alpha}(1)\label{eq:Lemma 2_2}
\end{equation}
(cf. \citet[proof of Thm 3.3]{pakes1989simulation}). Here $\lVert T^{1/2}\boldsymbol{\theta}(\alpha,\hat{\beta}_{T\alpha},\hat{\pi}_{T\alpha},\hat{\varpi}_{T\alpha})\rVert=\lVert\mathcal{I}_{\alpha}^{-1/2}\boldsymbol{\theta}_{T\alpha}\rVert$
so that, as $0<\inf_{\alpha\in A}\lambda_{\min}(\mathcal{I}_{\alpha})$
by Assumption \ref{assu:quadexp}(iii),
\begin{equation}
\lvert R_{T}(\alpha,\hat{\beta}_{T\alpha},\hat{\pi}_{T\alpha},\hat{\varpi}_{T\alpha})\rvert\leq o_{p\alpha}(1)+\left\Vert \boldsymbol{\theta}_{T\alpha}\right\Vert o_{p\alpha}(1)+\left\Vert \boldsymbol{\theta}_{T\alpha}\right\Vert ^{2}o_{p\alpha}(1).\label{eq:Lemma 2_2_2}
\end{equation}
Combining the results above (see (\ref{eq: Lemma 2_1}), the inequality
below (\ref{eq:Lemma 2 subresult}), and (\ref{eq:Lemma 2_2_2})),
organizing terms, and absorbing $\left\Vert \boldsymbol{\theta}_{T\alpha}\right\Vert o_{p\alpha}(1)$
into $\left\Vert \boldsymbol{\theta}_{T\alpha}\right\Vert O_{p\alpha}(1)$
leads to
\begin{equation}
0\leq\left\Vert \boldsymbol{\theta}_{T\alpha}\right\Vert O_{p\alpha}(1)+o_{p\alpha}(1)+(o_{p\alpha}(1)-\frac{1}{2})\left\Vert \boldsymbol{\theta}_{T\alpha}\right\Vert ^{2}.\label{eq:Lemma 2_3}
\end{equation}
In Supplementary Appendix D we show that the last term on the right
hand side of (\ref{eq:Lemma 2_3}) is dominated by $-\frac{1}{4}\left\Vert \boldsymbol{\theta}_{T\alpha}\right\Vert ^{2}+o_{p\alpha}(1)$
so that (absorbing constants into the $O_{p\alpha}(1)$ and $o_{p\alpha}(1)$
terms) $\left\Vert \boldsymbol{\theta}_{T\alpha}\right\Vert ^{2}\leq2\left\Vert \boldsymbol{\theta}_{T\alpha}\right\Vert O_{p\alpha}(1)+o_{p\alpha}(1)$.
Denoting the $O_{p\alpha}(1)$ term on the majorant side with $\xi_{T\alpha}$
and reorganizing one obtains 
\[
(\left\Vert \boldsymbol{\theta}_{T\alpha}\right\Vert -\xi_{T\alpha})^{2}=\left\Vert \boldsymbol{\theta}_{T\alpha}\right\Vert ^{2}-2\left\Vert \boldsymbol{\theta}_{T\alpha}\right\Vert \xi_{T\alpha}+\xi_{T\alpha}^{2}\leq\xi_{T\alpha}^{2}+o_{p\alpha}(1)=O_{p\alpha}(1).
\]
Taking square roots yields $\left\Vert \boldsymbol{\theta}_{T\alpha}\right\Vert =O_{p\alpha}(1)$
so that $\lVert T^{1/2}\boldsymbol{\theta}(\alpha,\hat{\beta}_{T\alpha},\hat{\pi}_{T\alpha},\hat{\varpi}_{T\alpha})\rVert=O_{p\alpha}(1)$
as claimed in assertion (i) of the lemma. Assertion (ii) of the lemma
follows from the result $\left\Vert \boldsymbol{\theta}_{T\alpha}\right\Vert =O_{p\alpha}(1)$
and (\ref{eq:Lemma 2_2_2}). Assertion (iii) follows directly from
assertion (ii).

To complete the proof of Lemma \ref{lem:AsInsigOfRemainder}, we now
justify the last equality in (\ref{eq:Lemma 2_1_2}). By Assumption
\ref{assu:quadexp}(ii), $T^{-1/2}S_{T\bullet}\Rightarrow S_{\bullet}$
in $\mathcal{C}(A,\mathbb{R}^{r})$, and by the continuous mapping
theorem (justification in Supplementary Appendix D), $Z_{T\alpha}=\mathcal{I}_{\alpha}^{-1}T^{-1/2}S_{T\alpha}$
converges weakly in $\mathcal{C}(A,\mathbb{R}^{r})$ to a mean zero
$\mathbb{R}^{r}$-valued Gaussian process $Z_{\alpha}=\mathcal{I}_{\alpha}^{-\text{1}}S_{\alpha}$
whose sample paths are continuous in $\alpha$ with probability one
and that has $E[Z_{\alpha}Z_{\alpha}']=\mathcal{I}_{\alpha}^{-1}$
for all $\alpha\in A$. A further application of the continuous mapping
theorem (justification in Supplementary Appendix D) implies that $\sup_{\alpha\in A}\left\Vert Z_{T\alpha}\right\Vert $
converges in distribution in $\mathbb{R}$ and, as all probability
measures on $\mathbb{R}$ are tight, the limit must be tight. This
justifies the last equality in (\ref{eq:Lemma 2_1_2}).
\end{proof}
\smallskip{}
\begin{proof}[\textbf{\emph{Proof of Lemma \ref{lem:MaxLikVsMinQuadForm}}}]
 By the definition of $\boldsymbol{\hat{\lambda}}_{T\alpha q}$,
the fact that $\mathbf{0}\in\Theta_{\alpha,T}$, and (\ref{eq:Lemma 2 subresult}),
\[
\bigl\Vert\mathcal{I}_{\alpha}^{1/2}(\boldsymbol{\hat{\lambda}}_{T\alpha q}-Z_{T\alpha})\bigr\Vert^{2}=(\boldsymbol{\hat{\lambda}}_{T\alpha q}-Z_{T\alpha})^{\prime}\mathcal{I}_{\alpha}(\boldsymbol{\hat{\lambda}}_{T\alpha q}-Z_{T\alpha})\leq Z_{T\alpha}^{\prime}\mathcal{I}_{\alpha}Z_{T\alpha}+o_{p\alpha}(1)=O_{p\alpha}(1),
\]
implying that $\mathcal{I}_{\alpha}^{1/2}(\boldsymbol{\hat{\lambda}}_{T\alpha q}-Z_{T\alpha})=O_{p\alpha}(1)$.
Thus, a further use of (\ref{eq:Lemma 2 subresult}) and the condition
$0<\inf_{\alpha\in A}\lambda_{\min}(\mathcal{I}_{\alpha})$ in Assumption
\ref{assu:quadexp}(iii) yields $\boldsymbol{\hat{\lambda}}_{T\alpha q}=O_{p\alpha}(1)$. 

Next, we establish that $R_{T}(\alpha,\hat{\beta}_{T\alpha q},\hat{\pi}_{T\alpha q},\hat{\varpi}_{T\alpha q})=o_{p\alpha}(1)$.
First, $\boldsymbol{\hat{\lambda}}_{T\alpha q}=T^{1/2}\boldsymbol{\theta}(\alpha,\hat{\beta}_{T\alpha q},\hat{\pi}_{T\alpha q},\hat{\varpi}_{T\alpha q})$
and $\boldsymbol{\hat{\lambda}}_{T\alpha q}=O_{p\alpha}(1)$ imply
that $\boldsymbol{\theta}(\alpha,\hat{\beta}_{T\alpha q},\hat{\pi}_{T\alpha q},\hat{\varpi}_{T\alpha q})=o_{p\alpha}(1)$.
Second, to show that $(\hat{\beta}_{T\alpha q},\hat{\pi}_{T\alpha q},\hat{\varpi}_{T\alpha q})=(\beta^{*},\pi^{\ast},0)+o_{p\alpha}(1)$,
pick arbitrary $\epsilon,\delta>0$, and conclude from Assumption
\ref{assu:quadexp}(i)(b) that\linebreak{}
$\inf_{\alpha\in A}\inf$$_{(\beta,\pi,\varpi)\in B\times\Pi_{\alpha}:\left\Vert (\beta,\pi,\varpi)-(\beta^{*},\pi^{\ast},0)\right\Vert \geq\epsilon}\left\Vert \boldsymbol{\theta}(\alpha,\beta,\pi,\varpi)\right\Vert \geq\delta_{\epsilon}>0$
for some $\delta_{\epsilon}$. Now, as\linebreak{}
$\boldsymbol{\theta}(\alpha,\hat{\beta}_{T\alpha q},\hat{\pi}_{T\alpha q},\hat{\varpi}_{T\alpha q})=o_{p\alpha}(1)$,
we can find a $T_{\delta,\delta_{\epsilon}}$ such that for all $T\geq T_{\delta,\delta_{\epsilon}}$,\linebreak{}
$P(\sup_{\alpha\in A}\lVert\boldsymbol{\theta}(\alpha,\hat{\beta}_{T\alpha q},\hat{\pi}_{T\alpha q},\hat{\varpi}_{T\alpha q})\rVert<\delta_{\epsilon})>1-\delta$.
Note that whenever the event\linebreak{}
$\{\sup_{\alpha\in A}\lVert\boldsymbol{\theta}(\alpha,\hat{\beta}_{T\alpha q},\hat{\pi}_{T\alpha q},\hat{\varpi}_{T\alpha q})\rVert<\delta_{\epsilon}\}$
occurs, the event $\{\sup_{\alpha\in A}\lVert(\hat{\beta}_{T\alpha q},\hat{\pi}_{T\alpha q},\hat{\varpi}_{T\alpha q})-(\beta^{*},\pi^{\ast},0)\rVert<\epsilon\}$
must also occur (if, on the contrary, $\lVert(\hat{\beta}_{T\alpha q},\hat{\pi}_{T\alpha q},\hat{\varpi}_{T\alpha q})-(\beta^{*},\pi^{\ast},0)\rVert\geq\epsilon$
for some $\alpha\in A$, then necessarily $\lVert\boldsymbol{\theta}(\alpha,\hat{\beta}_{T\alpha q},\hat{\pi}_{T\alpha q},\hat{\varpi}_{T\alpha q})\rVert\geq\delta_{\epsilon}$)).
Therefore for all $T\geq T_{\delta,\delta_{\epsilon}}$,\linebreak{}
$1-\delta<P(\sup_{\alpha\in A}\lVert\boldsymbol{\theta}(\alpha,\hat{\beta}_{T\alpha q},\hat{\pi}_{T\alpha q},\hat{\varpi}_{T\alpha q})\rVert<\delta_{\epsilon})\leq P(\sup_{\alpha\in A}\lVert(\hat{\beta}_{T\alpha q},\hat{\pi}_{T\alpha q},\hat{\varpi}_{T\alpha q})-(\beta^{*},\pi^{\ast},0)\rVert<\epsilon)$,
so that $\sup_{\alpha\in A}\lVert(\hat{\beta}_{T\alpha q},\hat{\pi}_{T\alpha q},\hat{\varpi}_{T\alpha q})-(\beta^{*},\pi^{\ast},0)\rVert=o_{p}(1)$,
as desired. 

Third, as $(\hat{\beta}_{T\alpha q},\hat{\pi}_{T\alpha q},\hat{\varpi}_{T\alpha q})=(\beta^{*},\pi^{\ast},0)+o_{p\alpha}(1)$,
using the same argument as in the proof of Lemma \ref{lem:AsInsigOfRemainder}
(see the derivation of equation (\ref{eq:Lemma 2_2})) now leads to
\[
\lvert R_{T}(\alpha,\hat{\beta}_{T\alpha q},\hat{\pi}_{T\alpha q},\hat{\varpi}_{T\alpha q})\rvert=(1+\lVert T^{1/2}\boldsymbol{\theta}(\alpha,\hat{\beta}_{T\alpha q},\hat{\pi}_{T\alpha q},\hat{\varpi}_{T\alpha q})\rVert)^{2}o_{p\alpha}(1)=(1+\bigl\Vert\boldsymbol{\hat{\lambda}}_{T\alpha q}\bigr\Vert)^{2}o_{p\alpha}(1).
\]
This, together with the result $\boldsymbol{\hat{\lambda}}_{T\alpha q}=O_{p\alpha}(1)$
established above, yields $R_{T}(\alpha,\hat{\beta}_{T\alpha q},\hat{\pi}_{T\alpha q},\hat{\varpi}_{T\alpha q})=o_{p\alpha}(1)$.

Now, by expansion (\ref{eq:Quadr-Exp_2}), the definitions of $(\hat{\beta}_{T\alpha},\hat{\pi}_{T\alpha},\hat{\varpi}_{T\alpha})$
and $\boldsymbol{\hat{\lambda}}_{T\alpha q}$, and making use of results
$R_{T}(\alpha,\hat{\beta}_{T\alpha},\hat{\pi}_{T\alpha},\hat{\varpi}_{T\alpha})=o_{p\alpha}(1)$
and $R_{T}(\alpha,\hat{\beta}_{T\alpha q},\hat{\pi}_{T\alpha q},\hat{\varpi}_{T\alpha q})=o_{p\alpha}(1)$,
\begin{align*}
o_{p\alpha}(1) & \leq L_{T}^{\pi}(\alpha,\hat{\beta}_{T\alpha},\hat{\pi}_{T\alpha},\hat{\varpi}_{T\alpha})-L_{T}^{\pi}(\alpha,\hat{\beta}_{T\alpha q},\hat{\pi}_{T\alpha q},\hat{\varpi}_{T\alpha q})\\
 & =\frac{1}{2}(\boldsymbol{\hat{\lambda}}_{T\alpha q}-Z_{T\alpha})^{\prime}\mathcal{I}_{\alpha}(\boldsymbol{\hat{\lambda}}_{T\alpha q}-Z_{T\alpha})\\
 & \quad-\frac{1}{2}\bigl[T^{1/2}\boldsymbol{\theta}(\alpha,\hat{\beta}_{T\alpha},\hat{\pi}_{T\alpha},\hat{\varpi}_{T\alpha})-Z_{T\alpha}\bigr]'\mathcal{I}_{\alpha}\bigl[T^{1/2}\boldsymbol{\theta}(\alpha,\hat{\beta}_{T\alpha},\hat{\pi}_{T\alpha},\hat{\varpi}_{T\alpha})-Z_{T\alpha}\bigr]+o_{p\alpha}(1)\\
 & \leq o_{p\alpha}(1),
\end{align*}
implying, by (\ref{eq:Lemma 2 kaava}), the desired result.
\end{proof}
\smallskip{}
\begin{proof}[\textbf{\emph{Proof of Lemma \ref{lem:Cone}}}]
\textbf{\emph{}} For any vectors $a,b\in\mathbb{R}^{r}$, we denote
$\left\Vert a-b\right\Vert _{\mathcal{I}_{\alpha}^{-1}}=[(a-b)^{\prime}\mathcal{I}_{\alpha}(a-b)]^{1/2}$,
and for any point $p\in\mathbb{R}^{r}$ and a set $S\subset\mathbb{R}^{r}$,
we define $\left\Vert p-S\right\Vert _{\mathcal{I}_{\alpha}^{-1}}$
via $\left\Vert p-S\right\Vert _{\mathcal{I}_{\alpha}^{-1}}^{2}=\inf_{s\in S}\left\Vert s-p\right\Vert _{\mathcal{I}_{\alpha}^{-1}}^{2}=\inf_{s\in S}[(s-p)^{\prime}\mathcal{I}_{\alpha}(s-p)]$.
With this notation, we need to prove that $\left\Vert Z_{T\alpha}-\Theta_{\alpha,T}\right\Vert _{\mathcal{I}_{\alpha}^{-1}}^{2}=\left\Vert Z_{T\alpha}-\Lambda\right\Vert _{\mathcal{I}_{\alpha}^{-1}}^{2}+o_{p\alpha}(1)$.
First note that, because $\Lambda$ is a cone, we have, for any $T$,
\[
\left\Vert Z_{T\alpha}-\Lambda\right\Vert _{\mathcal{I}_{\alpha}^{-1}}^{2}=T\inf_{\boldsymbol{\lambda}\in\Lambda}\bigl\{(T^{-1/2}\boldsymbol{\lambda}-T^{-1/2}Z_{T\alpha})^{\prime}\mathcal{I}_{\alpha}(T^{-1/2}\boldsymbol{\lambda}-T^{-1/2}Z_{T\alpha})\bigr\}=T\bigl\Vert T^{-1/2}Z_{T\alpha}-\Lambda\bigr\Vert_{\mathcal{I}_{\alpha}^{-1}}^{2}.
\]
Similarly, by the definitions of $\Theta_{\alpha,T}$ and $\Theta_{\alpha}$,
\[
\left\Vert Z_{T\alpha}-\Theta_{\alpha,T}\right\Vert _{\mathcal{I}_{\alpha}^{-1}}^{2}=\inf_{\boldsymbol{\lambda}\in\Theta_{\alpha}}\bigl\{(T^{1/2}\boldsymbol{\lambda}-Z_{T\alpha})^{\prime}\mathcal{I}_{\alpha}(T^{1/2}\boldsymbol{\lambda}-Z_{T\alpha})\bigr\}=T\bigl\Vert T^{-1/2}Z_{T\alpha}-\Theta_{\alpha}\bigr\Vert_{\mathcal{I}_{\alpha}^{-1}}^{2}.
\]

Now let $G_{T}(\alpha,\boldsymbol{x})=T\left\Vert \boldsymbol{x}-\Theta_{\alpha}\right\Vert _{\mathcal{I}_{\alpha}^{-1}}^{2}-T\left\Vert \boldsymbol{x}-\Lambda\right\Vert _{\mathcal{I}_{\alpha}^{-1}}^{2}$
define a (non-random) function on $A\times\mathbb{R}^{r}$. Because
$\{\Theta_{\alpha},\ \alpha\in A\}$ is locally uniformly equal to
the cone $\Lambda$, we can find a $\delta>0$ such that $\Theta_{\alpha}\cap(-\delta,\delta)^{r}=\Lambda\cap(-\delta,\delta)^{r}$
for all $\alpha\in A$. Furthermore, $\mathbf{0}\in\Theta_{\alpha}$
and $\mathbf{0}\in\Lambda$ (here $\mathbf{0}\in\mathbb{R}^{r}$).
Therefore, we can find a neighborhood $N_{\mathbf{0}}$ of $\mathbf{0}$
such that for all $(\alpha,\boldsymbol{x})\in A\times N_{\mathbf{0}}$,
\[
G_{T}(\alpha,\boldsymbol{x})=T\left\Vert \boldsymbol{x}-\Theta_{\alpha}\cap(-\delta,\delta)^{r}\right\Vert _{\mathcal{I}_{\alpha}^{-1}}^{2}-T\left\Vert \boldsymbol{x}-\Lambda\cap(-\delta,\delta)^{r}\right\Vert _{\mathcal{I}_{\alpha}^{-1}}^{2}=0.
\]
Now define $G_{T}(\alpha)$, a random function of $\alpha$, as $G_{T}(\alpha)=G_{T}(\alpha,T^{-1/2}Z_{T\alpha})$.
In the proof of Lemma \ref{lem:AsInsigOfRemainder} it was shown that
$\sup_{\alpha\in A}\lVert Z_{T\alpha}\rVert=O_{p}(1)$ (see (\ref{eq:Lemma 2_1_2}))
so that $T^{-1/2}\lVert Z_{T\alpha}\rVert=o_{p\alpha}(1)$. Therefore,
for all $\epsilon>0$,
\begin{eqnarray*}
P\bigl(\sup\nolimits _{\alpha\in A}\left\vert G_{T}\left(\alpha\right)\right\vert >\epsilon\bigr) & \leq & P\bigl(\sup\nolimits _{\alpha\in A}\left\vert G_{T}\left(\alpha\right)\right\vert >\epsilon\text{ ; }T^{-1/2}\sup\nolimits _{\alpha\in A}\left\Vert Z_{T\alpha}\right\Vert \in N_{0}\bigr)\\
 &  & +P\bigl(\sup\nolimits _{\alpha\in A}\left\vert G_{T}\left(\alpha\right)\right\vert >\epsilon\text{ ; }T^{-1/2}\sup\nolimits _{\alpha\in A}\left\Vert Z_{T\alpha}\right\Vert \notin N_{0}\bigr)\\
 & = & P\bigl(\sup\nolimits _{\alpha\in A}\left\vert G_{T}\left(\alpha\right)\right\vert >\epsilon\text{ ; }T^{-1/2}\sup\nolimits _{\alpha\in A}\left\Vert Z_{T\alpha}\right\Vert \notin N_{0}\bigr)\\
 & \leq & P\bigl(T^{-1/2}\sup\nolimits _{\alpha\in A}\left\Vert Z_{T\alpha}\right\Vert \notin N_{0}\bigr)\\
 & \rightarrow & 0,
\end{eqnarray*}
where the equality holds because $G_{T}(\alpha,\boldsymbol{x})=0$
for all $(\alpha,\boldsymbol{x})\in A\times N_{\mathbf{0}}$, and
the convergence holds because $T^{-1/2}\lVert Z_{T\alpha}\rVert=o_{p\alpha}(1)$.
Thus $\sup_{\alpha\in A}\left\vert G_{T}\left(\alpha\right)\right\vert =o_{p}(1)$,
implying the desired result $\left\Vert Z_{T\alpha}-\Theta_{\alpha,T}\right\Vert _{\mathcal{I}_{\alpha}^{-1}}^{2}=\left\Vert Z_{T\alpha}-\Lambda\right\Vert _{\mathcal{I}_{\alpha}^{-1}}^{2}+o_{p\alpha}(1)$.
\end{proof}
\smallskip{}
\begin{proof}[\textbf{\emph{Proof of Lemma \ref{lem:WeakConv}}}]
 It was shown in the proof of Lemma \ref{lem:AsInsigOfRemainder}
that $Z_{T\bullet}\Rightarrow Z_{\bullet}$ in $\mathcal{C}(A,\mathbb{R}^{r})$.
Therefore also $(Z_{T\bullet},\mathcal{I}_{\bullet})\Rightarrow(Z_{\bullet},\mathcal{I}_{\bullet})$
in $\mathcal{C}(A,\mathbb{R}^{r})\times\{\mathcal{I}_{\alpha}\}$
(\citet[Thm. 3.9]{billingsley1999convergence}). As the function $g:\mathcal{C}(A,\mathbb{R}^{r})\times\{\mathcal{I}_{\alpha}\}\to\mathcal{B}(A,\mathbb{R})$
mapping $(x_{\bullet},\mathcal{I}_{\bullet})$ ($\in\mathcal{C}(A,\mathbb{R}^{r})\times\{\mathcal{I}_{\alpha}\}$)
to $x_{\bullet}'\mathcal{I}_{\bullet}x_{\bullet}-\inf_{\boldsymbol{\lambda}\in\Lambda}\left\{ (\boldsymbol{\lambda}-x_{\bullet})^{\prime}\mathcal{I}_{\bullet}(\boldsymbol{\lambda}-x_{\bullet})\right\} $
is continuous (justification in Supplementary Appendix D), the continuous
mapping theorem is applicable. This, together with \citet[Thm 3.1]{billingsley1999convergence}
(for which it is necessary that the remainder term in (\ref{eq:BeforeLemmaWeakConv})
is $o_{p\alpha}(1)$ and not only $o_{p}(1)$), implies that 
\[
2[L_{T}^{\pi}(\bullet,\hat{\beta}_{T\bullet},\hat{\pi}_{T\bullet},\hat{\varpi}_{T\bullet})-L_{T}^{\pi}(\bullet,\beta^{*},\pi^{\ast},0)]\Rightarrow Z_{\bullet}^{\prime}\mathcal{I}_{\bullet}Z_{\bullet}-\inf\nolimits _{\boldsymbol{\lambda}\in\Lambda}\left\{ (\boldsymbol{\lambda}-Z_{\bullet})^{\prime}\mathcal{I}_{\bullet}(\boldsymbol{\lambda}-Z_{\bullet})\right\} ,
\]
establishing the desired result.
\end{proof}
\smallskip{}
\begin{proof}[\textbf{\emph{Proof of Lemma \ref{lem:WeakLimitSubvec}}}]
The proof consists of reasonably straightforward matrix algebra.
For details, see Supplementary Appendix D.
\end{proof}
\smallskip{}
\begin{proof}[\textbf{\emph{Proof of Lemma \ref{lem:WeakConvLinear}}}]
 The required arguments are standard but presented for completeness
and to contrast them with arguments that lead to Lemma \ref{lem:WeakConv}.
The reparameterization described in Assumption \ref{assu:pi-repam}
is unnecessary and the original $\tilde{\phi}$\textendash parameterization
may be used (alternatively, consider the identity mapping $\pi=\boldsymbol{\pi}(\tilde{\phi})=\tilde{\phi}$).
As for the quadratic expansion of the log-likelihood function, let
$\boldsymbol{\theta}(\tilde{\phi})=(\tilde{\phi}-\tilde{\phi}^{\ast})$
take the role of $\boldsymbol{\theta}(\alpha,\beta,\pi,\varpi)$,
and note that straightforward derivations (similar to those used in
the LMAR example in Section 3.3.1) yield 
\begin{align*}
L_{T}^{0}(\tilde{\phi})-L_{T}^{0}(\tilde{\phi}^{\ast}) & =(T^{-1/2}S_{T}^{0})'[T^{1/2}\boldsymbol{\theta}(\tilde{\phi})]-\frac{1}{2}[T^{1/2}\boldsymbol{\theta}(\tilde{\phi})]^{\prime}\mathcal{I}^{0}[T^{1/2}\boldsymbol{\theta}(\tilde{\phi})]+R_{T}(\tilde{\phi}),\\
R_{T}(\tilde{\phi}) & =\frac{1}{2}[T^{1/2}\boldsymbol{\theta}(\tilde{\phi})]^{\prime}[T^{-1}\nabla_{\phi\phi'}L_{T}^{0}(\dot{\phi})-(-\mathcal{I}^{0})][T^{1/2}\boldsymbol{\theta}(\tilde{\phi})],
\end{align*}
with $\dot{\phi}$ denoting a point between $\tilde{\phi}$ and $\tilde{\phi}^{*}$.
Validity of Assumption \ref{assu:quadexp} follows from the arguments
used in connection with the LMAR example together with Assumption
\ref{assu:ScoreEqualsLinearScore}. Assumption \ref{assu:cone} holds
with $\Lambda=\mathbb{R}^{p+2}$. Arguments analogous to those that
lead to Lemma \ref{lem:WeakConv} now yield the stated convergence
result, and the convergence is joint as in both cases it follows from
the weak convergence result $T^{-1/2}S_{T\bullet}\Rightarrow S_{\bullet}$. 
\end{proof}
\smallskip{}
\begin{proof}[\textbf{\emph{Proof of Theorem 1}}]
 Under Assumption \ref{assu:ScoreEqualsLinearScore}, the random
process $S_{\theta\alpha}'\mathcal{I}_{\theta\theta\alpha}^{-1}S_{\theta\alpha}$
in Lemma \ref{lem:WeakLimitSubvec} coincides with the random variable
$S^{0\prime}(\mathcal{I}^{0})^{-1}S^{0}$ in Lemma \ref{lem:WeakConvLinear}.
Therefore the expression of $LR_{T}(\alpha)$ in (\ref{eq:LR_1}),
Lemmas \ref{lem:WeakConv}, \ref{lem:WeakLimitSubvec}, and \ref{lem:WeakConvLinear},
and \citet[Thm 3.1]{billingsley1999convergence} (for which it is
necessary that the remainder term in (\ref{eq:LR_1}) is $o_{p\alpha}(1)$
and not only $o_{p}(1)$) imply the weak convergence result for $LR_{T}(\alpha)$.
The result for $LR_{T}$ follows from the continuous mapping theorem.
\end{proof}
\smallskip{}

\section{Details for the LMAR example}

In this appendix it appears convenient to denote $\alpha_{1,t}^{L}$
instead of $\alpha_{t}^{L}$ and to set $\alpha_{2,t}^{L}=1-\alpha_{1,t}^{L}$.
In some cases we also include the argument $\alpha$ and denote $\alpha_{1,t}^{L}(\alpha)$
and $\alpha_{2,t}^{L}(\alpha)$. The same notation is employed in
the Supplementary Appendix and a similar modification is used in the
case of the GMAR model. 

\paragraph{Assumptions 1\textendash 4.}

Assumption 1(i) is assumed to hold, 1(ii) holds as $A$ is compact,
and 1(iii) holds by the definition of the mixing weight. For the verification
of Assumption 2, see the GMAR example in Appendix C; the LMAR case
is treated there as well. To verify Assumption \ref{assu:pi-repam},
note first that conditions (i) and (ii) clearly hold, and for condition
(iii), we have 
\[
\boldsymbol{\pi}_{\alpha}(\phi,\varphi)-\boldsymbol{\pi}_{\alpha}(\phi^{\ast},\phi^{\ast})=(\phi,\phi-\varphi)-(\phi^{\ast},0)=\left(\phi-\phi^{\ast},\left(\phi-\phi^{\ast}\right)-\left(\varphi-\phi^{\ast}\right)\right)\mbox{\ensuremath{\alpha_{1,t}^{L}} }.
\]
Choosing $\left\Vert x\right\Vert _{*}=\left\Vert x\right\Vert _{1}=\sum_{i=1}^{2q}\left|x_{i}\right|$
and using the triangle inequality it is straightforward to check that
\[
\left\Vert \left(\phi-\phi^{\ast},\left(\phi-\phi^{\ast}\right)-\left(\varphi-\phi^{\ast}\right)\right)\right\Vert \leq2\left\Vert \phi-\phi^{\ast}\right\Vert +\left\Vert \varphi-\phi^{\ast}\right\Vert \leq2\left\Vert (\phi,\varphi)-(\phi^{\ast},\phi^{\ast})\right\Vert _{1}.
\]
Thus, Assumption \ref{assu:pi-repam}(iii) holds by Lemma \ref{lem:LipCondForUnifCons}.
Regarding Assumption \ref{assu:cont_differentiability}, as $\alpha_{1,t}^{L}$
does not depend on $(\phi,\varphi)$ and $\boldsymbol{\pi}_{\alpha}^{-1}(\pi,\varpi)=(\pi,\pi-\varpi)$,
the required differentiability conditions hold for all positive integers
$k$. 

\paragraph{Assumption 5: Computation of the required derivatives.}

As $\alpha_{1,t}^{L}$ does not depend on $(\phi,\varphi)$ and $\boldsymbol{\pi}_{\alpha}^{-1}(\pi,\varpi)=(\pi,\pi-\varpi)$,
the quantities $f_{2,t}^{\pi}(\alpha,\pi,\varpi)$ and $l_{t}^{\pi}(\alpha,\pi,\varpi)$
take the form 
\begin{align*}
f_{2,t}^{\pi}(\alpha,\pi,\varpi) & =\alpha_{1,t}^{L}(\alpha)f_{t}(\pi)+(1-\alpha_{1,t}^{L}(\alpha))f_{t}(\pi-\varpi),\\
l_{t}^{\pi}(\alpha,\pi,\varpi) & =\log[\alpha_{1,t}^{L}(\alpha)f_{t}(\pi)+(1-\alpha_{1,t}^{L}(\alpha))f_{t}(\pi-\varpi)].
\end{align*}
Straightforward differentiation yields the following expressions for
the first and second partial derivatives with respect to $\pi$ and
$\varpi$ (recall that $\nabla$ and $\nabla^{2}$ denote first and
second order differentiation with respect to the indicated parameters,
and $\nabla f_{t}(\cdot)$ denotes differentiation of $f_{t}(\cdot)$
in (\ref{eq:f_t}) with respect to $\tilde{\phi}=(\tilde{\phi}_{0},\tilde{\phi}_{1},\ldots,\tilde{\phi}_{p},\tilde{\sigma}_{1}^{2})$):
\begin{align*}
\nabla_{\pi}l_{t}^{\pi}(\alpha,\pi,\varpi) & =[\alpha_{1,t}^{L}(\alpha)\nabla f_{t}(\pi)+\alpha_{2,t}^{L}(\alpha)\nabla f_{t}(\pi-\varpi)]/f_{2,t}^{\pi}(\alpha,\pi,\varpi),\\
\nabla_{\varpi}l_{t}^{\pi}(\alpha,\pi,\varpi) & =[-\alpha_{2,t}^{L}(\alpha)\nabla f_{t}(\pi-\varpi)]/f_{2,t}^{\pi}(\alpha,\pi,\varpi),\\
\nabla_{\pi\pi'}^{2}l_{t}^{\pi}(\alpha,\pi,\varpi) & =\alpha_{1,t}^{L}(\alpha)\biggl(\frac{\nabla^{2}f_{t}(\pi)}{f_{2,t}^{\pi}(\alpha,\pi,\varpi)}-\frac{\nabla f_{t}(\pi)}{f_{2,t}^{\pi}(\alpha,\pi,\varpi)}\frac{\alpha_{1,t}^{L}(\alpha)\nabla'f_{t}(\pi)+\alpha_{2,t}^{L}(\alpha)\nabla'f_{t}(\pi-\varpi)}{f_{2,t}^{\pi}(\alpha,\pi,\varpi)}\biggr)\\
 & +\alpha_{2,t}^{L}(\alpha)\biggl(\frac{\nabla^{2}f_{t}(\pi-\varpi)}{f_{2,t}^{\pi}(\alpha,\pi,\varpi)}-\frac{\nabla f_{t}(\pi-\varpi)}{f_{2,t}^{\pi}(\alpha,\pi,\varpi)}\frac{\alpha_{1,t}^{L}(\alpha)\nabla'f_{t}(\pi)+\alpha_{2,t}^{L}(\alpha)\nabla'f_{t}(\pi-\varpi)}{f_{2,t}^{\pi}(\alpha,\pi,\varpi)}\biggr),\\
\nabla_{\pi\varpi'}^{2}l_{t}^{\pi}(\alpha,\pi,\varpi) & =\alpha_{1,t}^{L}(\alpha)\biggl(-\frac{\nabla f_{t}(\pi)}{f_{2,t}^{\pi}(\alpha,\pi,\varpi)}\frac{-\alpha_{2,t}^{L}(\alpha)\nabla'f_{t}(\pi-\varpi)}{f_{2,t}^{\pi}(\alpha,\pi,\varpi)}\biggr)\\
 & +\alpha_{2,t}^{L}(\alpha)\biggl(-\frac{\nabla^{2}f_{t}(\pi-\varpi)}{f_{2,t}^{\pi}(\alpha,\pi,\varpi)}+\frac{\nabla f_{t}(\pi-\varpi)}{f_{2,t}^{\pi}(\alpha,\pi,\varpi)}\frac{\alpha_{2,t}^{L}(\alpha)\nabla'f_{t}(\pi-\varpi)}{f_{2,t}^{\pi}(\alpha,\pi,\varpi)}\biggr),\\
\nabla_{\varpi\varpi'}^{2}l_{t}^{\pi}(\alpha,\pi,\varpi) & =-\alpha_{2,t}^{L}(\alpha)\biggl(-\frac{\nabla^{2}f_{t}(\pi-\varpi)}{f_{2,t}^{\pi}(\alpha,\pi,\varpi)}+\frac{\nabla f_{t}(\pi-\varpi)}{f_{2,t}^{\pi}(\alpha,\pi,\varpi)}\frac{\alpha_{2,t}^{L}(\alpha)\nabla'f_{t}(\pi-\varpi)}{f_{2,t}^{\pi}(\alpha,\pi,\varpi)}\biggr).
\end{align*}
The corresponding expressions evaluated at $(\alpha,\pi,\varpi)=(\alpha,\pi^{\ast},0)$
take the form 
\begin{align}
\nabla_{(\pi,\varpi)}l_{t}^{\pi}(\alpha,\pi^{\ast},0) & =(1,-(1-\alpha_{1,t}^{L}(\alpha)))\otimes\frac{\nabla f_{t}(\pi^{\ast})}{f_{t}(\pi^{\ast})},\label{eq:LMARDer1*}\\
\nabla_{\pi\pi'}^{2}l_{t}^{\pi}(\alpha,\pi^{\ast},0) & =\frac{\nabla^{2}f_{t}(\pi^{\ast})}{f_{t}(\pi^{\ast})}-\frac{\nabla f_{t}(\pi^{\ast})}{f_{t}(\pi^{\ast})}\frac{\nabla'f_{t}(\pi^{\ast})}{f_{t}(\pi^{\ast})},\label{eq:LMARDer2a*}\\
\nabla_{\pi\varpi'}^{2}l_{t}^{\pi}(\alpha,\pi^{\ast},0) & =-(1-\alpha_{1,t}^{L}(\alpha))\frac{\nabla^{2}f_{t}(\pi^{\ast})}{f_{t}(\pi^{\ast})}+(1-\alpha_{1,t}^{L}(\alpha))\frac{\nabla f_{t}(\pi^{\ast})}{f_{t}(\pi^{\ast})}\frac{\nabla'f_{t}(\pi^{\ast})}{f_{t}(\pi^{\ast})},\label{eq:LMARDer2b*}\\
\nabla_{\varpi\varpi'}^{2}l_{t}^{\pi}(\alpha,\pi^{\ast},0) & =(1-\alpha_{1,t}^{L}(\alpha))\frac{\nabla^{2}f_{t}(\pi^{\ast})}{f_{t}(\pi^{\ast})}-(1-\alpha_{1,t}^{L}(\alpha))^{2}\frac{\nabla f_{t}(\pi^{\ast})}{f_{t}(\pi^{\ast})}\frac{\nabla'f_{t}(\pi^{\ast})}{f_{t}(\pi^{\ast})}.\label{eq:LMARDer2c*}
\end{align}

\paragraph{Assumption 5: Verifying the assumption.}

Omitting the unnecessary $\beta$ from $\boldsymbol{\theta}(\alpha,\beta,\pi,\varpi)$,
we have $\boldsymbol{\theta}(\alpha,\pi,\varpi)=(\pi-\pi^{\ast},\varpi)=(\theta,\vartheta)$
so that part (i) is clearly satisfied with the parameter space 
\begin{align*}
\Theta_{\alpha}=\Theta & =\{\boldsymbol{\theta}=(\theta,\vartheta)\in\mathbb{R}^{2q_{2}}:\theta=\pi-\pi^{\ast},\vartheta=\varpi\text{ for some }(\pi,\varpi)\in\Pi\}\\
 & =\{\boldsymbol{\theta}=(\theta,\vartheta)\in\mathbb{R}^{2q_{2}}:\theta=\phi-\phi^{\ast},\vartheta=\phi-\varphi\text{ for some }(\phi,\varphi)\in\Phi\times\Phi\}
\end{align*}
independent of $\alpha$ and with $0$ ($\in\mathbb{R}^{2q_{2}}$)
an interior point of $\Theta$. The first two requirements in part
(ii) are similarly clear, whereas the third requirement follows from
the continuity of $\alpha_{1,t}^{L}(\alpha)$ in $\alpha$. The weak
convergence requirement in part (ii) is verified in Supplementary
Appendix E.1.

Now consider part (iii) of Assumption \ref{assu:quadexp}, and first
consider the positive definiteness of $\mathcal{I}_{\alpha}$ for
a fixed $\alpha\in A$. It suffices to show that
\begin{equation}
a'{\color{magenta}}\frac{\nabla f_{t}(\pi^{\ast})}{f_{t}(\pi^{\ast})}=\alpha_{2,t}^{L}(\alpha)b'\frac{\nabla f_{t}(\pi^{\ast})}{f_{t}(\pi^{\ast})}\,\,\,\textrm{a.s.}\label{eq:LMAR_PD_1}
\end{equation}
only if $a=(a_{1},\ldots,a_{q_{2}})=0$ and $b=(b_{1},\ldots,b_{q_{2}})=0$.
For brevity, denote $g_{t}(\pi)=[y_{t}-(\pi_{1}+\pi_{2}y_{t-1}+\cdots+\pi_{p+1}y_{t-p})]/\pi_{p+2}^{1/2}=[y_{t}-(\tilde{\phi}_{0}+\tilde{\phi}_{1}y_{t-1}+\cdots+\tilde{\phi}_{p}y_{t-p})]/\tilde{\sigma}_{1}$
and $\boldsymbol{z}_{t-1}=(1,\boldsymbol{y}_{t-1})$ so that $g_{t}(\pi^{*})=\varepsilon_{t}$.
Straightforward differentiation yields 
\begin{equation}
\frac{\nabla f_{t}(\pi)}{f_{t}(\pi)}=\left[\begin{array}{c}
\frac{1}{\tilde{\sigma}_{1}}\boldsymbol{z}_{t-1}g_{t}(\pi)\\
\frac{1}{2\tilde{\sigma}_{1}^{2}}(g_{t}^{2}(\pi)-1)
\end{array}\right]\,\,\,\textrm{so that}\,\,\,\frac{\nabla f_{t}(\pi^{*})}{f_{t}(\pi^{*})}=\left[\begin{array}{c}
\frac{1}{\tilde{\sigma}_{1}^{*}}\boldsymbol{z}_{t-1}\varepsilon_{t}\\
\frac{1}{2\tilde{\sigma}_{1}^{*2}}(\varepsilon_{t}^{2}-1)
\end{array}\right]\label{eq:LMAR_nablaf_per_f}
\end{equation}
(where $\tilde{\sigma}_{1}^{2}=\pi_{p+2}$). Multiplying both sides
of equation (\ref{eq:LMAR_PD_1}) by $(\varepsilon_{t}^{2}-1)2\tilde{\sigma}_{1}^{*2}$,
taking expectations conditional on $\mathcal{F}_{t-1}$, and making
use of the fact that odd moments of the normal distribution are zero,
yields $a_{q_{2}}E[(\varepsilon_{t}^{2}-1)^{2}]=\alpha_{2,t}^{L}(\alpha)b_{q_{2}}E[(\varepsilon_{t}^{2}-1)^{2}]$
a.s. Because $\alpha_{2,t}^{L}(\alpha)\neq0$ and not equal to a constant
(see Section 3.1.1), it follows that $a_{q_{2}}=b_{q_{2}}=0$. Therefore,
equation (\ref{eq:LMAR_PD_1}) (multiplied by $\sigma_{1}^{*}$) now
reduces to $(a_{1},\ldots,a_{q_{2}-1})'\boldsymbol{z}_{t-1}\varepsilon_{t}=\alpha_{2,t}^{L}(\alpha)(b_{1},\ldots,b_{q_{2}-1})'\boldsymbol{z}_{t-1}\varepsilon_{t}$
a.s. Multiplying this equation by $\varepsilon_{t}$, taking expectations
conditional on $\mathcal{F}_{t-1}$, and dividing by $E[\varepsilon_{t}^{2}]=\sigma_{1}^{*2}$
yields $(a_{1},\ldots,a_{q_{2}-1})'\boldsymbol{z}_{t-1}=\alpha_{2,t}^{L}(\alpha)(b_{1},\ldots,b_{q_{2}-1})'\boldsymbol{z}_{t-1}$
a.s. This is clearly impossible unless $a_{1}=\ldots=a_{q_{2}-1}=b_{1}=\ldots=b_{q_{2}-1}=0$,
because $\alpha_{2,t}^{L}(\alpha)$ is a positive and strictly decreasing
function of $\alpha'\boldsymbol{z}_{t-1}$ ($\neq\alpha_{0}$) and
because $(a_{1},\ldots,a_{q_{2}-1})'\boldsymbol{z}_{t-1}$ and $(b_{1},\ldots,b_{q_{2}-1})'\boldsymbol{z}_{t-1}$
are normally distributed or constants (if only $a_{1}$ and $b_{1}$
are nonzero). Therefore, $a=b=0$, so that $\mathcal{I}_{\alpha}$
is positive definite (for any fixed $\alpha\in A$). 

To complete the verification of part (iii), we show that $\mathcal{I}_{\alpha}$
is a continuous function of $\alpha$ and such that $0<\inf_{\alpha\in A}\lambda_{\min}(\mathcal{I}_{\alpha})$
and $\sup{}_{\alpha\in A}\lambda_{\max}(\mathcal{I}_{\alpha})<\infty$.
For continuity, let $\alpha_{n}$ be a sequence of points in $A$
converging to $\alpha_{\bullet}\in A$. It suffices to demonstrate
that $\mathrm{\textrm{\textrm{lim}}}{}_{n\rightarrow\infty}E\bigl[\alpha_{2,t}^{L}(\alpha_{n})\frac{\nabla f_{t}(\pi^{\ast})}{f_{t}(\pi^{\ast})}\frac{\nabla'f_{t}(\pi^{\ast})}{f_{t}(\pi^{\ast})}\bigr]=E\bigl[\alpha_{2,t}^{L}(\alpha_{\bullet})\frac{\nabla f_{t}(\pi^{\ast})}{f_{t}(\pi^{\ast})}\frac{\nabla'f_{t}(\pi^{\ast})}{f_{t}(\pi^{\ast})}\bigr]$
and similarly with $\alpha_{2,t}^{L}(\cdot)$ replaced by its square.
This, however, is an immediate consequence of the dominated convergence
theorem because $\alpha_{2,t}^{L}(\alpha)$ is a continuous positive
function of $\alpha$ and smaller than 1, and because $E\bigl[\bigl\|\frac{\nabla f_{t}(\pi^{\ast})}{f_{t}(\pi^{\ast})}\frac{\nabla'f_{t}(\pi^{\ast})}{f_{t}(\pi^{\ast})}\bigr\|\bigr]<\infty$
due to Lemma \ref{lem:XXX1} (in Supplementary Appendix F.5). The
statements on the eigenvalues follow from the continuity of $\mathcal{I}_{\alpha}$,
the compactness of its domain $A$, and the positive definiteness
of $\mathcal{I}_{\alpha}$ for all fixed $\alpha\in A$ shown above. 

As for part (iv) of Assumption \ref{assu:quadexp}, based on the expression
of the remainder term in (\ref{LMAR_RemainderTerm}) it suffices to
show that for all sequences of (non-random) positive scalars $\{\gamma_{T},\ T\geq1\}$
for which $\gamma_{T}\rightarrow0$ as $T\rightarrow\infty$, 
\begin{equation}
\sup{}_{(\pi,\varpi)\in\Pi:\left\Vert (\pi,\varpi)-(\pi^{\ast},0)\right\Vert \leq\gamma_{T}}\bigl\Vert T^{-1}\nabla_{(\pi,\varpi)(\pi,\varpi)'}^{2}L_{T}^{\pi}(\alpha,\pi,\varpi)-(-\mathcal{I}_{\alpha})\bigr\Vert=o_{p\alpha}(1),\label{eq:LMAR_RemainderTermDetails1}
\end{equation}
where the parameter space of $(\pi,\varpi)$, denoted by $\Pi$, is
independent of $\alpha$, as noted above. First we show that a uniform
law of large numbers applies to the matrix $T^{-1}\nabla_{(\pi,\varpi)(\pi,\varpi)'}^{2}L_{T}^{\pi}(\alpha,\pi,\varpi)$
on $A\times\Pi$, that is, \textbf{
\begin{equation}
\sup{}_{\alpha\in A}\sup\nolimits _{(\pi,\varpi)\in\Pi}\bigl\Vert T^{-1}\nabla_{(\pi,\varpi)(\pi,\varpi)'}^{2}L_{T}^{\pi}(\alpha,\pi,\varpi)-E[\nabla_{(\pi,\varpi)(\pi,\varpi)'}^{2}l_{t}^{\pi}(\alpha,\pi,\varpi)]\bigr\Vert=o_{p}(1).\label{eq:LMAR_RemainderTermDetails2}
\end{equation}
}As $T^{-1}\nabla_{(\pi,\varpi)(\pi,\varpi)'}^{2}L_{T}^{\pi}(\alpha,\pi,\varpi)=T^{-1}\sum_{t=1}^{T}\nabla_{(\pi,\varpi)(\pi,\varpi)'}^{2}l_{t}^{\pi}(\alpha,\pi,\varpi)$
with $\nabla_{(\pi,\varpi)(\pi,\varpi)'}^{2}l_{t}^{\pi}(\alpha,\pi,\varpi)$
a function of the stationary and ergodic process $(y_{t},\boldsymbol{y}_{t-1})$
(by Assumption 1(i)), we only need to establish that $E\bigl[\textrm{sup}{}_{\alpha\in A}\sup\nolimits _{(\pi,\varpi)\in\Pi}\lVert\nabla_{(\pi,\varpi)(\pi,\varpi)'}^{2}l_{t}^{\pi}(\alpha,\pi,\varpi)\rVert\bigr]<\infty$
(see \citet{rangarao1962relations}). Verification of this moment
condition is provided in Supplementary Appendix E.2. Furthermore,
using the dominated convergence theorem and arguments similar to those
used in part (iii), it can be shown that $E[\nabla_{(\pi,\varpi)(\pi,\varpi)'}^{2}l_{t}^{\pi}(\alpha,\pi,\varpi)]$
is a (uniformly) continuous function of $(\alpha,\pi,\varpi)$ on
$A\times\Pi$. Now note that the left hand side of (\ref{eq:LMAR_RemainderTermDetails1})
is dominated by
\begin{align*}
 & \sup\nolimits _{(\pi,\varpi)\in\Pi}\bigl\Vert T^{-1}\nabla_{(\pi,\varpi)(\pi,\varpi)'}^{2}L_{T}^{\pi}(\alpha,\pi,\varpi)-E[\nabla_{(\pi,\varpi)(\pi,\varpi)'}^{2}l_{t}^{\pi}(\alpha,\pi,\varpi)]\bigr\Vert\\
 & \qquad+\sup\nolimits _{(\pi,\varpi)\in\Pi:\left\Vert (\pi,\varpi)-(\pi^{\ast},0)\right\Vert \leq\gamma_{T}}\bigl\Vert E[\nabla_{(\pi,\varpi)(\pi,\varpi)'}^{2}l_{t}^{\pi}(\alpha,\pi,\varpi)]-(-\mathcal{I}_{\alpha})\bigr\Vert.
\end{align*}
The former term is, due to (\ref{eq:LMAR_RemainderTermDetails2}),
of order $o_{p\alpha}(1)$. Regarding the latter term, the supremum
of it over $\alpha\in A$ converges to zero due to the uniform continuity
of $E[\nabla_{(\pi,\varpi)(\pi,\varpi)'}^{2}l_{t}^{\pi}(\alpha,\pi,\varpi)]$
and the fact that $E[\nabla_{(\pi,\varpi)(\pi,\varpi)'}^{2}l_{t}^{\pi}(\alpha,\pi^{\ast},0)]=-\mathcal{I}_{\alpha}$
(this fact follows from the expression of $\nabla_{(\pi,\varpi)(\pi,\varpi)'}^{2}l_{t}^{\pi}(\alpha,\pi^{\ast},0)$
given in (\ref{eq:LMARDer2a*})\textendash (\ref{eq:LMARDer2c*})
and Lemma \ref{lem:XXX3}). Thus, we have verified (\ref{eq:LMAR_RemainderTermDetails1}),
and hence Assumption 5(iv).

\paragraph{Assumptions 6\textendash 8.}

That $\Theta$ is locally (uniformly) equal to the cone $\Lambda=\mathbb{R}^{2q}$
follows from the expression of the set $\Theta$ given in the verification
of Assumption 5 above and the fact that $0$ ($\in\mathbb{R}^{2q}$)
is an interior point of $\Theta$. Assumption 7 is clear, as Assumption
\ref{assu:cone} holds with the cone $\Lambda=\mathbb{R}^{2q}$. Assumption
8 is clear from the verification of Assumption \ref{assu:quadexp}.

\paragraph{Expression of $\hat{s}_{t\alpha}$ in Section 4.1.}

Let $\hat{\varepsilon}_{t}$ denote the OLS residuals rescaled by
the estimated standard deviation, i.e., $\hat{\varepsilon}_{t}=(y_{t}-\hat{\tilde{\phi}}_{T,0}-\hat{\tilde{\phi}}_{T,1}y_{t-1}-\cdots-\hat{\tilde{\phi}}_{T,p}y_{t-p})/\hat{\tilde{\sigma}}_{T}$,
and set $\hat{s}_{t\alpha}=\bigl(\nabla f_{t}(\hat{\tilde{\phi}}_{T})/f_{t}(\hat{\tilde{\phi}}_{T}),-(1-\alpha_{1,t}^{L}(\alpha))\nabla f_{t}(\hat{\tilde{\phi}}_{T})/f_{t}(\hat{\tilde{\phi}}_{T})\bigr)$
with $\nabla f_{t}(\hat{\tilde{\phi}}_{T})/f_{t}(\hat{\tilde{\phi}}_{T})=(\frac{1}{\hat{\tilde{\sigma}}_{T}}\boldsymbol{z}_{t-1}\hat{\varepsilon}_{t},\frac{1}{2\hat{\tilde{\sigma}}_{T}^{2}}(\hat{\varepsilon}_{t}^{2}-1))$
(see (\ref{eq:LMAR_S_Ta}) and (\ref{eq:LMAR_nablaf_per_f})).

\section{Details for the GMAR example}

\paragraph{Assumption 1.}

Assumption 1(i) is assumed to hold. Assumption 1(ii) holds as $A$
is a compact subset of $(0,1)$. Assumption 1(iii) holds by the definition
of the mixing weight.

\paragraph{Assumption 2.}

For each fixed $\alpha\in A$, compactness of $B$ and $\Phi$ together
with the continuity of $L_{T}(\alpha,\beta,\phi,\varphi)=\sum_{t=1}^{T}l_{t}(\alpha,\beta,\phi,\varphi)$
ensures the existence of a measurable maximizer $(\hat{\beta}_{T\alpha},\hat{\phi}_{T\alpha},\hat{\varphi}_{T\alpha})$.
Hence part (i) holds (with the $o_{p\alpha}(1)$ term equal to zero).
(In the GMAR case, this maximizer is not unique when $\alpha=1/2$,
but this does not matter for Assumption 2.)

To prove that $\sup_{\alpha\in A}||(\hat{\beta}_{T\alpha},\hat{\phi}_{T\alpha},\hat{\varphi}_{T\alpha})-(\beta^{\ast},\phi^{\ast},\phi^{\ast})||\stackrel{p}{\to}0$,
by \citet[Lemma A1]{andrews1993tests} it suffices to show that (a)
$\sup_{(\alpha,\beta,\phi,\varphi)\in A\times B\times\Phi\times\Phi}\left\vert T^{-1}L_{T}(\alpha,\beta,\phi,\varphi)-E[l_{t}(\alpha,\beta,\phi,\varphi)]\right\vert \stackrel{p}{\to}0$
as $T\rightarrow\infty$ and that (b) for every neighborhood $N(\beta^{\ast},\phi^{\ast},\phi^{\ast})$
of $(\beta^{\ast},\phi^{\ast},\phi^{\ast})$, 
\[
\sup_{\alpha\in A}\sup_{(\beta,\phi,\varphi)\in B\times\Phi\times\Phi\backslash N(\beta^{\ast},\phi^{\ast},\phi^{\ast})}(E[l_{t}(\alpha,\beta,\phi,\varphi)]-E[l_{t}(\alpha,\beta^{\ast},\phi^{\ast},\phi^{\ast})])<0.
\]

Property (a) can be verified by using the uniform law of large numbers
given in \citet{rangarao1962relations}. As $T^{-1}L_{T}(\alpha,\beta,\phi,\varphi)=T^{-1}\sum_{t=1}^{T}l_{t}(\alpha,\beta,\phi,\varphi)$
with $l_{t}(\alpha,\beta,\phi,\varphi)$ a function of the stationary
and ergodic process $(y_{t},\boldsymbol{y}_{t-1})$, we only need
to show that $E\bigl[\sup_{(\alpha,\beta,\phi,\varphi)\in A\times B\times\Phi\times\Phi}|l_{t}(\alpha,\beta,\phi,\varphi)|\bigr]<\infty$.
Making use of Assumption 1, it is easy to show that $C_{1}\exp\{-C_{2}(1+y_{t}^{2}+\cdots+y_{t-p}^{2})\})\leq f_{t}(\beta,\phi)\leq C_{2}$
for some $0<C_{1},C_{2}<\infty$ and for all $(\beta,\phi)\in B\times\Phi$,
so that $\log(C_{1})-C_{2}(1+y_{t}^{2}+\cdots+y_{t-p}^{2})\leq l_{t}(\alpha,\beta,\phi,\varphi)\leq\log(C_{2})$
for all $(\alpha,\beta,\phi,\varphi)\in A\times B\times\Phi\times\Phi$
(cf. \citet[pp. 264--265]{kalliovirta2015gaussian}); this holds in
both the LMAR and GMAR cases. The required moment condition follows
from this. 

As for property (b), the uniform law of large numbers used above also
delivers the continuity of  the limit function $E[l_{t}(\alpha,\beta,\phi,\varphi)]$
on the compact set $A\times B\times\Phi\times\Phi$. Therefore it
suffices to show that, for each fixed $\alpha\in A$, $E[l_{t}(\alpha,\beta,\phi,\varphi)]-E[l_{t}(\alpha,\beta^{\ast},\phi^{\ast},\phi^{\ast})]\leq0$
with equality if and only if $(\beta,\phi,\varphi)=(\beta^{*},\phi^{\ast},\phi^{\ast})$.
For the GMAR model, this can be straightforwardly shown with arguments
used in the proof of Theorem 2 in \citet{kalliovirta2015gaussian}.
To this end, define 
\[
\mathsf{n_{1}}(\nu{}_{1,t}\mid\boldsymbol{\nu}_{1,t-1};(\beta,\phi))=(2\bar{\pi}\tilde{\sigma}_{1}^{2})^{-1/2}\exp\biggl(-\frac{(\nu{}_{1,t}-\tilde{\phi}_{0}-\sum_{i=1}^{p}\tilde{\phi}_{i}\nu{}_{1,t-i})^{2}}{2\tilde{\sigma}_{1}^{2}}\biggr),
\]
where $\nu{}_{1,t}$ is the auxiliary Gaussian AR($p$) process introduced
in the GMAR example of Section 2.2. Clearly, $\mathsf{n_{1}}(\nu{}_{1,t}\mid\boldsymbol{\nu}_{1,t-1};(\beta,\phi))$
is the conditional density of $\nu{}_{1,t}$ given $\boldsymbol{\nu}_{1,t-1}=(\nu{}_{1,t-1},\ldots,\nu{}_{1,t-p})$.
The notation $\mathsf{n_{1}}(\nu{}_{2,t}\mid\boldsymbol{\nu}_{2,t-1};(\beta,\varphi))$
is defined similarly by using the parameters $\tilde{\varphi}_{i}$
and $\tilde{\sigma}_{2}^{2}$ instead of $\tilde{\phi}_{i}$ and $\tilde{\sigma}_{1}^{2}$.

Now, in the same way as in the above-mentioned proof of \citet{kalliovirta2015gaussian}
we can use arguments based on the Kullback-Leibler divergence and
conclude that, for each fixed $\alpha\in A$, $E[l_{t}(\alpha,\beta,\phi,\varphi)]-E[l_{t}(\alpha,\beta^{\ast},\phi^{\ast},\phi^{\ast})]\leq0$
with equality if and only if for almost all $(y,\boldsymbol{y})\in\mathbb{R}\times\mathbb{R}^{p}$
\begin{equation}
\alpha_{1}^{G}\mathsf{n_{1}}(y\mid\boldsymbol{y};(\beta,\phi))+\alpha_{2}^{G}\mathsf{n_{1}}(y\mid\boldsymbol{y};(\beta,\varphi))=\mathsf{n_{1}}(y\mid\boldsymbol{y};(\beta^{\ast},\phi^{\ast})),\label{eq:GMARLMAR_Ident}
\end{equation}
where we use $\alpha_{m}^{G}$ to stand for $\alpha_{m,t}^{G}$ but
with $\boldsymbol{y}_{t-1}$ therein replaced by $\boldsymbol{y}$
($m=1,2$). Using well-known results on identification of finite mixtures
of Gaussian distributions we find that, for each fixed $\alpha\in A$,
and for each fixed $\boldsymbol{y}\in\mathbb{R}^{p}$ at a time, $\mathsf{n_{1}}(y\mid\boldsymbol{y};(\beta,\phi))=\mathsf{n_{1}}(y\mid\boldsymbol{y};(\beta,\varphi))=\mathsf{n_{1}}(y\mid\boldsymbol{y};(\beta^{\ast},\phi^{\ast}))$
for almost all $y$. Using the arguments following equation (A.4)
in \citet{kalliovirta2015gaussian} we can now establish the desired
result $(\beta,\phi,\varphi)=(\beta^{*},\phi^{\ast},\phi^{\ast})$.

The arguments used for the GMAR model above can also be used for the
LMAR model, but two things are worth noting. First, the proof given
for the GMAR model above goes through even when there are no common
parameters so that $\phi$ and $\varphi$ could be used in place of
$(\beta,\phi)$ and $(\beta,\varphi)$. Second, equation (\ref{eq:GMARLMAR_Ident})
can be obtained in the same way as in the GMAR case even though the
derivation of the related equation (A.4) in \citet{kalliovirta2015gaussian}
made use of the explicit expression of the stationary density of $(y_{t},\boldsymbol{y}_{t-1})$
which is known for the GMAR model but, in general, unknown for the
LMAR model. The reason for this is that the null hypothesis is here
assumed to hold so that $y_{t}$ is a linear Gaussian AR($p$) process,
implying that $(y_{t},\boldsymbol{y}_{t-1})$ is normally distributed
with density function a $p+1$ dimensional counterpart of the $p$
dimensional normal density function $\mathsf{n}_{p}(\mathbf{\boldsymbol{\nu}}_{1,t};\tilde{\phi})$
defined in equation (\ref{Normal p}) (see the GMAR example of Section
2.2). After observing these two facts we can proceed in the same way
as in the GMAR case and conclude that equation (\ref{eq:GMARLMAR_Ident})
holds also for the LMAR model as long as we replace the mixing weights
of the GMAR model with those of the LMAR model. As the arguments employed
in the proof of the GMAR case after equation (\ref{eq:GMARLMAR_Ident})
made no use of the mixing weights they apply also to the LMAR model
and can be used to complete the proof.

\paragraph{Assumptions 3 and 4.}

Conditions (i) and (ii) of Assumption \ref{assu:pi-repam} clearly
hold and condition (iii) can be verified in the same way as in the
case of the LMAR model. Specifically, we have
\[
\boldsymbol{\pi}_{\alpha}(\phi,\varphi)-\boldsymbol{\pi}_{\alpha}(\phi^{\ast},\phi^{\ast})=(\alpha\phi+(1-\alpha)\varphi,\phi-\varphi)-(\phi^{\ast},0)=\left(\alpha(\phi-\phi^{\ast})+(1-\alpha)(\varphi-\phi^{\ast}),\left(\phi-\phi^{\ast}\right)-\left(\varphi-\phi^{\ast}\right)\right),
\]
and choosing $\left\Vert x\right\Vert _{*}=\left\Vert x\right\Vert _{1}=\sum_{i=1}^{2q}\left|x_{i}\right|$
it can straightforwardly be seen that condition (iii) holds by Lemma
\ref{lem:LipCondForUnifCons}. Regarding Assumption \ref{assu:cont_differentiability},
based on the expression of $\alpha_{t}^{G}$ and the definition of
$\boldsymbol{\pi}_{\alpha}^{-1}(\pi,\varpi)=(\pi+(1-\alpha)\varpi,\pi-\alpha\varpi)$,
the required differentiability holds for all positive integers $k$.

\paragraph{Assumption 5: Derivation of expansion (\ref{eq:Quadr-Exp_1}).}

First note that, as $\boldsymbol{\pi}_{\alpha}^{-1}(\pi,\varpi)=(\pi+(1-\alpha)\varpi,\pi-\alpha\varpi)$,
the reparameterized mixing weight in the GMAR model is given by 

\[
\alpha_{1,t}^{G\pi}(\alpha,\beta,\pi,\varpi)=\frac{\alpha\mathsf{n}_{p}(\boldsymbol{y}_{t-1};(\beta,\pi+(1-\alpha)\varpi))}{\alpha\mathsf{n}_{p}(\boldsymbol{y}_{t-1};(\beta,\pi+(1-\alpha)\varpi))+(1-\alpha)\mathsf{n}_{p}(\boldsymbol{y}_{t-1};(\beta,\pi-\alpha\varpi))}
\]
and the quantities $f_{2,t}^{\pi}(\alpha,\beta,\pi,\varpi)$ and $l_{t}^{\pi}(\alpha,\beta,\pi,\varpi)$
take the form 
\begin{align*}
f_{2,t}^{\pi}(\alpha,\beta,\pi,\varpi) & =\alpha_{1,t}^{G\pi}(\alpha,\beta,\pi,\varpi)f_{t}(\beta,\pi+(1-\alpha)\varpi)+(1-\alpha_{1,t}^{G\pi}(\alpha,\beta,\pi,\varpi))f_{t}(\beta,\pi-\alpha\varpi),\\
l_{t}^{\pi}(\alpha,\beta,\pi,\varpi) & =\log[\alpha_{1,t}^{G\pi}(\alpha,\beta,\pi,\varpi)f_{t}(\beta,\pi+(1-\alpha)\varpi)+(1-\alpha_{1,t}^{G\pi}(\alpha,\beta,\pi,\varpi))f_{t}(\beta,\pi-\alpha\varpi)].
\end{align*}
Partial derivatives of the (reparameterized) log-likelihood function
(with respect to $(\beta,\pi,\varpi)$) can be obtained with straightforward
differentation but, as the calculations are somewhat lengthy, they
are relegated to Supplementary Appendix F.1. 

Now consider, for an arbitrary fixed $\alpha\in A$, a standard fourth-order
Taylor expansion of $L_{T}^{\pi}(\alpha,\beta,\pi,\varpi)=\sum_{t=1}^{T}l_{t}^{\pi}(\alpha,\beta,\pi,\varpi)$
around $(\beta^{*},\pi^{*},0)$ with respect to the parameters $(\beta,\pi,\varpi)$.
For brevity, we write $\tilde{\pi}=(\beta,\pi)$ and $\tilde{\pi}^{*}=(\beta^{*},\pi^{*})$.
Collecting terms that turn out to be asymptotically negligible into
a remainder term yields 
\begin{align}
 & L_{T}^{\pi}(\alpha,\tilde{\pi},\varpi)-L_{T}^{\pi}(\alpha,\tilde{\pi}^{*},0)\nonumber \\
 & =(\tilde{\pi}-\tilde{\pi}^{\ast})^{\prime}\nabla_{\tilde{\pi}}L_{T}^{\pi}(\alpha,\tilde{\pi}^{*},0)+\frac{1}{2!}(\tilde{\pi}-\tilde{\pi}^{\ast})^{\prime}\nabla_{\tilde{\pi}\tilde{\pi}^{\prime}}^{2}L_{T}^{\pi}(\alpha,\tilde{\pi}^{*},0)(\tilde{\pi}-\tilde{\pi}^{\ast})\nonumber \\
 & +\frac{1}{2!}\varpi^{\prime}\nabla_{\varpi\varpi^{\prime}}^{2}L_{T}^{\pi}(\alpha,\tilde{\pi}^{*},0)\varpi+\frac{3}{3!}\sum_{i=1}^{q_{1}+q_{2}}\sum_{j=1}^{q_{2}}\sum_{k=1}^{q_{2}}\nabla_{\tilde{\pi}_{i}\varpi_{j}\varpi_{k}}^{3}L_{T}^{\pi}(\alpha,\tilde{\pi}^{*},0)(\tilde{\pi}_{i}-\tilde{\pi}_{i}^{\ast})\varpi_{j}\varpi_{k}\nonumber \\
 & +\frac{1}{4!}\sum_{i=1}^{q_{2}}\sum_{j=1}^{q_{2}}\sum_{k=1}^{q_{2}}\sum_{l=1}^{q_{2}}\nabla_{\varpi_{i}\varpi_{j}\varpi_{k}\varpi_{l}}^{4}L_{T}^{\pi}(\alpha,\tilde{\pi}^{*},0)\varpi_{i}\varpi_{j}\varpi_{k}\varpi_{l}+R_{T}^{(1)}(\alpha,\tilde{\pi},\varpi)\label{QuarticExpansion}
\end{align}
with an explicit expression of the remainder term $R_{T}^{(1)}(\alpha,\tilde{\pi},\varpi)$
available in Supplementary Appendix F.2. Therein we also demonstrate
that this fourth-order Taylor expansion can be written as a quadratic
expansion of the form (\ref{eq:Quadr-Exp_1}) given by 
\begin{align}
 & L_{T}^{\pi}(\alpha,\beta,\pi,\varpi)-L_{T}^{\pi}(\alpha,\beta^{*},\pi^{*},0)\nonumber \\
 & =S_{T}^{\prime}\boldsymbol{\theta}(\alpha,\beta,\pi,\varpi)-\frac{1}{2}[T^{1/2}\boldsymbol{\theta}(\alpha,\beta,\pi,\varpi)]^{\prime}\mathcal{I}[T^{1/2}\boldsymbol{\theta}(\alpha,\beta,\pi,\varpi)]+R_{T}(\alpha,\beta,\pi,\varpi)\label{QuadraticExpansion}
\end{align}
or, setting $Z_{T}=\mathcal{I}^{-1}T^{-1/2}S_{T}$, in an alternative
form corresponding to (\ref{eq:Quadr-Exp_2}), with an explicit expression
of the remainder term $R_{T}(\alpha,\beta,\pi,\varpi)$ available
in Supplementary Appendix F.2. 

The required derivatives are available in Supplementary Appendix F.1.
Here we only present the derivatives that appear in the expression
of $S_{T}$ in (\ref{eq:GMAR_S_T}), that is, the components of $\tilde{\nabla}_{\boldsymbol{\theta}}l_{t}^{\pi\ast}=(\tilde{\nabla}_{\theta}l_{t}^{\pi\ast},\tilde{\nabla}_{\vartheta}l_{t}^{\pi\ast})$.
From Supplementary Appendix F.1 we obtain (here $\nabla_{i}$ denotes
the $i$th component of a derivative, and $\nabla\mathsf{n}_{p}(\cdot)$
denotes differentiation of $\mathsf{n}_{p}(\cdot)$ in (\ref{Normal p})
with respect to $\tilde{\phi}=(\tilde{\phi}_{0},\tilde{\phi}_{1},\ldots,\tilde{\phi}_{p},\tilde{\sigma}_{1}^{2})$)
\begin{align}
\nabla_{\beta}l_{t}^{\pi}(\alpha,\beta^{*},\pi^{*},0) & =\frac{\nabla_{1}f_{t}^{*}}{f_{t}^{*}}\label{eq:AppBAss5Derivs1}\\
\nabla_{\pi_{i}}l_{t}^{\pi}(\alpha,\beta^{*},\pi^{*},0) & =\frac{\nabla_{i+1}f_{t}^{*}}{f_{t}^{*}}\label{eq:AppBAss5Derivs2}\\
\nabla_{\varpi_{i}\varpi_{j}}^{2}l_{t}^{\pi}(\alpha,\beta^{*},\pi^{*},0) & =\alpha(1-\alpha)\biggl[\frac{\nabla_{i+1}\mathsf{n}_{p}^{*}}{\mathsf{n}_{p}^{*}}\frac{\nabla_{j+1}f_{t}^{*}}{f_{t}^{*}}+\frac{\nabla_{i+1}f_{t}^{*}}{f_{t}^{*}}\frac{\nabla_{j+1}\mathsf{n}_{p}^{*}}{\mathsf{n}_{p}^{*}}+\frac{\nabla_{i+1,j+1}^{2}f_{t}^{*}}{f_{t}^{*}}\biggr]\label{eq:AppBAss5Derivs3}
\end{align}
where $i,j=1,\ldots,p+1$, and, for brevity, we denote $f_{t}^{*}=f_{t}(\beta^{*},\pi^{*})$,
$\mathsf{n}_{p}^{*}=\mathsf{n}_{p}(\beta^{*},\pi^{*})$, and similarly
for their derivatives. Explicit expressions for the derivatives of
$f_{t}$ and $\mathsf{n}_{p}$ are given in Supplementary Appendix
F.3.

\paragraph{Assumption 5: Verifying the assumption.}

As $\boldsymbol{\theta}(\alpha,\beta,\pi,\varpi)=(\beta-\beta^{\ast},\pi-\pi^{\ast},\alpha(1-\alpha)v(\varpi))$,
part (i.a) is clearly satisfied. Assumption 1 requires $\alpha$ to
be bounded away from zero and one, so that also part (i.b) is satisfied. 

For part (ii), notice from (\ref{eq:GMAR_S_T}), (\ref{eq:GMAR_S_T_component3}),
and (\ref{eq:AppBAss5Derivs1})\textendash (\ref{eq:AppBAss5Derivs3})
that $S_{T}=\sum_{t=1}^{T}s_{t}$ with
\begin{equation}
s_{t}=\left(\frac{\nabla f_{t}^{*}}{f_{t}^{*}},c_{11}X_{t,1,1}^{*},\ldots,c_{q_{2}-1,q_{2}}X_{t,q_{2}-1,q_{2}}^{*}\right)\label{eq:AppBAss5_s_t}
\end{equation}
where, for $i,j\in\{1,\ldots,q_{2}\}$, the $c_{ij}$'s are as in
Section 3.3.1 ($c_{ij}=1/2$ if $i=j$ and $c_{ij}=1$ if $i\neq j$)
and 
\begin{equation}
X_{t,i,j}^{*}=\frac{\nabla_{i+1}\mathsf{n}_{p}^{*}}{\mathsf{n}_{p}^{*}}\frac{\nabla_{j+1}f_{t}^{*}}{f_{t}^{*}}+\frac{\nabla_{i+1}f_{t}^{*}}{f_{t}^{*}}\frac{\nabla_{j+1}\mathsf{n}_{p}^{*}}{\mathsf{n}_{p}^{*}}+\frac{\nabla_{i+1,j+1}^{2}f_{t}^{*}}{f_{t}^{*}}\label{eq:Notation_X*}
\end{equation}
(note that $s_{t}$, and hence $S_{T}$, does not involve $\alpha$
as it cancels out from the expressions in (\ref{eq:GMAR_S_T_component3})
and (\ref{eq:AppBAss5Derivs3})). Therefore the first three requirements
in part (ii) are clearly satisfied. For the weak convergence requirement
in part (ii) it now suffices to show that $T^{-1/2}S_{T}\overset{d}{\to}S$
in $\mathbb{R}^{r}$ for some multivariate Gaussian random vector
$S$ with mean zero and $E[SS']=\mathcal{I}$. To this end, $s_{t}$
clearly forms a stationary and ergodic process. Moreover, due to Lemma
\ref{lem:XXX3} in Supplementary Appendix F.5, $E[\nabla_{i}f_{t}^{*}/f_{t}^{*}\mid\boldsymbol{y}_{t-1}]=E[\nabla_{ij}^{2}f_{t}^{*}/f_{t}^{*}\mid\boldsymbol{y}_{t-1}]=0$
for any $i,j\in\{1,\ldots,q_{2}\}$ so that $s_{t}$ is a martingale
difference sequence. From the expression of $\mathcal{I}$ in (\ref{MatrixI})
in Supplementary Appendix F.2 it is clear that $E[s_{t}s_{t}^{\prime}]=\mathcal{I}$.
Positive definiteness of $\mathcal{I}$ is proven in Supplementary
Appendix F.3. The stated convergence result now follows from the central
limit theorem of \citet{billingsley1961lindeberg} in conjunction
with the Cramér-Wold device. 

For part (iii), it suffices to show the finiteness and positive definiteness
of $\mathcal{I}$; these are proven in Supplementary Appendices F.2
and F.3. Part (iv) is proven in Supplementary Appendix F.4.

\paragraph{Assumption 6.}

By the definition of the set $\Theta_{\alpha}$ and the transformation
$(\phi,\varphi)\rightarrow(\pi,\varpi)$, the set $\Theta_{\alpha}$
(see Sec 3.3.1) can equivalently be expressed as 
\begin{align*}
\Theta_{\alpha} & =\{\boldsymbol{\theta}=(\theta,\vartheta)\in\mathbb{R}^{q_{1}+q_{2}+q_{\vartheta}}:\theta=(\beta-\beta^{\ast},\alpha(\phi-\phi^{\ast})+(1-\alpha)(\varphi-\phi^{\ast})),\vartheta=\alpha(1-\alpha)v(\phi-\varphi)\\
 & \qquad\text{ for some }(\beta,\phi,\varphi)\in B\times\Phi^{2}\}.
\end{align*}
We aim to show that the collection of sets $\{\Theta_{\alpha},\ \alpha\in A\}$
is locally uniformly equal to the cone $\Lambda=\mathbb{\mathbb{R}}^{q_{1}+q_{2}}\times v(\mathbb{\mathbb{R}}^{q_{2}})$
where $v(\mathbb{\mathbb{R}}^{q_{2}})=\{v(\varpi):\varpi\in\mathbb{\mathbb{R}}^{q_{2}}\}$. 

Let $\bar{S}((\beta^{*},\phi^{\ast}),\delta)$ denote a closed $(q_{1}+q_{2})$-sphere
centered at $(\beta^{*},\phi^{\ast})$ and with radius $\delta$,
and $\bar{S}(\phi^{\ast},\delta)$ a similar $q_{2}$-sphere. As $(\beta^{*},\phi^{\ast})$
is an interior point of $B\times\Phi$, we can find a $\delta_{1}>0$
such that $\bar{S}((\beta^{*},\phi^{\ast}),\delta_{1})\times\bar{S}(\phi^{\ast},\delta_{1})\subset B\times\Phi^{2}$.
By the definitions of the transformations $(\alpha,\beta,\phi,\varphi)\rightarrow\theta(\alpha,\beta,\phi,\varphi)$
and $(\alpha,\beta,\phi,\varphi)\rightarrow\vartheta(\alpha,\beta,\phi,\varphi)$
(defined implicitly by the definition of $\Theta_{\alpha}$ above),
we can find a $\delta_{2}>0$ such that 
\[
(-\delta_{2},\delta_{2})^{q_{1}+q_{2}}\times v((-\delta_{2},\delta_{2})^{q_{2}})\subset{\textstyle \bigcap\nolimits _{\alpha\in A}}\Theta_{\alpha}(\delta_{1})\subset{\textstyle \bigcap\nolimits _{\alpha\in A}}\Theta_{\alpha}
\]
where 
\begin{align*}
\Theta_{\alpha}(\delta_{1}) & =\{\boldsymbol{\theta}=(\theta,\vartheta)\in\mathbb{R}^{q_{1}+q_{2}+q_{\vartheta}}:\theta=(\beta-\beta^{\ast},\alpha(\phi-\phi^{\ast})+(1-\alpha)(\varphi-\phi^{\ast})),\vartheta=\alpha(1-\alpha)v(\phi-\varphi)\\
 & \text{ \qquad for some }(\beta,\phi,\varphi)\in\bar{S}((\beta^{*},\phi^{\ast}),\delta_{1})\times\bar{S}(\phi^{\ast},\delta_{1})\}.
\end{align*}
Thus $(-\delta_{2},\delta_{2})^{q_{1}+q_{2}}\times v((-\delta_{2},\delta_{2})^{q_{2}})\subset\Theta_{\alpha}$
for all $\alpha\in A$ so that (assuming, without loss of generality,
that $\delta_{2}<1$) 
\[
(-\delta_{2}^{2},\delta_{2}^{2})^{q_{1}+q_{2}+q_{\vartheta}}\cap\Theta_{\alpha}=(-\delta_{2}^{2},\delta_{2}^{2})^{q_{1}+q_{2}}\times v((-\delta_{2},\delta_{2})^{q_{2}})=(-\delta_{2}^{2},\delta_{2}^{2})^{q_{1}+q_{2}+q_{\vartheta}}\cap[\mathbb{\mathbb{R}}^{q_{1}+q_{2}}\times v(\mathbb{\mathbb{R}}^{q_{2}})]
\]
for all $\alpha\in A$. Thus the collection of sets $\{\Theta_{\alpha},\ \alpha\in A\}$
is locally uniformly equal to the cone $\Lambda=\mathbb{\mathbb{R}}^{q_{1}+q_{2}}\times v(\mathbb{\mathbb{R}}^{q_{2}})$.

\paragraph{Assumptions 7 and 8.}

Assumption 7 is satisfied as Assumption \ref{assu:cone} holds with
the cone $\Lambda=\mathbb{\mathbb{R}}^{q_{1}+q_{2}}\times v(\mathbb{\mathbb{R}}^{q_{2}})$
(note that now $q_{\vartheta}=q_{2}(q_{2}+1)/2$). Assumption 8 is
clear from the verification of Assumption \ref{assu:quadexp}.

\paragraph{Expression of $\hat{s}_{t\alpha}$ in Section 4.1.}

Let $\hat{\varepsilon}_{t}$ and $\nabla f_{t}(\hat{\tilde{\phi}}_{T})/f_{t}(\hat{\tilde{\phi}}_{T})$
be as in the LMAR example (see Appendix B) and set (see (\ref{eq:AppBAss5_s_t})
and (\ref{eq:Notation_X*})) $\hat{s}_{t}=\bigl(\nabla f_{t}(\hat{\tilde{\phi}}_{T})/f_{t}(\hat{\tilde{\phi}}_{T}),c_{11}X_{t,1,1}(\hat{\tilde{\phi}}_{T}),\ldots,c_{q_{2}-1,q_{2}}X_{t,q_{2}-1,q_{2}}(\hat{\tilde{\phi}}_{T})\bigr)$
where, for $i,j\in\{1,\ldots,q_{2}\}$, the $c_{ij}$'s are as in
Section 3.3.1 ($c_{ij}=1/2$ if $i=j$ and $c_{ij}=1$ if $i\neq j$)
and 
\[
X_{t,i,j}(\hat{\tilde{\phi}}_{T})=\frac{\nabla_{i+1}\mathsf{n}_{p}(\hat{\tilde{\phi}}_{T})}{\mathsf{n}_{p}(\hat{\tilde{\phi}}_{T})}\frac{\nabla_{j+1}f_{t}(\hat{\tilde{\phi}}_{T})}{f_{t}(\hat{\tilde{\phi}}_{T})}+\frac{\nabla_{i+1}f_{t}(\hat{\tilde{\phi}}_{T})}{f_{t}(\hat{\tilde{\phi}}_{T})}\frac{\nabla_{j+1}\mathsf{n}_{p}(\hat{\tilde{\phi}}_{T})}{\mathsf{n}_{p}(\hat{\tilde{\phi}}_{T})}+\frac{\nabla_{i+1,j+1}^{2}f_{t}(\hat{\tilde{\phi}}_{T})}{f_{t}(\hat{\tilde{\phi}}_{T})}.
\]
Explicit expressions for the elements of $\nabla^{2}f_{t}(\hat{\tilde{\phi}}_{T})/f_{t}(\hat{\tilde{\phi}}_{T})$
can be obtained from (\ref{eq:AppF3_nabla2f_f}) in Supplementary
Appendix F.3 by replacing $\varepsilon_{t}$ and $\tilde{\sigma}_{1}^{*}$
therein with $\hat{\varepsilon}_{t}$ and $\hat{\tilde{\sigma}}_{T}$,
respectively. Expressions for the elements of $\nabla\mathsf{n}_{p}(\hat{\tilde{\phi}}_{T})/\mathsf{n}_{p}(\hat{\tilde{\phi}}_{T})$
can be obtained by evaluating (\ref{eq:AppF3_nabla_np_np}) in Supplementary
Appendix F.3 at $\tilde{\phi}=\hat{\tilde{\phi}}_{T}$. 

\bibliographystyle{elsarticle-harv}
\bibliography{mikabib}

\begin{thebibliography}{43}
\expandafter\ifx\csname natexlab\endcsname\relax\def\natexlab#1{#1}\fi
\expandafter\ifx\csname url\endcsname\relax
  \def\url#1{\texttt{#1}}\fi
\expandafter\ifx\csname urlprefix\endcsname\relax\def\urlprefix{URL }\fi

\bibitem[{Andrews(1992)}]{andrews1992generic}
Andrews, D.~W., 1992. Generic uniform convergence. Econometric Theory 8,
  241--257.

\bibitem[{Andrews(1993)}]{andrews1993tests}
Andrews, D.~W., 1993. Tests for parameter instability and structural change
  with unknown change point. Econometrica 61, 821--856.

\bibitem[{Andrews(1999)}]{andrews1999estimation}
Andrews, D.~W., 1999. Estimation when a parameter is on a boundary.
  Econometrica 67, 1341--1383.

\bibitem[{Andrews(2000)}]{andrews2000inconsistency}
Andrews, D.~W., 2000. Inconsistency of the bootstrap when a parameter is on the
  boundary of the parameter space. Econometrica 68, 399--405.

\bibitem[{Andrews(2001)}]{andrews2001testing}
Andrews, D.~W., 2001. Testing when a parameter is on the boundary of the
  maintained hypothesis. Econometrica 69, 683--734.

\bibitem[{Andrews and Ploberger(1995)}]{andrews1995admissibility}
Andrews, D.~W., Ploberger, W., 1995. Admissibility of the likelihood ratio test
  when a nuisance parameter is present only under the alternative. Annals of
  Statistics 23, 1609--1629.

\bibitem[{Bec et~al.(2008)Bec, Rahbek, and Shephard}]{bec2008acr}
Bec, F., Rahbek, A., Shephard, N., 2008. The {ACR} model: a multivariate
  dynamic mixture autoregression. Oxford Bulletin of Economics and Statistics
  70, 583--618.

\bibitem[{Billingsley(1961)}]{billingsley1961lindeberg}
Billingsley, P., 1961. The {L}indeberg-{L}evy theorem for martingales.
  Proceedings of the American Mathematical Society 12, 788--792.

\bibitem[{Billingsley(1999)}]{billingsley1999convergence}
Billingsley, P., 1999. Convergence of Probability Measures, 2nd Edition. Wiley.

\bibitem[{Carrasco et~al.(2014)Carrasco, Hu, and
  Ploberger}]{carrasco2014optimal}
Carrasco, M., Hu, L., Ploberger, W., 2014. Optimal test for {M}arkov switching
  parameters. Econometrica 82, 765--784.

\bibitem[{Cho and White(2007)}]{cho2007testing}
Cho, J.~S., White, H., 2007. Testing for regime switching. Econometrica 75,
  1671--1720.

\bibitem[{Davies(1977)}]{davies1977hypothesis}
Davies, R.~B., 1977. Hypothesis testing when a nuisance parameter is present
  only under the alternative. Biometrika 64, 247--254.

\bibitem[{Davies(1987)}]{davies1987hypothesis}
Davies, R.~B., 1987. Hypothesis testing when a nuisance parameter is present
  only under the alternative. Biometrika 74, 33--43.

\bibitem[{Diebold et~al.(1994)Diebold, Lee, and Weinbach}]{diebold1994regime}
Diebold, F.~X., Lee, J.-H., Weinbach, G.~C., 1994. Regime switching with
  time-varying transition probabilities. In: Hargreaves, C. (Ed.),
  Nonstationary Time Series Analysis and Cointegration. Oxford University
  Press, pp. 283--302.

\bibitem[{Dueker et~al.(2011)Dueker, Psaradakis, Sola, and
  Spagnolo}]{dueker2011multivariate}
Dueker, M.~J., Psaradakis, Z., Sola, M., Spagnolo, F., 2011. Multivariate
  contemporaneous-threshold autoregressive models. Journal of Econometrics 160,
  311--325.

\bibitem[{Dueker et~al.(2007)Dueker, Sola, and
  Spagnolo}]{dueker2007contemporaneous}
Dueker, M.~J., Sola, M., Spagnolo, F., 2007. Contemporaneous threshold
  autoregressive models: estimation, testing and forecasting. Journal of
  Econometrics 141, 517--547.

\bibitem[{Filardo(1994)}]{filardo1994business}
Filardo, A.~J., 1994. Business-cycle phases and their transitional dynamics.
  Journal of Business \& Economic Statistics 12, 299--308.

\bibitem[{Fr{\"u}hwirth-Schnatter(2006)}]{fruhwirth2006finite}
Fr{\"u}hwirth-Schnatter, S., 2006. Finite Mixture and Markov Switching Models.
  Springer.

\bibitem[{Garcia(1998)}]{garcia1998asymptotic}
Garcia, R., 1998. Asymptotic null distribution of the likelihood ratio test in
  {M}arkov switching models. International Economic Review 39, 763--788.

\bibitem[{Gouri{\'e}roux and Robert(2006)}]{gourieroux2006stochastic}
Gouri{\'e}roux, C., Robert, C.~Y., 2006. Stochastic unit root models.
  Econometric Theory 22, 1052--1090.

\bibitem[{Hallin and Ley(2014)}]{hallin2014skew}
Hallin, M., Ley, C., 2014. Skew-symmetric distributions and {F}isher
  information: the double sin of the skew-normal. Bernoulli 20, 1432--1453.

\bibitem[{Hamilton(2016)}]{hamilton2016macroeconomic}
Hamilton, J.~D., 2016. Macroeconomic regimes and regime shifts. In: Uhlig, H.,
  Taylor, J. (Eds.), Handbook of Macroeconomics. Vol.~2A. Elsevier, pp.
  163--201.

\bibitem[{Hansen(1992)}]{hansen1992likelihood}
Hansen, B.~E., 1992. The likelihood ratio test under nonstandard conditions:
  testing the {M}arkov switching model of {GNP}. Journal of Applied
  Econometrics 7, S61--S82.

\bibitem[{Hansen(1996)}]{hansen1996inference}
Hansen, B.~E., 1996. Inference when a nuisance parameter is not identified
  under the null hypothesis. Econometrica 64, 413--430.

\bibitem[{Jeffries(1998)}]{jeffries1998logistic}
Jeffries, N.~O., 1998. Logistic mixtures of generalized linear model times
  series. Ph.D. thesis, University of Maryland, College Park.

\bibitem[{Kalliovirta et~al.(2015)Kalliovirta, Meitz, and
  Saikkonen}]{kalliovirta2015gaussian}
Kalliovirta, L., Meitz, M., Saikkonen, P., 2015. A {G}aussian mixture
  autoregressive model for univariate time series. Journal of Time Series
  Analysis 36, 247--266.

\bibitem[{Kalliovirta et~al.(2016)Kalliovirta, Meitz, and
  Saikkonen}]{kalliovirta2016gaussian}
Kalliovirta, L., Meitz, M., Saikkonen, P., 2016. Gaussian mixture vector
  autoregression. Journal of Econometrics 192, 485--498.

\bibitem[{Kasahara and Shimotsu(2012)}]{kasahara2012testing}
Kasahara, H., Shimotsu, K., 2012. Testing the number of components in finite
  mixture models, unpublished working paper.

\bibitem[{Kasahara and Shimotsu(2015)}]{kasahara2015testing}
Kasahara, H., Shimotsu, K., 2015. Testing the number of components in normal
  mixture regression models. Journal of the American Statistical Association
  110, 1632--1645.

\bibitem[{Kasahara and Shimotsu(2017)}]{Kasahara2017Testing}
Kasahara, H., Shimotsu, K., 2017. Testing the number of regimes in {M}arkov
  regime switching models, unpublished working paper.

\bibitem[{Kim et~al.(2008)Kim, Piger, and Startz}]{kim2008estimation}
Kim, C.-J., Piger, J., Startz, R., 2008. Estimation of {M}arkov
  regime-switching regression models with endogenous switching. Journal of
  Econometrics 143, 263--273.

\bibitem[{Lanne and Saikkonen(2003)}]{lanne2003modeling}
Lanne, M., Saikkonen, P., 2003. Modeling the {US} short-term interest rate by
  mixture autoregressive processes. Journal of Financial Econometrics 1,
  96--125.

\bibitem[{L{\"u}tkepohl(2005)}]{lutkepohl2005new}
L{\"u}tkepohl, H., 2005. New Introduction to Multiple Time Series Analysis.
  Springer.

\bibitem[{McLachlan and Peel(2000)}]{mclachlan2000finite}
McLachlan, G., Peel, D., 2000. Finite Mixture Models. Wiley.

\bibitem[{Pakes and Pollard(1989)}]{pakes1989simulation}
Pakes, A., Pollard, D., 1989. Simulation and the asymptotics of optimization
  estimators. Econometrica 57, 1027--1057.

\bibitem[{Qu and Zhuo(2017)}]{Qu2017Likelihood}
Qu, Z., Zhuo, F., 2017. Likelihood ratio based tests for {M}arkov regime
  switching, unpublished working paper.

\bibitem[{Ranga~Rao(1962)}]{rangarao1962relations}
Ranga~Rao, R., 1962. Relations between weak and uniform convergence of measures
  with applications. Annals of Mathematical Statistics 33, 659--680.

\bibitem[{Rotnitzky et~al.(2000)Rotnitzky, Cox, Bottai, and
  Robins}]{rotnitzky2000likelihood}
Rotnitzky, A., Cox, D.~R., Bottai, M., Robins, J., 2000. Likelihood-based
  inference with singular information matrix. Bernoulli 6, 243--284.

\bibitem[{Shen and He(2015)}]{shen2015inference}
Shen, J., He, X., 2015. Inference for subgroup analysis with a structured
  logistic-normal mixture model. Journal of the American Statistical
  Association 110, 303--312.

\bibitem[{Wong and Li(2000)}]{wong2000mixture}
Wong, C.~S., Li, W.~K., 2000. On a mixture autoregressive model. Journal of the
  Royal Statistical Society: Series B 62, 95--115.

\bibitem[{Wong and Li(2001)}]{wong2001logistic}
Wong, C.~S., Li, W.~K., 2001. On a logistic mixture autoregressive model.
  Biometrika 88, 833--846.

\bibitem[{Zhu and Zhang(2004)}]{zhu2004hypothesis}
Zhu, H., Zhang, H., 2004. Hypothesis testing in mixture regression models.
  Journal of the Royal Statistical Society: Series B 66, 3--16.

\bibitem[{Zhu and Zhang(2006)}]{zhu2006asymptotics}
Zhu, H., Zhang, H., 2006. Asymptotics for estimation and testing procedures
  under loss of identifiability. Journal of Multivariate Analysis 97, 19--45.

\end{thebibliography}

\paragraph*{\protect\pagebreak{}}

\section*{Supplementary Appendix to \protect \\
`Testing for observation-dependent regime switching in mixture autoregressive
models' by Meitz and Saikkonen (not meant for publication)}

\bigskip{}

\section{Further details for the general results}
\begin{proof}[\textbf{\emph{Proof of Lemma \ref{lem:AsInsigOfRemainder}, further
details}}]
To justify that the last term on the right hand side of (\ref{eq:Lemma 2_3})
is dominated by $-\frac{1}{4}\left\Vert \boldsymbol{\theta}_{T\alpha}\right\Vert ^{2}+o_{p\alpha}(1)$,
note first that 
\[
\left(o_{p\alpha}\left(1\right)-\frac{1}{2}\right)\left\Vert \boldsymbol{\theta}_{T\alpha}\right\Vert ^{2}=\left(o_{p\alpha}\left(1\right)-\frac{1}{4}\right)\left\Vert \boldsymbol{\theta}_{T\alpha}\right\Vert ^{2}-\frac{1}{4}\left\Vert \boldsymbol{\theta}_{T\alpha}\right\Vert ^{2}:=W_{T\alpha}\left\Vert \boldsymbol{\theta}_{T\alpha}\right\Vert ^{2}-\frac{1}{4}\left\Vert \boldsymbol{\theta}_{T\alpha}\right\Vert ^{2},
\]
where $W_{T\alpha}=-\frac{1}{4}+o_{p\alpha}\left(1\right)$. Thus,
$P\left(\sup_{\alpha\in A}W_{T\alpha}\leq0\right)\rightarrow1$ and
(here $\mathbf{1}(\cdot)$ denotes the indicator function)
\begin{align*}
\sup_{\alpha\in A}W_{T\alpha}\left\Vert \boldsymbol{\theta}_{T\alpha}\right\Vert ^{2} & =\sup_{\alpha\in A}W_{T\alpha}\left\Vert \boldsymbol{\theta}_{T\alpha}\right\Vert ^{2}\mathbf{1}\left(\sup_{\alpha\in A}W_{T\alpha}\leq0\right)+\sup_{\alpha\in A}W_{T\alpha}\left\Vert \boldsymbol{\theta}_{T\alpha}\right\Vert ^{2}\mathbf{1}\left(\sup_{\alpha\in A}W_{T\alpha}>0\right)\\
 & \leq\sup_{\alpha\in A}W_{T\alpha}\left\Vert \boldsymbol{\theta}_{T\alpha}\right\Vert ^{2}\mathbf{1}\left(\sup_{\alpha\in A}W_{T\alpha}>0\right),
\end{align*}
where the last term is non-negative and positive with probability
that is at most $P\left(\sup_{\alpha\in A}W_{T\alpha}>0\right)\rightarrow0$.
Thus, combining the above derivations yields the desired result $\left(o_{p\alpha}\left(1\right)-\frac{1}{2}\right)\left\Vert \boldsymbol{\theta}_{T\alpha}\right\Vert ^{2}\leq-\frac{1}{4}\left\Vert \boldsymbol{\theta}_{T\alpha}\right\Vert ^{2}+o_{p\alpha}\left(1\right)$.

To justify the use of the continuous mapping theorem, note that in
the first instance it is applied with the function $g:\mathcal{C}(A,\mathbb{R}^{r})\times\{\mathcal{I}_{\alpha}\}\to\mathcal{C}(A,\mathbb{R}^{r})$
mapping $(x_{\bullet},\mathcal{I}_{\bullet})$ to $\mathcal{I}_{\bullet}^{-1}x_{\bullet}$.
Here $\mathcal{I}_{\alpha}^{-\text{1}}x_{\alpha}$ is continuous in
$\alpha$ by Assumption \ref{assu:quadexp}(iii). Also, the latter
set in the product $\mathcal{C}(A,\mathbb{R}^{r})\times\{\mathcal{I}_{\alpha}\}$
contains only the non-random function $\mathcal{I}_{\alpha}$; this
product space can be equipped with essentially the same metric as
$\mathcal{C}(A,\mathbb{R}^{r})$; cf. Andrews and Ploberger (1994,
p. 1392 and 1407) and \citet[proof of Theorem 5]{zhu2006asymptotics}.
In the second instance, the continuous mapping theorem is applied
with the function $g:\mathcal{B}(A,\mathbb{R}^{r})\to\mathbb{R}$
mapping $x_{\bullet}$ ($\in\mathcal{B}(A,\mathbb{R}^{r})$) to $\sup_{\alpha\in A}\left\Vert x_{\alpha}\right\Vert $.
For continuity, we need to establish that if a sequence $x_{n\bullet}$
converges to $x_{\bullet}$ in $\mathcal{B}(A,\mathbb{R}^{r})$, then
$g(x_{n\bullet})\to g(x_{\bullet})$ in $\mathbb{R}$ (i.e., if $\sup_{\alpha\in A}\left\Vert x_{n\alpha}-x_{\alpha}\right\Vert \to0$,
then $\left|\sup_{\alpha\in A}\left\Vert x_{n\alpha}\right\Vert -\sup_{\alpha\in A}\left\Vert x_{\alpha}\right\Vert \right|\to0$).
The triangle inequality implies that $\sup_{\alpha\in A}\left\Vert x_{n\alpha}\right\Vert \leq\sup_{\alpha\in A}\left\Vert x_{n\alpha}-x_{\alpha}\right\Vert +\sup_{\alpha\in A}\left\Vert x_{\alpha}\right\Vert $,
as well as the same result with $x_{n\alpha}$ and $x_{\alpha}$ interchanged,
and the desired result follows from these inequalities. 
\end{proof}
\bigskip{}
\begin{proof}[\textbf{\emph{Proof of Lemma \ref{lem:WeakConv}, further details}}]
It remains to verify the continuity mentioned in the proof. For simplicity,
consider the continuity of the functions $g_{1}:\mathcal{C}(A,\mathbb{R}^{r})\times\{\mathcal{I}_{\alpha}\}\to\mathcal{B}(A,\mathbb{R})$
mapping $(x_{\bullet},\mathcal{I}_{\bullet})$ to $x_{\bullet}'\mathcal{I}_{\bullet}x_{\bullet}$
and $g_{2}:\mathcal{C}(A,\mathbb{R}^{r})\times\{\mathcal{I}_{\alpha}\}\to\mathcal{B}(A,\mathbb{R})$
mapping $(x_{\bullet},\mathcal{I}_{\bullet})$ to $\inf_{\boldsymbol{\lambda}\in\Lambda}\left\{ (\boldsymbol{\lambda}-x_{\bullet})^{\prime}\mathcal{I}_{\bullet}(\boldsymbol{\lambda}-x_{\bullet})\right\} $
separately. For $g_{1}$, continuity is rather clear, for if a sequence
$(x_{n\bullet},\mathcal{I}_{\bullet})$ converges to $(x_{\bullet},\mathcal{I}_{\bullet})$
in $\mathcal{C}(A,\mathbb{R}^{r})\times\{\mathcal{I}_{\alpha}\}$,
then $g_{1}((x_{n\bullet},\mathcal{I}_{\bullet}))\to g_{1}((x_{\bullet},\mathcal{I}_{\bullet}))$
in $\mathcal{B}(A,\mathbb{R})$ (i.e., if $\sup_{\alpha\in A}\left\Vert x_{n\alpha}-x_{\alpha}\right\Vert \to0$,
then $\sup_{\alpha\in A}\left|x_{n\alpha}'\mathcal{I}_{\alpha}x_{n\alpha}-x_{\alpha}'\mathcal{I}_{\alpha}x_{\alpha}\right|\to0$).
For the continuity of $g_{2}$, suppose that $\sup_{\alpha\in A}\left\Vert x_{n\alpha}-x_{\alpha}\right\Vert \to0$,
and consider $\sup_{\alpha\in A}\left|\inf_{\boldsymbol{\lambda}\in\Lambda}\left\{ (\boldsymbol{\lambda}-x_{n\alpha})^{\prime}\mathcal{I}_{\alpha}(\boldsymbol{\lambda}-x_{n\alpha})\right\} -\inf_{\boldsymbol{\lambda}\in\Lambda}\left\{ (\boldsymbol{\lambda}-x_{\alpha})^{\prime}\mathcal{I}_{\alpha}(\boldsymbol{\lambda}-x_{\alpha})\right\} \right|$.
Noting that
\[
\inf_{\boldsymbol{\lambda}\in\Lambda}\left\{ (\boldsymbol{\lambda}-x_{n\alpha})^{\prime}\mathcal{I}_{\alpha}(\boldsymbol{\lambda}-x_{n\alpha})\right\} =\bigl\{\inf_{\boldsymbol{\lambda}\in\Lambda}\lVert\mathcal{I}_{\alpha}^{1/2}(\boldsymbol{\lambda}-x_{n\alpha})\rVert\bigr\}^{2}
\]
 and similarly for the other infimum, we need to consider
\begin{equation}
\sup_{\alpha\in A}\Bigl\{\bigl|\inf_{\boldsymbol{\lambda}\in\Lambda}\lVert\mathcal{I}_{\alpha}^{1/2}(\boldsymbol{\lambda}-x_{n\alpha})\rVert-\inf_{\boldsymbol{\lambda}\in\Lambda}\lVert\mathcal{I}_{\alpha}^{1/2}(\boldsymbol{\lambda}-x_{\alpha})\rVert\bigr|\bigl(\inf_{\boldsymbol{\lambda}\in\Lambda}\lVert\mathcal{I}_{\alpha}^{1/2}(\boldsymbol{\lambda}-x_{n\alpha})\rVert+\inf_{\boldsymbol{\lambda}\in\Lambda}\lVert\mathcal{I}_{\alpha}^{1/2}(\boldsymbol{\lambda}-x_{\alpha})\rVert\bigr)\Bigr\}.\label{eq:LemWeakConvSuppApp}
\end{equation}
Using the triangle inequality and properties of the Euclidean vector
norm, 
\[
\lVert\mathcal{I}_{\alpha}^{1/2}(\boldsymbol{\lambda}-x_{n\alpha})\rVert\leq\lVert\mathcal{I}_{\alpha}^{1/2}(\boldsymbol{\lambda}-x_{\alpha})\rVert+\lVert\mathcal{I}_{\alpha}^{1/2}(x_{n\alpha}-x_{\alpha})\rVert\leq\lVert\mathcal{I}_{\alpha}^{1/2}(\boldsymbol{\lambda}-x_{\alpha})\rVert+\left(\lambda_{\max}\left(\mathcal{I}_{\alpha}\right)\right)^{1/2}\left\Vert x_{n\alpha}-x_{\alpha}\right\Vert ,
\]
and similarly with $x_{n\alpha}$ and $x_{\alpha}$ exchanged, so
that 
\[
\Bigl|\inf_{\boldsymbol{\lambda}\in\Lambda}\lVert\mathcal{I}_{\alpha}^{1/2}(\boldsymbol{\lambda}-x_{n\alpha})\rVert-\inf_{\boldsymbol{\lambda}\in\Lambda}\lVert\mathcal{I}_{\alpha}^{1/2}(\boldsymbol{\lambda}-x_{\alpha})\rVert\Bigr|\leq\left(\lambda_{\max}\left(\mathcal{I}_{\alpha}\right)\right)^{1/2}\left\Vert x_{n\alpha}-x_{\alpha}\right\Vert .
\]
As was noted after Assumption 6, the cone $\Lambda$ contains the
origin, so that the term in (\ref{eq:LemWeakConvSuppApp}) in parentheses
is dominated by $\left(\lambda_{\max}\left(\mathcal{I}_{\alpha}\right)\right)^{1/2}(\left\Vert x_{n\alpha}\right\Vert +\left\Vert x_{\alpha}\right\Vert )$.
Now, due to Assumption 5(iii), the fact that $x_{n\bullet},x_{\bullet}$
are bounded, and the assumed $\sup_{\alpha\in A}\left\Vert x_{n\alpha}-x_{\alpha}\right\Vert \to0$,
the quantity in (\ref{eq:LemWeakConvSuppApp}) converges to zero. 
\end{proof}
\bigskip{}
\begin{proof}[\textbf{\emph{Proof of Lemma \ref{lem:WeakLimitSubvec}}}]
For brevity and clarity, within this proof we use somewhat simplified
notation and let
\[
\mathcal{I}_{\alpha}^{-1}=\begin{bmatrix}A & B\\
B' & C
\end{bmatrix}
\]
denote the partition of $\mathcal{I}_{\alpha}^{-1}$ (so that, e.g.,
$C$ is shorthand for $(\mathcal{I}_{\alpha}^{-1})_{\vartheta\vartheta}$).
This implies that $\mathcal{I}_{\alpha}$ can be expressed as
\[
\mathcal{I}_{\alpha}=\begin{bmatrix}D^{-1} & -D^{-1}BC^{-1}\\
-C^{-1}B'D^{-1} & C^{-1}+C^{-1}B'D^{-1}BC^{-1}
\end{bmatrix}
\]
where $D=A-BC^{-1}B'$ (thus, e.g., $D^{-1}=\mathcal{I}_{\theta\theta\alpha}$).
Note also that $A$, $C$, and $D$ are symmetric (as $\mathcal{I}_{\alpha}$
is symmetric). 

First note that $S_{\alpha}=\mathcal{I}_{\alpha}Z_{\alpha}$ can be
expressed as
\[
\hspace{-10pt}S_{\alpha}=\hspace{-3pt}\begin{bmatrix}D^{-1} & -D^{-1}BC^{-1}\\
-C^{-1}B'D^{-1} & C^{-1}+C^{-1}B'D^{-1}BC^{-1}
\end{bmatrix}\begin{bmatrix}Z_{\theta\alpha}\\
Z_{\vartheta\alpha}
\end{bmatrix}=\begin{bmatrix}D^{-1}Z_{\theta\alpha}-D^{-1}BC^{-1}Z_{\vartheta\alpha}\\
-C^{-1}B'D^{-1}Z_{\theta\alpha}+C^{-1}Z_{\vartheta\alpha}+C^{-1}B'D^{-1}BC^{-1}Z_{\vartheta\alpha}
\end{bmatrix}
\]
so that $S_{\theta\alpha}'DS_{\theta\alpha}$ equals
\begin{align*}
S_{\theta\alpha}'DS_{\theta\alpha} & =(D^{-1}Z_{\theta\alpha}-D^{-1}BC^{-1}Z_{\vartheta\alpha})'D(D^{-1}Z_{\theta\alpha}-D^{-1}BC^{-1}Z_{\vartheta\alpha})\\
 & =Z_{\theta\alpha}'D^{-1}Z_{\theta\alpha}-Z_{\theta\alpha}'D^{-1}BC^{-1}Z_{\vartheta\alpha}-Z_{\vartheta\alpha}'C^{-1}B'D^{-1}Z_{\theta\alpha}+Z_{\vartheta\alpha}'C^{-1}B'D^{-1}BC^{-1}Z_{\vartheta\alpha}.
\end{align*}
Now, since $Z_{\alpha}'\mathcal{I}_{\alpha}Z_{\alpha}$ can be written
as 
\begin{align*}
Z_{\alpha}'\mathcal{I}_{\alpha}Z_{\alpha} & =\begin{bmatrix}Z_{\theta\alpha}\\
Z_{\vartheta\alpha}
\end{bmatrix}'\begin{bmatrix}D^{-1} & -D^{-1}BC^{-1}\\
-C^{-1}B'D^{-1} & C^{-1}+C^{-1}B'D^{-1}BC^{-1}
\end{bmatrix}\begin{bmatrix}Z_{\theta\alpha}\\
Z_{\vartheta\alpha}
\end{bmatrix}\\
 & =Z_{\theta\alpha}'D^{-1}Z_{\theta\alpha}-Z_{\theta\alpha}'D^{-1}BC^{-1}Z_{\vartheta\alpha}-Z_{\vartheta\alpha}'C^{-1}B'D^{-1}Z_{\theta\alpha}\\
 & +Z_{\vartheta\alpha}'C^{-1}Z_{\vartheta\alpha}+Z_{\vartheta\alpha}'C^{-1}B'D^{-1}BC^{-1}Z_{\vartheta\alpha},
\end{align*}
we obtain 
\begin{equation}
Z_{\alpha}'\mathcal{I}_{\alpha}Z_{\alpha}=Z_{\vartheta\alpha}'C^{-1}Z_{\vartheta\alpha}+S_{\theta\alpha}'DS_{\theta\alpha}.\label{eq:ProofLemWeakLimitSubvec1}
\end{equation}

Now consider $\inf_{\boldsymbol{\lambda}\in\Lambda}\left\{ (\boldsymbol{\lambda}-Z_{\alpha})^{\prime}\mathcal{I}_{\alpha}(\boldsymbol{\lambda}-Z_{\alpha})\right\} $.
Similarly as above,
\begin{align*}
 & (\boldsymbol{\lambda}-Z_{\alpha})^{\prime}\mathcal{I}_{\alpha}(\boldsymbol{\lambda}-Z_{\alpha})\\
 & =(\boldsymbol{\lambda}_{\theta}-Z_{\theta\alpha})'D^{-1}(\boldsymbol{\lambda}_{\theta}-Z_{\theta\alpha})-(\boldsymbol{\lambda}_{\theta}-Z_{\theta\alpha})'D^{-1}BC^{-1}(\boldsymbol{\lambda}_{\vartheta}-Z_{\vartheta\alpha})-(\boldsymbol{\lambda}_{\vartheta}-Z_{\vartheta\alpha})'C^{-1}B'D^{-1}(\boldsymbol{\lambda}_{\theta}-Z_{\theta\alpha})\\
 & +(\boldsymbol{\lambda}_{\vartheta}-Z_{\vartheta\alpha})'C^{-1}(\boldsymbol{\lambda}_{\vartheta}-Z_{\vartheta\alpha})+(\boldsymbol{\lambda}_{\vartheta}-Z_{\vartheta\alpha})'C^{-1}B'D^{-1}BC^{-1}(\boldsymbol{\lambda}_{\vartheta}-Z_{\vartheta\alpha})\\
 & =(\boldsymbol{\lambda}_{\theta}-Z_{\theta\alpha})'D^{-1}(\boldsymbol{\lambda}_{\theta}-Z_{\theta\alpha})-(\boldsymbol{\lambda}_{\theta}-Z_{\theta\alpha})'D^{-1}[BC^{-1}(\boldsymbol{\lambda}_{\vartheta}-Z_{\vartheta\alpha})]-[BC^{-1}(\boldsymbol{\lambda}_{\vartheta}-Z_{\vartheta\alpha})]'D^{-1}(\boldsymbol{\lambda}_{\theta}-Z_{\theta\alpha})\\
 & +(\boldsymbol{\lambda}_{\vartheta}-Z_{\vartheta\alpha})'C^{-1}(\boldsymbol{\lambda}_{\vartheta}-Z_{\vartheta\alpha})+[BC^{-1}(\boldsymbol{\lambda}_{\vartheta}-Z_{\vartheta\alpha})]'D^{-1}[BC^{-1}(\boldsymbol{\lambda}_{\vartheta}-Z_{\vartheta\alpha})]\\
 & =(\boldsymbol{\lambda}_{\vartheta}-Z_{\vartheta\alpha})'C^{-1}(\boldsymbol{\lambda}_{\vartheta}-Z_{\vartheta\alpha})+\{(\boldsymbol{\lambda}_{\theta}-Z_{\theta\alpha})-[BC^{-1}(\boldsymbol{\lambda}_{\vartheta}-Z_{\vartheta\alpha})]\}'D^{-1}\{(\boldsymbol{\lambda}_{\theta}-Z_{\theta\alpha})-[BC^{-1}(\boldsymbol{\lambda}_{\vartheta}-Z_{\vartheta\alpha})]\}.
\end{align*}
Now, for any fixed $\boldsymbol{\lambda}_{\vartheta}\in\mathbb{R}^{q_{\vartheta}}$,
Assumption \ref{assu:cone2} implies that 
\[
\inf_{\boldsymbol{\lambda}_{\theta}\in\mathbb{R}^{q_{\theta}}}\{(\boldsymbol{\lambda}_{\theta}-Z_{\theta\alpha})-[BC^{-1}(\boldsymbol{\lambda}_{\vartheta}-Z_{\vartheta\alpha})]\}'D^{-1}\{(\boldsymbol{\lambda}_{\theta}-Z_{\theta\alpha})-[BC^{-1}(\boldsymbol{\lambda}_{\vartheta}-Z_{\vartheta\alpha})]\}=0
\]
(cf. \citet[eqn. (7.35)]{andrews1999estimation}) so that

\begin{equation}
\inf_{\boldsymbol{\lambda}\in\Lambda}\left\{ (\boldsymbol{\lambda}-Z_{\alpha})^{\prime}\mathcal{I}_{\alpha}(\boldsymbol{\lambda}-Z_{\alpha})\right\} =\inf_{\boldsymbol{\lambda}_{\vartheta}\in\Lambda_{\vartheta}}\left\{ (\boldsymbol{\lambda}_{\vartheta}-Z_{\vartheta\alpha})'C^{-1}(\boldsymbol{\lambda}_{\vartheta}-Z_{\vartheta\alpha})\right\} .\label{eq:ProofLemWeakLimitSubvec2}
\end{equation}
Combining (\ref{eq:ProofLemWeakLimitSubvec1}) and (\ref{eq:ProofLemWeakLimitSubvec2})
and recalling that $C^{-1}=(\mathcal{I}_{\alpha}^{-1})_{\vartheta\vartheta}^{-1}$
and $D=\mathcal{I}_{\theta\theta\alpha}^{-1}$ yields the equality
stated in the lemma.

Finally, $(\mathcal{I}_{\alpha}^{-1})_{\vartheta\vartheta}$ and $Z_{\vartheta\alpha}$
can be expressed as 
\begin{align*}
\hspace{-10pt}(\mathcal{I}_{\alpha}^{-1})_{\vartheta\vartheta} & =(\mathcal{I}_{\vartheta\vartheta\alpha}-\mathcal{I}_{\vartheta\theta\alpha}\mathcal{I}_{\theta\theta\alpha}^{-1}\mathcal{I}_{\theta\vartheta\alpha})^{-1}\ [=\mathcal{I}_{\vartheta\vartheta\alpha}^{-1}+\mathcal{I}_{\vartheta\vartheta\alpha}^{-1}\mathcal{I}_{\vartheta\theta\alpha}(\mathcal{I}_{\theta\theta\alpha}-\mathcal{I}_{\theta\vartheta\alpha}\mathcal{I}_{\vartheta\vartheta\alpha}^{-1}\mathcal{I}_{\vartheta\theta\alpha})^{-1}\mathcal{I}_{\theta\vartheta\alpha}\mathcal{I}_{\vartheta\vartheta\alpha}^{-1}],\\
Z_{\vartheta\alpha} & =(\mathcal{I}_{\alpha}^{-1})_{\vartheta\vartheta}(S_{\vartheta\alpha}-\mathcal{I}_{\vartheta\theta\alpha}\mathcal{I}_{\theta\theta\alpha}^{-1}S_{\theta\alpha})\ [=\mathcal{I}_{\vartheta\vartheta\alpha}^{-1}S_{\vartheta\alpha}+\mathcal{I}_{\vartheta\vartheta\alpha}^{-1}\mathcal{I}_{\vartheta\theta\alpha}(\mathcal{I}_{\theta\theta\alpha}-\mathcal{I}_{\theta\vartheta\alpha}\mathcal{I}_{\vartheta\vartheta\alpha}^{-1}\mathcal{I}_{\vartheta\theta\alpha})^{-1}(\mathcal{I}_{\theta\vartheta\alpha}\mathcal{I}_{\vartheta\vartheta\alpha}^{-1}S_{\vartheta\alpha}-S_{\theta\alpha})],
\end{align*}
where the two different expressions result from two different ways
of writing the inverse of a partitioned matrix.
\end{proof}
\bigskip{}

\pagebreak{}

\section{Further details for the LMAR example}

\subsection{Verification of Assumption 5(ii), further details}

As for the weak convergence requirement in part (ii), we rely on Theorem
2 (and the remarks that follow it) in \citet{andrews1995admissibility}.
As can be seen from the proof of their Theorem 2, it suffices to verify
their conditions EP1(a), EP1(e), and EP4 (omitting the weakly exogeneous
$X_{t}$ variables therein). Under the null hypothesis, $y_{t}$ is
a linear Gaussian AR($p$) process so that condition EP1(a) is satisfied
with geometrically declining mixing numbers. To check condition EP1(e),
we show that $E\bigl[\textrm{sup}{}_{\alpha\in A}\sup\nolimits _{(\pi,\varpi)\in\Pi}\lvert l_{t}^{\pi}(\alpha,\pi,\varpi)\rvert\bigr]<\infty$,
$E\bigl[\textrm{sup}{}_{\alpha\in A}\lVert\nabla_{(\pi,\varpi)}l_{t}^{\pi}(\alpha,\pi^{*},0)\rVert^{r}\bigr]<\infty$
for any positive $r$, and $E\bigl[\textrm{sup}{}_{\alpha\in A}\sup\nolimits _{(\pi,\varpi)\in\Pi}\lVert\nabla_{(\pi,\varpi)(\pi,\varpi)'}^{2}l_{t}^{\pi}(\alpha,\pi,\varpi)\rVert\bigr]<\infty$.
The first of these moment conditions follows from the arguments used
to verify our Assumption 2 (see the verification of this Assumption
for the GMAR model; the details for the LMAR model are presented there).
The second holds due to the expression of $\nabla_{(\pi,\varpi)}l_{t}^{\pi}\left(\alpha,\pi^{\ast},0\right)$
in (\ref{eq:LMARDer1*}), the fact that $0<\alpha_{1,t}^{L}(\alpha)<1$,
and Lemma \ref{lem:XXX1}. The third is verified below in Supplementary
Appedix E.2. As the compactness requirement of condition EP4(a) holds
by our Assumption 1(ii), it remains to verify EP4(b). To this end,
note that for arbitrary $a,b\in A$, 
\[
\lVert\nabla_{(\pi,\varpi)}l_{t}^{\pi}(a,\pi^{*},0)-\nabla_{(\pi,\varpi)}l_{t}^{\pi}(b,\pi^{*},0)\rVert^{r}=\lvert\alpha_{1,t}^{L}(a)-\alpha_{1,t}^{L}(b)\rvert^{r}\left\Vert \frac{\nabla f_{t}(\pi^{\ast})}{f_{t}(\pi^{\ast})}\right\Vert ^{r}.
\]
By straightforward differentiation, $\nabla_{\alpha}\alpha_{1,t}^{L}(\alpha)=\alpha_{1,t}^{L}(\alpha)(1-\alpha_{1,t}^{L}(\alpha))(1,y_{t-1},\ldots,y_{t-m})$,
so that by the mean value theorem
\[
\alpha_{1,t}^{L}(a)-\alpha_{1,t}^{L}(b)=\alpha_{1,t}^{L}(c_{a,b})(1-\alpha_{1,t}^{L}(c_{a,b}))(1,y_{t-1},\ldots,y_{t-m})'(a-b)
\]
for some $c_{a,b}\in\mathbb{R}^{m+1}$ between $a$ and $b$ (as $A$
is not necessarily convex, $c_{a,b}$ does not necessarily belong
to $A$, but this has no effect in what follows as the expression
$\alpha_{1,t}^{L}(c_{a,b})$ is nevertheless well defined for all
$c_{a,b}\in\mathbb{R}^{m+1}$). Setting $B_{t}=1+\lvert y_{t-1}\rvert+\ldots+\lvert y_{t-m}\rvert$
and noting that $0<\alpha_{1,t}^{L}(c_{a,b})<1$ this implies that
\[
\lvert\alpha_{1,t}^{L}(a)-\alpha_{1,t}^{L}(b)\rvert\leq\lvert(1,y_{t-1},\ldots,y_{t-m})'(a-b)\rvert\leq(1+\lvert y_{t-1}\rvert+\ldots+\lvert y_{t-m}\rvert)\lVert a-b\rVert=B_{t}\lVert a-b\rVert.
\]
Hence
\[
E\Bigl[\sup_{a,b\in A,\lVert a-b\rVert<\delta}\lVert\nabla_{(\pi,\varpi)}l_{t}^{\pi}(a,\pi^{*},0)-\nabla_{(\pi,\varpi)}l_{t}^{\pi}(b,\pi^{*},0)\rVert^{r}\Bigr]<\delta^{r}E\biggl[B_{t}^{r}\left\Vert \frac{\nabla f_{t}(\pi^{\ast})}{f_{t}(\pi^{\ast})}\right\Vert ^{r}\biggr]
\]
where on the majorant side the expectation is finite (due to the fact
that the $y_{t}$'s possess moments of all orders, see also the proof
of Lemma \ref{lem:XXX1}). Hence condition EP4(b) holds, and the desired
weak convergence follows. 

\subsection{Verification of Assumption 5(iv), further details}

It remains to show that $E\bigl[\textrm{sup}{}_{\alpha\in A}\sup\nolimits _{(\pi,\varpi)\in\Pi}\lVert\nabla_{(\pi,\varpi)(\pi,\varpi)'}^{2}l_{t}^{\pi}(\alpha,\pi,\varpi)\rVert\bigr]<\infty$.
This in turn follows if we show the same with $\nabla_{(\pi,\varpi)(\pi,\varpi)'}^{2}l_{t}^{\pi}(\alpha,\pi,\varpi)$
replaced by $\nabla_{\pi\pi'}^{2}l_{t}^{\pi}(\alpha,\pi,\varpi)$,
$\nabla_{\varpi\varpi'}^{2}l_{t}^{\pi}(\alpha,\pi,\varpi)$, and $\nabla_{\pi\varpi'}^{2}l_{t}^{\pi}(\alpha,\pi,\varpi)$.
Consider the expression of $\nabla_{\pi\pi'}^{2}l_{t}^{\pi}(\alpha,\pi,\varpi)$
given in Appendix B and recall that $0<\alpha_{1,t}^{L}(\alpha),\alpha_{2,t}^{L}(\alpha)<1$
and $\nabla f_{t}(\pi)=f_{t}(\pi)\nabla_{\pi}l_{t}^{0}(\pi)$ with
$l_{t}^{0}(\pi)=\log[f_{t}(\pi)]$. Then we can, for instance, write

\[
\biggl\|\frac{\alpha_{1,t}^{L}(\alpha)\nabla f_{t}(\pi)}{f_{2,t}^{\pi}(\alpha,\pi,\varpi)}\biggr\|=\biggl\|\frac{\alpha_{1,t}^{L}(\alpha)f_{t}(\pi)\nabla_{\pi}l_{t}^{0}(\pi)}{\alpha_{1,t}^{L}(\alpha)f_{t}(\pi)+\alpha_{2,t}^{L}(\alpha)f_{t}(\pi-\varpi)}\biggr\|\leq\left\Vert \nabla_{\pi}l_{t}^{0}(\pi)\right\Vert 
\]
and
\begin{align*}
\biggl\|\frac{\alpha_{1,t}^{L}(\alpha)\nabla'f_{t}(\pi)+\alpha_{2,t}^{L}(\alpha)\nabla'f_{t}(\pi-\varpi)}{f_{2,t}^{\pi}(\alpha,\pi,\varpi)}\biggr\| & =\biggl\|\frac{\alpha_{1,t}^{L}(\alpha)f_{t}(\pi)\nabla_{\pi}l_{t}^{0}(\pi)+\alpha_{2,t}^{L}(\alpha)f_{t}(\pi-\varpi)\nabla_{\pi}l_{t}^{0}(\pi-\varpi)}{\alpha_{1,t}^{L}(\alpha)f_{t}(\pi)+\alpha_{2,t}^{L}(\alpha)f_{t}(\pi-\varpi)}\biggr\|\\
 & \leq\left\Vert \nabla_{\pi}l_{t}^{0}(\pi)\right\Vert +\left\Vert \nabla_{\pi}l_{t}^{0}(\pi-\varpi)\right\Vert .
\end{align*}
As similar inequalities can be obtained for the second term of $\nabla_{\pi\pi'}^{2}l_{t}^{\pi}(\alpha,\pi,\varpi)$,
we get
\begin{align*}
\left\Vert \nabla_{\pi\pi'}^{2}l_{t}^{\pi}(\alpha,\pi,\varpi)\right\Vert  & \leq\biggl\|\alpha_{1,t}^{L}(\alpha)\frac{\nabla^{2}f_{t}(\pi)}{f_{2,t}^{\pi}(\alpha,\pi,\varpi)}\biggr\|+\left\Vert \nabla_{\pi}l_{t}^{0}(\pi)\right\Vert ^{2}+\left\Vert \nabla_{\pi}l_{t}^{0}(\pi)\right\Vert \left\Vert \nabla_{\pi}l_{t}^{0}(\pi-\varpi)\right\Vert \\
 & +\biggl\|\alpha_{2,t}^{L}(\alpha)\frac{\nabla^{2}f_{t}(\pi-\varpi)}{f_{2,t}^{\pi}(\alpha,\pi,\varpi)}\biggr\|+\left\Vert \nabla_{\pi}l_{t}^{0}(\pi-\varpi)\right\Vert ^{2}+\left\Vert \nabla_{\pi}l_{t}^{0}(\pi)\right\Vert \left\Vert \nabla_{\pi}l_{t}^{0}(\pi-\varpi)\right\Vert .
\end{align*}
Next note that 
\begin{align*}
\frac{\nabla^{2}f_{t}(\pi)}{f_{2,t}^{\pi}(\alpha,\pi,\varpi)} & =\frac{\nabla(f_{t}(\pi)\nabla_{\pi}l_{t}^{0}(\pi))}{f_{2,t}^{\pi}(\alpha,\pi,\varpi)}=\frac{f_{t}(\pi)\nabla_{\pi}l_{t}^{0}(\pi)\nabla_{\pi'}l_{t}^{0}(\pi)}{f_{2,t}^{\pi}(\alpha,\pi,\varpi)}+\frac{f_{t}(\pi)\nabla_{\pi\pi'}^{2}l_{t}^{0}(\pi)}{f_{2,t}^{\pi}(\alpha,\pi,\varpi)},
\end{align*}
so that arguments similar to those already used above give
\[
\biggl\|\alpha_{1,t}^{L}(\alpha)\frac{\nabla^{2}f_{t}(\pi)}{f_{2,t}^{\pi}(\alpha,\pi,\varpi)}\biggr\|\leq\left\Vert \nabla_{\pi}l_{t}^{0}(\pi)\right\Vert ^{2}+\left\Vert \nabla_{\pi\pi'}^{2}l_{t}^{0}(\pi)\right\Vert 
\]
and 
\[
\biggl\|\alpha_{2,t}^{L}(\alpha)\frac{\nabla^{2}f_{t}(\pi-\varpi)}{f_{2,t}^{\pi}(\alpha,\pi,\varpi)}\biggr\|\leq\left\Vert \nabla_{\pi}l_{t}^{0}(\pi-\varpi)\right\Vert ^{2}+\left\Vert \nabla_{\pi\pi'}^{2}l_{t}^{0}(\pi-\varpi)\right\Vert .
\]
Hence, we can conclude that
\begin{align*}
\left\Vert \nabla_{\pi\pi'}^{2}l_{t}^{\pi}(\alpha,\pi,\varpi)\right\Vert  & \leq2\left\Vert \nabla_{\pi}l_{t}^{0}(\pi)\right\Vert ^{2}+2\left\Vert \nabla_{\pi}l_{t}^{0}(\pi-\varpi)\right\Vert ^{2}+2\left\Vert \nabla_{\pi}l_{t}^{0}(\pi)\right\Vert \left\Vert \nabla_{\pi}l_{t}^{0}(\pi-\varpi)\right\Vert \\
 & \qquad\mbox{+\ensuremath{\left\Vert \nabla_{\pi\pi'}^{2}\mathit{l}_{t}^{0}(\pi)\right\Vert }}+\left\Vert \nabla_{\pi\pi'}^{2}l_{t}^{0}(\pi-\varpi)\right\Vert .
\end{align*}
To bound the expression on the dominant side, note that $\nabla_{\pi}l_{t}^{0}(\pi)=\frac{\nabla f_{t}(\pi)}{f_{t}(\pi)}$
and $\nabla_{\pi\pi'}^{2}\mathit{l}_{t}^{0}(\pi)=\frac{\nabla^{2}f_{t}(\pi)}{f_{t}(\pi)}-\frac{\nabla f_{t}(\pi)}{f_{t}(\pi)}\frac{\nabla'f_{t}(\pi)}{f_{t}(\pi)}$
so that Lemma \ref{lem:XXX1} ensures that $E\left[\textrm{sup}{}_{\alpha\in A}\sup\nolimits _{(\pi,\varpi)\in\Pi}\left\Vert \nabla_{\pi\pi'}^{2}l_{t}^{\pi}(\alpha,\pi,\varpi)\right\Vert \right]<\infty$.
An inspection of the expressions of $\nabla_{\pi\varpi'}^{2}l_{t}^{\pi}(\alpha,\pi,\varpi)$
and $\nabla_{\varpi\varpi'}^{2}l_{t}^{\pi}(\alpha,\pi,\varpi)$ in
Appendix B shows that a similar result can be obtained with $\nabla_{\pi\pi'}^{2}l_{t}^{\pi}(\alpha,\pi,\varpi)$
replaced by $\nabla_{\pi\varpi'}^{2}l_{t}^{\pi}(\alpha,\pi,\varpi)$
and $\nabla_{\varpi\varpi'}^{2}l_{t}^{\pi}(\alpha,\pi,\varpi)$, yielding
the desired result. \pagebreak{}

\section{Further details for the GMAR example}

\subsection{Partial derivatives of the reparameterized log-likelihood function}

Here we present certain partial derivatives of $l_{t}^{\pi}(\alpha,\beta,\pi,\varpi)$
with respect to $(\beta,\pi,\varpi)$. For brevity, set $\tilde{\pi}=(\beta,\pi)$
(and similarly $\tilde{\pi}^{*}=(\beta^{*},\pi^{*})$), so that the
desired derivatives are with respect to $\tilde{\pi}$ and $\varpi$
or, elementwise, with respect to $\tilde{\pi}_{i}$ and $\varpi_{j}$
for $i=1,\ldots,p+2$ and $j=1,\ldots,p+1$. In the derivative expressions
below, the subindices in $\tilde{\pi}$ and $\varpi$ are tacitly
assumed to be withing these ranges. For brevity, denote $l_{t}^{\pi*}=l_{t}^{\pi}(\alpha,\tilde{\pi}^{*},0)$,
$f_{t}^{*}=f_{t}(\tilde{\pi}^{*})$, $\mathsf{n}_{p}^{*}=\mathsf{n}_{p}(\tilde{\pi}^{*})$,
and similarly for their partial derivatives. 

The following derivatives are obtained with straightforward (but tedious
and lengthy) differentiation. The necessary calculations for the first-
and second-order derivatives are presented in Supplementary Appendix
F.7, but for brevity we omit the detailed calculations for the third-
and fourth-order derivatives. 

First- and second-order derivatives:
\begin{align*}
\nabla_{\tilde{\pi}_{i}}l_{t}^{\pi*} & =\frac{\nabla_{i}f_{t}^{*}}{f_{t}^{*}}\\
\nabla_{\varpi_{j}}l_{t}^{\pi*} & =0\\
\nabla_{\tilde{\pi}_{i}\tilde{\pi}_{j}}^{2}l_{t}^{\pi*} & =\frac{\nabla_{ij}^{2}f_{t}^{*}}{f_{t}^{*}}-\frac{\nabla_{i}f_{t}^{*}}{f_{t}^{*}}\frac{\nabla_{j}f_{t}^{*}}{f_{t}^{*}}.\\
\nabla_{\tilde{\pi}_{i}\varpi_{j}}^{2}l_{t}^{\pi*} & =0\\
\nabla_{\varpi_{i}\varpi_{j}}^{2}l_{t}^{\pi*} & =\alpha_{1}\alpha_{2}\left[\frac{\nabla_{i+1}\mathsf{n}_{p}^{*}}{\mathsf{n}_{p}^{*}}\frac{\nabla_{j+1}f_{t}^{*}}{f_{t}^{*}}+\frac{\nabla_{i+1}f_{t}^{*}}{f_{t}^{*}}\frac{\nabla_{j+1}\mathsf{n}_{p}^{*}}{\mathsf{n}_{p}^{*}}+\frac{\nabla_{i+1,j+1}^{2}f_{t}^{*}}{f_{t}^{*}}\right]
\end{align*}
Third-order derivatives:
\begin{align*}
\nabla_{\tilde{\pi}_{i}\tilde{\pi}_{j}\tilde{\pi}_{k}}^{3}l_{t}^{\pi*} & =\frac{\nabla_{ijk}^{3}f_{t}^{*}}{f_{t}^{*}}-\frac{\nabla_{ij}^{2}f_{t}^{*}}{f_{t}^{*}}\frac{\nabla_{k}f_{t}^{*}}{f_{t}^{*}}-\frac{\nabla_{ik}^{2}f_{t}^{*}}{f_{t}^{*}}\frac{\nabla_{j}f_{t}^{*}}{f_{t}^{*}}-\frac{\nabla_{jk}^{2}f_{t}^{*}}{f_{t}^{*}}\frac{\nabla_{i}f_{t}^{*}}{f_{t}^{*}}+2\frac{\nabla_{i}f_{t}^{*}}{f_{t}^{*}}\frac{\nabla_{j}f_{t}^{*}}{f_{t}^{*}}\frac{\nabla_{k}f_{t}^{*}}{f_{t}^{*}}\\
\nabla_{\tilde{\pi}_{i}\tilde{\pi}_{j}\varpi_{k}}^{3}l_{t}^{\pi*} & =-\alpha_{1}\alpha_{2}\frac{\nabla_{i}f_{t}^{*}}{f_{t}^{*}}\left(\frac{\nabla_{j,k+1}^{2}\mathsf{n}_{p}^{*}}{\mathsf{n}_{p}^{*}}-\frac{\nabla_{j}\mathsf{n}_{p}^{*}}{\mathsf{n}_{p}^{*}}\frac{\nabla_{k+1}\mathsf{n}_{p}^{*}}{\mathsf{n}_{p}^{*}}\right)\\
\nabla_{\tilde{\pi}_{i}\varpi_{j}\varpi_{k}}^{3}l_{t}^{\pi*} & =\alpha_{1}\alpha_{2}\biggl[\frac{\nabla_{k+1}\mathsf{n}_{p}^{*}}{\mathsf{n}_{p}^{*}}\left(\frac{\nabla_{i,j+1}^{2}f_{t}^{*}}{f_{t}^{*}}-\frac{\nabla_{i}f_{t}^{*}}{f_{t}^{*}}\frac{\nabla_{j+1}f_{t}^{*}}{f_{t}^{*}}\right)+\frac{\nabla_{j+1}\mathsf{n}_{p}^{*}}{\mathsf{n}_{p}^{*}}\left(\frac{\nabla_{i,k+1}^{2}f_{t}^{*}}{f_{t}^{*}}-\frac{\nabla_{i}f_{t}^{*}}{f_{t}^{*}}\frac{\nabla_{k+1}f_{t}^{*}}{f_{t}^{*}}\right)\\
 & \qquad\qquad+\left(\frac{\nabla_{i,j+1,k+1}^{3}f_{t}^{*}}{f_{t}^{*}}-\frac{\nabla_{i}f_{t}^{*}}{f_{t}^{*}}\frac{\nabla_{j+1,k+1}^{2}f_{t}^{*}}{f_{t}^{*}}\right)+\left(\frac{\nabla_{i,k+1}^{2}\mathsf{n}_{p}^{*}}{\mathsf{n}_{p}^{*}}-\frac{\nabla_{i}\mathsf{n}_{p}^{*}}{\mathsf{n}_{p}^{*}}\frac{\nabla_{k+1}\mathsf{n}_{p}^{*}}{\mathsf{n}_{p}^{*}}\right)\frac{\nabla_{j+1}f_{t}^{*}}{f_{t}^{*}}\\
 & \qquad\qquad+\left(\frac{\nabla_{i,j+1}^{2}\mathsf{n}_{p}^{*}}{\mathsf{n}_{p}^{*}}-\frac{\nabla_{i}\mathsf{n}_{p}^{*}}{\mathsf{n}_{p}^{*}}\frac{\nabla_{j+1}\mathsf{n}_{p}^{*}}{\mathsf{n}_{p}^{*}}\right)\frac{\nabla_{k+1}f_{t}^{*}}{f_{t}^{*}}\biggr]\\
\nabla_{\varpi_{i}\varpi_{j}\varpi_{k}}^{3}l_{t}^{\pi*} & =\alpha_{1}\alpha_{2}(\alpha_{2}-\alpha_{1})\biggl[\frac{\nabla_{i+1,j+1}^{2}\mathsf{n}_{p}^{*}}{\mathsf{n}_{p}^{*}}\frac{\nabla_{k+1}f_{t}^{*}}{f_{t}^{*}}+\frac{\nabla_{i+1,k+1}^{2}\mathsf{n}_{p}^{*}}{\mathsf{n}_{p}^{*}}\frac{\nabla_{j+1}f_{t}^{*}}{f_{t}^{*}}+\frac{\nabla_{j+1,k+1}^{2}\mathsf{n}_{p}^{*}}{\mathsf{n}_{p}^{*}}\frac{\nabla_{i+1}f_{t}^{*}}{f_{t}^{*}}\\
 & \qquad\qquad\qquad\qquad+\frac{\nabla_{i+1}\mathsf{n}_{p}^{*}}{\mathsf{n}_{p}^{*}}\frac{\nabla_{j+1,k+1}^{2}f_{t}^{*}}{f_{t}^{*}}+\frac{\nabla_{j+1}\mathsf{n}_{p}^{*}}{\mathsf{n}_{p}^{*}}\frac{\nabla_{i+1,k+1}^{2}f_{t}^{*}}{f_{t}^{*}}+\frac{\nabla_{k+1}\mathsf{n}_{p}^{*}}{\mathsf{n}_{p}^{*}}\frac{\nabla_{i+1,j+1}^{2}f_{t}^{*}}{f_{t}^{*}}\\
 & \qquad\qquad\qquad\qquad+\frac{\nabla_{i+1,j+1,k+1}^{3}f_{t}^{*}}{f_{t}^{*}}\biggr]
\end{align*}
Fourth-order derivative (fourth-order derivatives with respect to
$\tilde{\pi}$ will not be explicitly needed, and thus we omit their
expressions):

\begin{align*}
 & \nabla_{\varpi_{i}\varpi_{j}\varpi_{k}\varpi_{l}}^{4}l_{t}^{\pi*}\\
 & =-\alpha_{1}^{2}\alpha_{2}^{2}\Biggl[\left(\frac{\nabla_{i+1}\mathsf{n}_{p}^{*}}{\mathsf{n}_{p}^{*}}\frac{\nabla_{j+1,k+1}^{2}\mathsf{n}_{p}^{*}}{\mathsf{n}_{p}^{*}}+\frac{\nabla_{j+1}\mathsf{n}_{p}^{*}}{\mathsf{n}_{p}^{*}}\frac{\nabla_{i+1,k+1}^{2}\mathsf{n}_{p}^{*}}{\mathsf{n}_{p}^{*}}+\frac{\nabla_{k+1}\mathsf{n}_{p}^{*}}{\mathsf{n}_{p}^{*}}\frac{\nabla_{i+1,j+1}^{2}\mathsf{n}_{p}^{*}}{\mathsf{n}_{p}^{*}}\right)\frac{\nabla_{l+1}f_{t}^{*}}{f_{t}^{*}}\\
 & \qquad\qquad+\left(\frac{\nabla_{i+1}\mathsf{n}_{p}^{*}}{\mathsf{n}_{p}^{*}}\frac{\nabla_{j+1,l+1}^{2}\mathsf{n}_{p}^{*}}{\mathsf{n}_{p}^{*}}+\frac{\nabla_{j+1}\mathsf{n}_{p}^{*}}{\mathsf{n}_{p}^{*}}\frac{\nabla_{i+1,l+1}^{2}\mathsf{n}_{p}^{*}}{\mathsf{n}_{p}^{*}}+\frac{\nabla_{l+1}\mathsf{n}_{p}^{*}}{\mathsf{n}_{p}^{*}}\frac{\nabla_{i+1,j+1}^{2}\mathsf{n}_{p}^{*}}{\mathsf{n}_{p}^{*}}\right)\frac{\nabla_{k+1}f_{t}^{*}}{f_{t}^{*}}\\
 & \qquad\qquad+\left(\frac{\nabla_{i+1}\mathsf{n}_{p}^{*}}{\mathsf{n}_{p}^{*}}\frac{\nabla_{k+1,l+1}^{2}\mathsf{n}_{p}^{*}}{\mathsf{n}_{p}^{*}}+\frac{\nabla_{k+1}\mathsf{n}_{p}^{*}}{\mathsf{n}_{p}^{*}}\frac{\nabla_{i+1,l+1}^{2}\mathsf{n}_{p}^{*}}{\mathsf{n}_{p}^{*}}+\frac{\nabla_{l+1}\mathsf{n}_{p}^{*}}{\mathsf{n}_{p}^{*}}\frac{\nabla_{i+1,k+1}^{2}\mathsf{n}_{p}^{*}}{\mathsf{n}_{p}^{*}}\right)\frac{\nabla_{j+1}f_{t}^{*}}{f_{t}^{*}}\\
 & \qquad\qquad+\left(\frac{\nabla_{j+1}\mathsf{n}_{p}^{*}}{\mathsf{n}_{p}^{*}}\frac{\nabla_{k+1,l+1}^{2}\mathsf{n}_{p}^{*}}{\mathsf{n}_{p}^{*}}+\frac{\nabla_{k+1}\mathsf{n}_{p}^{*}}{\mathsf{n}_{p}^{*}}\frac{\nabla_{j+1,l+1}^{2}\mathsf{n}_{p}^{*}}{\mathsf{n}_{p}^{*}}+\frac{\nabla_{l+1}\mathsf{n}_{p}^{*}}{\mathsf{n}_{p}^{*}}\frac{\nabla_{j+1,k+1}^{2}\mathsf{n}_{p}^{*}}{\mathsf{n}_{p}^{*}}\right)\frac{\nabla_{i+1}f_{t}^{*}}{f_{t}^{*}}\\
 & \qquad\qquad+\left(\frac{\nabla_{i+1}\mathsf{n}_{p}^{*}}{\mathsf{n}_{p}^{*}}\frac{\nabla_{j+1}f_{t}^{*}}{f_{t}^{*}}+\frac{\nabla_{j+1}\mathsf{n}_{p}^{*}}{\mathsf{n}_{p}^{*}}\frac{\nabla_{i+1}f_{t}^{*}}{f_{t}^{*}}+\frac{\nabla_{i+1,j+1}^{2}f_{t}^{*}}{f_{t}^{*}}\right)\\
 & \qquad\qquad\qquad\qquad\times\left(\frac{\nabla_{k+1}\mathsf{n}_{p}^{*}}{\mathsf{n}_{p}^{*}}\frac{\nabla_{l+1}f_{t}^{*}}{f_{t}^{*}}+\frac{\nabla_{l+1}\mathsf{n}_{p}^{*}}{\mathsf{n}_{p}^{*}}\frac{\nabla_{k+1}f_{t}^{*}}{f_{t}^{*}}+\frac{\nabla_{k+1,l+1}^{2}f_{t}^{*}}{f_{t}^{*}}\right)\\
 & \qquad\qquad+\left(\frac{\nabla_{i+1}\mathsf{n}_{p}^{*}}{\mathsf{n}_{p}^{*}}\frac{\nabla_{k+1}f_{t}^{*}}{f_{t}^{*}}+\frac{\nabla_{k+1}\mathsf{n}_{p}^{*}}{\mathsf{n}_{p}^{*}}\frac{\nabla_{i+1}f_{t}^{*}}{f_{t}^{*}}+\frac{\nabla_{i+1,k+1}^{2}f_{t}^{*}}{f_{t}^{*}}\right)\\
 & \qquad\qquad\qquad\qquad\times\left(\frac{\nabla_{j+1}\mathsf{n}_{p}^{*}}{\mathsf{n}_{p}^{*}}\frac{\nabla_{l+1}f_{t}^{*}}{f_{t}^{*}}+\frac{\nabla_{l+1}\mathsf{n}_{p}^{*}}{\mathsf{n}_{p}^{*}}\frac{\nabla_{j+1}f_{t}^{*}}{f_{t}^{*}}+\frac{\nabla_{j+1,l+1}^{2}f_{t}^{*}}{f_{t}^{*}}\right)\\
 & \qquad\qquad+\left(\frac{\nabla_{i+1}\mathsf{n}_{p}^{*}}{\mathsf{n}_{p}^{*}}\frac{\nabla_{l+1}f_{t}^{*}}{f_{t}^{*}}+\frac{\nabla_{l+1}\mathsf{n}_{p}^{*}}{\mathsf{n}_{p}^{*}}\frac{\nabla_{i+1}f_{t}^{*}}{f_{t}^{*}}+\frac{\nabla_{i+1,l+1}^{2}f_{t}^{*}}{f_{t}^{*}}\right)\\
 & \qquad\qquad\qquad\qquad\times\left(\frac{\nabla_{j+1}\mathsf{n}_{p}^{*}}{\mathsf{n}_{p}^{*}}\frac{\nabla_{k+1}f_{t}^{*}}{f_{t}^{*}}+\frac{\nabla_{k+1}\mathsf{n}_{p}^{*}}{\mathsf{n}_{p}^{*}}\frac{\nabla_{j+1}f_{t}^{*}}{f_{t}^{*}}+\frac{\nabla_{j+1,k+1}^{2}f_{t}^{*}}{f_{t}^{*}}\right)\Biggr]\\
 & +\alpha_{1}\alpha_{2}(1-3\alpha_{1}\alpha_{2})\Biggl[\frac{\nabla_{i+1,j+1,k+1}^{3}\mathsf{n}_{p}^{*}}{\mathsf{n}_{p}^{*}}\frac{\nabla_{l+1}f_{t}^{*}}{f_{t}^{*}}+\frac{\nabla_{i+1,j+1,l+1}^{3}\mathsf{n}_{p}^{*}}{\mathsf{n}_{p}^{*}}\frac{\nabla_{k+1}f_{t}^{*}}{f_{t}^{*}}+\frac{\nabla_{i+1,k+1,l+1}^{3}\mathsf{n}_{p}^{*}}{\mathsf{n}_{p}^{*}}\frac{\nabla_{j+1}f_{t}^{*}}{f_{t}^{*}}\\
 & \qquad\qquad\qquad\qquad+\frac{\nabla_{j+1,k+1,l+1}^{3}\mathsf{n}_{p}^{*}}{\mathsf{n}_{p}^{*}}\frac{\nabla_{i+1}f_{t}^{*}}{f_{t}^{*}}+\frac{\nabla_{i+1}\mathsf{n}_{p}^{*}}{\mathsf{n}_{p}^{*}}\frac{\nabla_{j+1,k+1,l+1}^{3}f_{t}^{*}}{f_{t}^{*}}+\frac{\nabla_{j+1}\mathsf{n}_{p}^{*}}{\mathsf{n}_{p}^{*}}\frac{\nabla_{i+1,k+1,l+1}^{3}f_{t}^{*}}{f_{t}^{*}}\\
 & \qquad\qquad\qquad\qquad+\frac{\nabla_{k+1}\mathsf{n}_{p}^{*}}{\mathsf{n}_{p}^{*}}\frac{\nabla_{i+1,j+1,l+1}^{3}f_{t}^{*}}{f_{t}^{*}}+\frac{\nabla_{l+1}\mathsf{n}_{p}^{*}}{\mathsf{n}_{p}^{*}}\frac{\nabla_{i+1,j+1,k+1}^{3}f_{t}^{*}}{f_{t}^{*}}+\frac{\nabla_{i+1,j+1,k+1,l+1}^{4}f_{t}^{*}}{f_{t}^{*}}\Biggr]\\
 & +\alpha_{1}\alpha_{2}(\alpha_{2}-\alpha_{1})^{2}\Biggl[\frac{\nabla_{i+1,j+1}^{2}\mathsf{n}_{p}^{*}}{\mathsf{n}_{p}^{*}}\frac{\nabla_{k+1,l+1}^{2}f_{t}^{*}}{f_{t}^{*}}+\frac{\nabla_{i+1,k+1}^{2}\mathsf{n}_{p}^{*}}{\mathsf{n}_{p}^{*}}\frac{\nabla_{j+1,l+1}^{2}f_{t}^{*}}{f_{t}^{*}}+\frac{\nabla_{i+1,l+1}^{2}\mathsf{n}_{p}^{*}}{\mathsf{n}_{p}^{*}}\frac{\nabla_{j+1,k+1}^{2}f_{t}^{*}}{f_{t}^{*}}\\
 & \qquad\qquad\qquad\qquad+\frac{\nabla_{j+1,k+1}^{2}\mathsf{n}_{p}^{*}}{\mathsf{n}_{p}^{*}}\frac{\nabla_{i+1,l+1}^{2}f_{t}^{*}}{f_{t}^{*}}+\frac{\nabla_{j+1,l+1}^{2}\mathsf{n}_{p}^{*}}{\mathsf{n}_{p}^{*}}\frac{\nabla_{i+1,k+1}^{2}f_{t}^{*}}{f_{t}^{*}}+\frac{\nabla_{k+1,l+1}^{2}\mathsf{n}_{p}^{*}}{\mathsf{n}_{p}^{*}}\frac{\nabla_{i+1,j+1}^{2}f_{t}^{*}}{f_{t}^{*}}\Biggr]
\end{align*}

\pagebreak{}

\subsection{Fourth-order expansion of the log-likelihood function}

\noindent \textbf{Justification of (\ref{QuarticExpansion}) and expression
of }$R_{T}^{(1)}(\alpha,\tilde{\pi},\varpi)$\textbf{.} Straightforward
calculation yields the fourth-order Taylor expansion (\ref{QuarticExpansion})
with the remainder term (for brevity, we again write $\tilde{\pi}=(\beta,\pi)$
and $\tilde{\pi}^{*}=(\beta^{*},\pi^{*})$) 
\begin{align}
R_{T}^{(1)}(\alpha,\tilde{\pi},\varpi) & =\varpi^{\prime}\nabla_{\varpi}L_{T}^{\pi}(\alpha,\tilde{\pi}^{\ast},0)+\frac{2}{2!}(\tilde{\pi}-\tilde{\pi}^{\ast})^{\prime}\nabla_{\tilde{\pi}\varpi^{\prime}}^{2}L_{T}^{\pi}(\alpha,\tilde{\pi}^{\ast},0)\varpi\nonumber \\
 & +\frac{1}{3!}\sum_{i=1}^{q_{1}+q_{2}}\sum_{j=1}^{q_{1}+q_{2}}\sum_{k=1}^{q_{1}+q_{2}}\nabla_{\tilde{\pi}_{i}\tilde{\pi}_{j}\tilde{\pi}_{k}}^{3}L_{T}^{\pi}(\alpha,\tilde{\pi}^{\ast},0)(\tilde{\pi}_{i}-\tilde{\pi}_{i}^{\ast})(\tilde{\pi}_{j}-\tilde{\pi}_{j}^{\ast})(\tilde{\pi}_{k}-\tilde{\pi}_{k}^{\ast})\nonumber \\
 & +\frac{3}{3!}\sum_{i=1}^{q_{1}+q_{2}}\sum_{j=1}^{q_{1}+q_{2}}\sum_{k=1}^{q_{2}}\nabla_{\tilde{\pi}_{i}\tilde{\pi}_{j}\varpi_{k}}^{3}L_{T}^{\pi}(\alpha,\tilde{\pi}^{\ast},0)(\tilde{\pi}_{i}-\tilde{\pi}_{i}^{\ast})(\tilde{\pi}_{j}-\tilde{\pi}_{j}^{\ast})\varpi_{k}\nonumber \\
 & +\frac{1}{3!}\sum_{i=1}^{q_{2}}\sum_{j=1}^{q_{2}}\sum_{k=1}^{q_{2}}\nabla_{\varpi_{i}\varpi_{j}\varpi_{k}}^{3}L_{T}^{\pi}(\alpha,\tilde{\pi}^{\ast},0)\varpi_{i}\varpi_{j}\varpi_{k}\nonumber \\
 & +\frac{1}{4!}\sum_{i=1}^{q_{1}+q_{2}}\sum_{j=1}^{q_{1}+q_{2}}\sum_{k=1}^{q_{1}+q_{2}}\sum_{l=1}^{q_{1}+q_{2}}\nabla_{\tilde{\pi}_{i}\tilde{\pi}_{j}\tilde{\pi}_{k}\tilde{\pi}_{l}}^{4}L_{T}^{\pi}(\alpha,\dot{\tilde{\pi}},\dot{\varpi})(\tilde{\pi}_{i}-\tilde{\pi}_{i}^{\ast})(\tilde{\pi}_{j}-\tilde{\pi}_{j}^{\ast})(\tilde{\pi}_{k}-\tilde{\pi}_{k}^{\ast})(\tilde{\pi}_{l}-\tilde{\pi}_{l}^{\ast})\nonumber \\
 & +\frac{4}{4!}\sum_{i=1}^{q_{1}+q_{2}}\sum_{j=1}^{q_{1}+q_{2}}\sum_{k=1}^{q_{1}+q_{2}}\sum_{l=1}^{q_{2}}\nabla_{\tilde{\pi}_{i}\tilde{\pi}_{j}\tilde{\pi}_{k}\varpi_{l}}^{4}L_{T}^{\pi}(\alpha,\dot{\tilde{\pi}},\dot{\varpi})(\tilde{\pi}_{i}-\tilde{\pi}_{i}^{\ast})(\tilde{\pi}_{j}-\tilde{\pi}_{j}^{\ast})(\tilde{\pi}_{k}-\tilde{\pi}_{k}^{\ast})\varpi_{l}\nonumber \\
 & +\frac{6}{4!}\sum_{i=1}^{q_{1}+q_{2}}\sum_{j=1}^{q_{1}+q_{2}}\sum_{k=1}^{q_{2}}\sum_{l=1}^{q_{2}}\nabla_{\tilde{\pi}_{i}\tilde{\pi}_{j}\varpi_{k}\varpi_{l}}^{4}L_{T}^{\pi}(\alpha,\dot{\tilde{\pi}},\dot{\varpi})(\tilde{\pi}_{i}-\tilde{\pi}_{i}^{\ast})(\tilde{\pi}_{j}-\tilde{\pi}_{j}^{\ast})\varpi_{k}\varpi_{l}\nonumber \\
 & +\frac{4}{4!}\sum_{i=1}^{q_{1}+q_{2}}\sum_{j=1}^{q_{2}}\sum_{k=1}^{q_{2}}\sum_{l=1}^{q_{2}}\nabla_{\tilde{\pi}_{i}\varpi_{j}\varpi_{k}\varpi_{l}}^{4}L_{T}^{\pi}(\alpha,\dot{\tilde{\pi}},\dot{\varpi})(\tilde{\pi}_{i}-\tilde{\pi}_{i}^{\ast})\varpi_{j}\varpi_{k}\varpi_{l}\nonumber \\
 & +\frac{1}{4!}\sum_{i=1}^{q_{2}}\sum_{j=1}^{q_{2}}\sum_{k=1}^{q_{2}}\sum_{l=1}^{q_{2}}\bigl(\nabla_{\varpi_{i}\varpi_{j}\varpi_{k}\varpi_{l}}^{4}L_{T}^{\pi}(\alpha,\dot{\tilde{\pi}},\dot{\varpi})-\nabla_{\varpi_{i}\varpi_{j}\varpi_{k}\varpi_{l}}L_{T}^{\pi}(\alpha,\tilde{\pi}^{\ast},0)\bigr)\varpi_{i}\varpi_{j}\varpi_{k}\varpi_{l},\label{RemainderTerm}
\end{align}
where $(\dot{\tilde{\pi}},\dot{\varpi})$ denotes a point between
$(\tilde{\pi},\varpi)$ and $(\tilde{\pi}^{\ast},0)$.

\bigskip{}

\noindent \textbf{Justification of (\ref{QuadraticExpansion}) and
expression of }$R_{T}(\alpha,\tilde{\pi},\varpi)$\textbf{.} We begin
with some useful notation. Let $\mathfrak{I}$ denote the index set
\[
\mathfrak{I}=((1,1),(2,2),\ldots,(q_{2},q_{2}),(1,2),(1,3),\ldots,(1,q_{2}),(2,3),\ldots,(q_{2}-1,q_{2})).
\]
For any scalars (or $d\times1$ column vectors) $A_{ij}$ indexed
by $i$ and $j$ (here and elsewhere it is tacitly assumed these indices
belong to $\{1,\ldots,q_{2}\}$), let $\left[A_{ij}\right]_{(i,j)\in\mathfrak{I}}$
denote the following $1\times q_{2}(q_{2}+1)/2$ row vector (or $d\times q_{2}(q_{2}+1)/2$
matrix): 
\[
\left[A_{ij}\right]_{(i,j)\in\mathfrak{I}}=\left[A_{11}:\cdots:A_{q_{2}q_{2}}:A_{12}:\cdots:A_{q_{2}-1,q_{2}}\right].
\]
For instance, $v(\varpi)=\left[\varpi_{i}\varpi_{j}\right]_{(i,j)\in\mathfrak{I}}^{\prime}$.
Similarly, for any scalars $A_{ijkl}$ indexed by $i,j,k,l$, let
$\left[A_{ijkl}\right]_{(i,j,k,l)\in\mathfrak{I\times I}}$\ denote
the following $q_{2}(q_{2}+1)/2\times q_{2}(q_{2}+1)/2$ matrix 
\[
\left[A_{ijkl}\right]_{(i,j,k,l)\in\mathfrak{I}\times\mathfrak{I}}=\begin{bmatrix}A_{1111} & \cdots & A_{11q_{2}q_{2}} & A_{1112} & \cdots & A_{1,1,q_{2}-1,q_{2}}\\
\vdots & \ddots & \vdots & \vdots & \ddots & \vdots\\
A_{q_{2}q_{2}11} & \cdots & A_{q_{2}q_{2}q_{2}q_{2}} & A_{q_{2}q_{2}12} & \cdots & A_{q_{2},q_{2},q_{2}-1,q_{2}}\\
A_{1211} & \cdots & A_{12q_{2}q_{2}} & A_{1212} & \cdots & A_{1,2,q_{2}-1,q_{2}}\\
\vdots & \ddots & \vdots & \vdots & \ddots & \vdots\\
A_{q_{2}-1,q_{2},1,1} & \cdots & A_{q_{2}-1,q_{2},q_{2},q_{2}} & A_{q_{2}-1,q_{2},1,2} & \cdots & A_{q_{2}-1,q_{2},q_{2}-1,q_{2}}
\end{bmatrix}.
\]
With this notation, and for any scalars $A_{ijkl}$ and $B_{ij}$
such that $A_{ijkl}=A_{jikl}$, $A_{ijkl}=A_{ijlk}$, and $B_{ij}=B_{ji}$
for all $i,j,k,l$, it holds that\footnote{\noindent To justify (\ref{PropertyIndexNotation}), partition the
index set $\mathfrak{I}$ as $\mathfrak{I=(I}_{1}\mathfrak{,I}_{2}\mathfrak{)}$
with $\mathfrak{I}_{1}=((1,1),(2,2),\ldots(q_{2},q_{2}))$ and $\mathfrak{I}_{2}=((1,2),(1,3),\ldots,(1,q_{2}),(2,3),\ldots,(q_{2}-1,q_{2}))$.
With straightforward algebra, 
\begin{align*}
 & \left[B_{ij}\right]_{(i,j)\in\mathfrak{I}}\left[c_{ij}c_{kl}A_{ijkl}\right]_{(i,j,k,l)\in\mathfrak{I}\times\mathfrak{I}}\left[B_{kl}\right]_{(k,l)\in\mathfrak{I}}^{\prime}\\
 & =\sum_{(i,j)\in\mathfrak{I}}\sum_{(k,l)\in\mathfrak{I}}c_{ij}c_{kl}A_{ijkl}B_{ij}B_{kl}\\
 & =\frac{1}{4}\sum_{(i,j)\in\mathfrak{I}_{1}}\sum_{(k,l)\in\mathfrak{I}_{1}}A_{ijkl}B_{ij}B_{kl}+\frac{1}{2}\sum_{(i,j)\in\mathfrak{I}_{1}}\sum_{(k,l)\in\mathfrak{I}_{2}}A_{ijkl}B_{ij}B_{kl}+\frac{1}{2}\sum_{(i,j)\in\mathfrak{I}_{2}}\sum_{(k,l)\in\mathfrak{I}_{1}}A_{ijkl}B_{ij}B_{kl}+\sum_{(i,j)\in\mathfrak{I}_{2}}\sum_{(k,l)\in\mathfrak{I}_{2}}A_{ijkl}B_{ij}B_{kl}\\
 & =\frac{1}{4}\left[\sum_{i=j}\sum_{k=l}A_{ijkl}B_{ij}B_{kl}+2\sum_{i=j}\sum_{k<l}A_{ijkl}B_{ij}B_{kl}+2\sum_{i<j}\sum_{k=l}A_{ijkl}B_{ij}B_{kl}+4\sum_{i<j}\sum_{k<l}A_{ijkl}B_{ij}B_{kl}\right]\\
 & =\frac{1}{4}\sum_{i=1}^{q_{2}}\sum_{j=1}^{q_{2}}\sum_{k=1}^{q_{2}}\sum_{l=1}^{q_{2}}A_{ijkl}B_{ij}B_{kl},
\end{align*}
where the properties $A_{ijkl}=A_{jikl}$, $A_{ijkl}=A_{ijlk}$, and
$B_{ij}=B_{ji}$ for all $i,j,k,l$, are used in the last equality.} 
\begin{equation}
\left[B_{ij}\right]_{(i,j)\in\mathfrak{I}}\left[c_{ij}c_{kl}A_{ijkl}\right]_{(i,j,k,l)\in\mathfrak{I}\times\mathfrak{I}}\left[B_{kl}\right]_{(k,l)\in\mathfrak{I}}^{\prime}=\frac{1}{4}\sum_{i=1}^{q_{2}}\sum_{j=1}^{q_{2}}\sum_{k=1}^{q_{2}}\sum_{l=1}^{q_{2}}A_{ijkl}B_{ij}B_{kl},\label{PropertyIndexNotation}
\end{equation}
where the $c_{ij}$'s are as in Section 3.3.1 ($c_{ij}=1/2$ if $i=j$
and $c_{ij}=1$ if $i\neq j$).

Now, to obtain (\ref{QuadraticExpansion}), introduce the matrix 
\[
\mathcal{J}_{T}=\begin{bmatrix}\mathcal{J}_{T,\tilde{\pi}\tilde{\pi}} & \mathcal{J}_{T,\tilde{\pi}\varpi\varpi}^{\prime}\\
\mathcal{J}_{T,\tilde{\pi}\varpi\varpi} & \mathcal{J}_{T,\varpi\varpi\varpi\varpi}
\end{bmatrix}
\]
where the matrices $\mathcal{J}_{T,\tilde{\pi}\tilde{\pi}}$ ($(q_{1}+q_{2})\times(q_{1}+q_{2})$),
$\mathcal{J}_{T,\tilde{\pi}\varpi\varpi}^{\prime}$ ($(q_{1}+q_{2})\times q_{\vartheta}$),
and $\mathcal{J}_{T,\varpi\varpi\varpi\varpi}$ ($q_{\vartheta}\times q_{\vartheta}$)
are defined as follows (here $\nabla_{\tilde{\pi}\tilde{\pi}^{\prime}}^{2}L_{T}^{\pi*}$
stands for $\nabla_{\tilde{\pi}\tilde{\pi}^{\prime}}^{2}L_{T}^{\pi}(\alpha,\tilde{\pi}^{\ast},0)$
etc.) 
\begin{align*}
\mathcal{J}_{T,\tilde{\pi}\tilde{\pi}} & =-T^{-1}\nabla_{\tilde{\pi}\tilde{\pi}^{\prime}}^{2}L_{T}^{\pi*},\\
\mathcal{J}_{T,\tilde{\pi}\varpi\varpi}^{\prime} & =-T^{-1}\frac{1}{\alpha_{1}\alpha_{2}}\left[c_{ij}\nabla_{\tilde{\pi}\varpi_{i}\varpi_{j}}^{3}L_{T}^{\pi*}\right]_{(i,j)\in\mathfrak{I}}\\
 & =-T^{-1}\frac{1}{\alpha_{1}\alpha_{2}}[c_{11}\nabla_{\tilde{\pi}\varpi_{1}\varpi_{1}}^{3}L_{T}^{\pi*}:\cdots:c_{q_{2}-1,q_{2}}\nabla_{\tilde{\pi}\varpi_{q_{2}-1}\varpi_{q_{2}}}^{3}L_{T}^{\pi*}],\\
\mathcal{J}_{T,\varpi\varpi\varpi\varpi} & =-T^{-1}\frac{8}{4!}\frac{1}{\alpha_{1}^{2}\alpha_{2}^{2}}\left[c_{ij}c_{kl}\nabla_{\varpi_{i}\varpi_{j}\varpi_{k}\varpi_{l}}^{4}L_{T}^{\pi*}\right]_{(i,j,k,l)\in\mathfrak{I}\times\mathfrak{I}}\\
 & =-T^{-1}\frac{8}{4!}\frac{1}{\alpha_{1}^{2}\alpha_{2}^{2}}\begin{bmatrix}c_{11}c_{11}\nabla_{\varpi_{1}\varpi_{1}\varpi_{1}\varpi_{1}}^{4}L_{T}^{\pi*} & \cdots & c_{q_{2}-1,q_{2}}c_{11}\nabla_{\varpi_{q_{2}-1}\varpi_{q_{2}}\varpi_{1}\varpi_{1}}^{4}L_{T}^{\pi*}\\
\vdots & \ddots & \vdots\\
c_{11}c_{q_{2}-1,q_{2}}\nabla_{\varpi_{1}\varpi_{1}\varpi_{q_{2}-1}\varpi_{q_{2}}}^{4}L_{T}^{\pi*} & \cdots & c_{q_{2}-1,q_{2}}c_{q_{2}-1,q_{2}}\nabla_{\varpi_{q_{2}-1}\varpi_{q_{2}}\varpi_{q_{2}-1}\varpi_{q_{2}}}^{4}L_{T}^{\pi*}
\end{bmatrix}.
\end{align*}
Straightforward computations (for the third one, use property (\ref{PropertyIndexNotation}))
now show that
\[
-\frac{1}{2}T^{1/2}(\tilde{\pi}-\tilde{\pi}^{\ast})^{\prime}\mathcal{J}_{T,\tilde{\pi}\tilde{\pi}}T^{1/2}(\tilde{\pi}-\tilde{\pi}^{\ast})=\frac{1}{2}(\tilde{\pi}-\tilde{\pi}^{\ast})^{\prime}\nabla_{\tilde{\pi}\tilde{\pi}^{\prime}}^{2}L_{T}^{\pi}(\alpha,\tilde{\pi}^{\ast},0)(\tilde{\pi}-\tilde{\pi}^{\ast}),
\]
\begin{align*}
 & -T^{1/2}(\tilde{\pi}-\tilde{\pi}^{\ast})^{\prime}\mathcal{J}_{T,\tilde{\pi}\varpi\varpi}^{\prime}T^{1/2}\alpha_{1}\alpha_{2}v(\varpi)\\
 & \quad\quad=(\tilde{\pi}-\tilde{\pi}^{\ast})^{\prime}\left[c_{11}\nabla_{\tilde{\pi}\varpi_{1}\varpi_{1}}^{3}L_{T}^{\pi}(\alpha,\tilde{\pi}^{\ast},0)\varpi_{1}^{2}+...+c_{q_{2}-1,q_{2}}\nabla_{\tilde{\pi}\varpi_{q_{2}-1}\varpi_{q_{2}}}^{3}L_{T}^{\pi}(\alpha,\tilde{\pi}^{\ast},0)\varpi_{q_{2}-1}\varpi_{q_{2}}\right]\\
 & \quad\quad=\frac{3}{3!}\sum_{i=1}^{q}\sum_{j=1}^{q_{2}}\sum_{k=1}^{q_{2}}\nabla_{\tilde{\pi}_{i}\varpi_{j}\varpi_{k}}^{3}L_{T}^{\pi}(\alpha,\tilde{\pi}^{\ast},0)(\tilde{\pi}_{i}-\tilde{\pi}_{i}^{\ast})\varpi_{j}\varpi_{k},
\end{align*}
\[
-\frac{1}{2}T\alpha_{1}^{2}\alpha_{2}^{2}v(\varpi)^{\prime}\mathcal{J}_{T,\varpi\varpi\varpi\varpi}v(\varpi)=\frac{1}{4!}\sum_{i=1}^{q_{2}}\sum_{j=1}^{q_{2}}\sum_{k=1}^{q_{2}}\sum_{l=1}^{q_{2}}\nabla_{\varpi_{i}\varpi_{j}\varpi_{k}\varpi_{l}}^{4}L_{T}^{\pi}(\alpha,\tilde{\pi}^{\ast},0)\varpi_{i}\varpi_{j}\varpi_{k}\varpi_{l}.
\]
 Therefore the fourth-order Taylor expansion of $L_{T}^{\pi}(\alpha,\tilde{\pi},\varpi)$
in (\ref{QuarticExpansion}) can be written as a quadratic expansion
given by 
\begin{equation}
L_{T}^{\pi}(\alpha,\tilde{\pi},\varpi)-L_{T}^{\pi}(\alpha,\tilde{\pi}^{\ast},0)=S_{T}^{\prime}\boldsymbol{\theta}(\alpha,\tilde{\pi},\varpi)-\frac{1}{2}[T^{1/2}\boldsymbol{\theta}(\alpha,\tilde{\pi},\varpi)]^{\prime}\mathcal{J}_{T}[T^{1/2}\boldsymbol{\theta}(\alpha,\tilde{\pi},\varpi)]+R_{T}^{(1)}(\alpha,\tilde{\pi},\varpi).\label{QuadraticExpansionIntermediateForm}
\end{equation}
Next, define 
\begin{equation}
\mathcal{I}=\begin{bmatrix}\mathcal{I}_{\tilde{\pi}\tilde{\pi}} & \mathcal{I}_{\tilde{\pi}\varpi\varpi}^{\prime}\\
\mathcal{I}_{\tilde{\pi}\varpi\varpi} & \mathcal{I}_{\varpi\varpi\varpi\varpi}
\end{bmatrix},\label{MatrixI}
\end{equation}
where the matrices $\mathcal{I}_{\tilde{\pi}\tilde{\pi}}$ ($(q_{1}+q_{2})\times(q_{1}+q_{2})$),
$\mathcal{I}_{\tilde{\pi}\varpi\varpi}^{\prime}$ ($(q_{1}+q_{2})\times q_{\vartheta}$),
and $\mathcal{I}_{\varpi\varpi\varpi\varpi}$ ($q_{\vartheta}\times q_{\vartheta}$)
are defined as follows 
\begin{align*}
\mathcal{I}_{\tilde{\pi}\tilde{\pi}} & =E\left[\frac{\nabla f_{t}(\tilde{\pi}^{*})}{f_{t}(\tilde{\pi}^{*})}\frac{\nabla'f_{t}(\tilde{\pi}^{*})}{f_{t}(\tilde{\pi}^{*})}\right],\\
\mathcal{I}_{\tilde{\pi}\varpi\varpi}^{\prime} & =\left[c_{ij}E\left[\frac{\nabla f_{t}(\tilde{\pi}^{*})}{f_{t}(\tilde{\pi}^{*})}X_{t,i,j}^{*}\right]\right]_{(i,j)\in\mathfrak{I}},\\
\mathcal{I}_{\varpi\varpi\varpi\varpi} & =\left[c_{ij}c_{kl}E\left[X_{t,i,j}^{*}X_{t,k,l}^{*}\right]\right]_{(i,j,k,l)\in\mathfrak{I}\times\mathfrak{I}},
\end{align*}
and where we have used the short-hand notation $X_{t,i,j}^{*}$, $i,j\in\{1,\ldots,q_{2}\}$
(see (\ref{eq:Notation_X*})). Finiteness of $\mathcal{I}$ follows
from Lemma \ref{lem:XXX1}. Now, defining 
\[
R_{T}^{(2)}(\alpha,\tilde{\pi},\varpi)=-\frac{1}{2}[T^{1/2}\boldsymbol{\theta}(\alpha,\tilde{\pi},\varpi)]^{\prime}(\mathcal{J}_{T}-\mathcal{I})[T^{1/2}\boldsymbol{\theta}(\alpha,\tilde{\pi},\varpi)]
\]
and adding and subtracting terms, expansion (\ref{QuadraticExpansionIntermediateForm})
can be written as (\ref{QuadraticExpansion}) with 
\[
R_{T}(\alpha,\tilde{\pi},\varpi)=R_{T}^{(1)}(\alpha,\tilde{\pi},\varpi)+R_{T}^{(2)}(\alpha,\tilde{\pi},\varpi).
\]

\subsection{Some more explicit derivatives and the verification of Assumption
5(iii)}

\paragraph{Some more explicit derivative expressions. }

We will require more explicit expressions for the components of $s_{t}$
in (\ref{eq:AppBAss5_s_t}) (see also (\ref{eq:AppBAss5Derivs1})\textendash (\ref{eq:AppBAss5Derivs3})
and (\ref{eq:Notation_X*})). Straightforward computation shows that
(as before, $\nabla f_{t}(\cdot)$ denotes differentiation of $f_{t}(\cdot)$
in (\ref{eq:f_t}) with respect to $\tilde{\phi}=(\tilde{\phi}_{0},\tilde{\phi}_{1},\ldots,\tilde{\phi}_{p},\tilde{\sigma}_{1}^{2})$)
\[
\frac{\nabla f_{t}(\tilde{\phi})}{f_{t}(\tilde{\phi})}=\nabla\log(f_{t}(\tilde{\phi})),\qquad\frac{\nabla^{2}f_{t}(\tilde{\phi})}{f_{t}(\tilde{\phi})}=\nabla^{2}\log(f_{t}(\tilde{\phi}))+\nabla\log(f_{t}(\tilde{\phi}))\nabla'\log(f_{t}(\tilde{\phi})),
\]
where (as $\log(f_{t}(\tilde{\phi}))=-\frac{1}{2}\log(2\bar{\pi})-\frac{1}{2}\log(\tilde{\sigma}_{1}^{2})-\frac{1}{2}g_{t}^{2}(\tilde{\phi})$
with $g_{t}(\tilde{\phi})=[y_{t}-(\tilde{\phi}_{0}+\tilde{\phi}_{1}y_{t-1}+\cdots+\tilde{\phi}_{p}y_{t-p})]/\tilde{\sigma}_{1}$)
\[
\nabla\log(f_{t}(\tilde{\phi}))=\begin{bmatrix}\frac{1}{\tilde{\sigma}_{1}}g_{t}(\tilde{\phi})\\
\frac{1}{\tilde{\sigma}_{1}}\boldsymbol{y}_{t-1}g_{t}(\tilde{\phi})\\
\frac{1}{2\tilde{\sigma}_{1}^{2}}(g_{t}^{2}(\tilde{\phi})-1)
\end{bmatrix},\qquad\nabla^{2}\log(f_{t}(\tilde{\phi}))=\begin{bmatrix}-\frac{1}{\tilde{\sigma}_{1}^{2}} & -\frac{1}{\tilde{\sigma}_{1}^{2}}\boldsymbol{y}_{t-1}' & -\frac{1}{\tilde{\sigma}_{1}^{3}}g_{t}(\tilde{\phi})\\
-\frac{1}{\tilde{\sigma}_{1}^{2}}\boldsymbol{y}_{t-1} & -\frac{1}{\tilde{\sigma}_{1}^{2}}\boldsymbol{y}_{t-1}\boldsymbol{y}_{t-1}' & -\frac{1}{\tilde{\sigma}_{1}^{3}}g_{t}(\tilde{\phi})\boldsymbol{y}_{t-1}\\
-\frac{1}{\tilde{\sigma}_{1}^{3}}g_{t}(\tilde{\phi}) & -\frac{1}{\tilde{\sigma}_{1}^{3}}g_{t}(\tilde{\phi})\boldsymbol{y}_{t-1}' & -\frac{1}{2\tilde{\sigma}_{1}^{4}}(2g_{t}^{2}(\tilde{\phi})-1)
\end{bmatrix},
\]
so that 
\begin{align}
\frac{\nabla f_{t}(\pi^{*})}{f_{t}(\pi^{*})} & =\begin{bmatrix}\frac{1}{\tilde{\sigma}_{1}^{*}}\varepsilon_{t}\\
\frac{1}{\tilde{\sigma}_{1}^{*}}\boldsymbol{y}_{t-1}\varepsilon_{t}\\
\frac{1}{2\tilde{\sigma}_{1}^{*2}}(\varepsilon_{t}^{2}-1)
\end{bmatrix},\nonumber \\
\frac{\nabla^{2}f_{t}(\pi^{*})}{f_{t}(\pi^{*})} & =\begin{bmatrix}-\frac{1}{\tilde{\sigma}_{1}^{*2}} & -\frac{1}{\tilde{\sigma}_{1}^{*2}}\boldsymbol{y}_{t-1}' & -\frac{1}{\tilde{\sigma}_{1}^{*3}}\varepsilon_{t}\\
-\frac{1}{\tilde{\sigma}_{1}^{*2}}\boldsymbol{y}_{t-1} & -\frac{1}{\tilde{\sigma}_{1}^{*2}}\boldsymbol{y}_{t-1}\boldsymbol{y}_{t-1}' & -\frac{1}{\tilde{\sigma}_{1}^{*3}}\boldsymbol{y}_{t-1}\varepsilon_{t}\\
-\frac{1}{\tilde{\sigma}_{1}^{*3}}\varepsilon_{t} & -\frac{1}{\tilde{\sigma}_{1}^{*3}}\boldsymbol{y}_{t-1}'\varepsilon_{t} & -\frac{1}{2\tilde{\sigma}_{1}^{*4}}(2\varepsilon_{t}^{2}-1)
\end{bmatrix}+\begin{bmatrix}\frac{1}{\tilde{\sigma}_{1}^{*}}\varepsilon_{t}\\
\frac{1}{\tilde{\sigma}_{1}^{*}}\boldsymbol{y}_{t-1}\varepsilon_{t}\\
\frac{1}{2\tilde{\sigma}_{1}^{*2}}(\varepsilon_{t}^{2}-1)
\end{bmatrix}\begin{bmatrix}\frac{1}{\tilde{\sigma}_{1}^{*}}\varepsilon_{t}\\
\frac{1}{\tilde{\sigma}_{1}^{*}}\boldsymbol{y}_{t-1}\varepsilon_{t}\\
\frac{1}{2\tilde{\sigma}_{1}^{*2}}(\varepsilon_{t}^{2}-1)
\end{bmatrix}'\nonumber \\
 & =\begin{bmatrix}\frac{1}{\tilde{\sigma}_{1}^{*2}}(\varepsilon_{t}^{2}-1) & \frac{1}{\tilde{\sigma}_{1}^{*2}}\boldsymbol{y}_{t-1}'(\varepsilon_{t}^{2}-1) & \frac{1}{2\tilde{\sigma}_{1}^{*3}}(\varepsilon_{t}^{3}-3\varepsilon_{t})\\
\frac{1}{\tilde{\sigma}_{1}^{*2}}\boldsymbol{y}_{t-1}(\varepsilon_{t}^{2}-1) & \frac{1}{\tilde{\sigma}_{1}^{*2}}\boldsymbol{y}_{t-1}\boldsymbol{y}_{t-1}'(\varepsilon_{t}^{2}-1) & \frac{1}{2\tilde{\sigma}_{1}^{*3}}\boldsymbol{y}_{t-1}(\varepsilon_{t}^{3}-3\varepsilon_{t})\\
\frac{1}{2\tilde{\sigma}_{1}^{*3}}(\varepsilon_{t}^{3}-3\varepsilon_{t}) & \frac{1}{2\tilde{\sigma}_{1}^{*3}}\boldsymbol{y}_{t-1}'(\varepsilon_{t}^{3}-3\varepsilon_{t}) & \frac{1}{4\tilde{\sigma}_{1}^{*4}}(\varepsilon_{t}^{4}-6\varepsilon_{t}^{2}+3)
\end{bmatrix}.\label{eq:AppF3_nabla2f_f}
\end{align}

Similar formulas hold for $\mathsf{n}_{p}(\boldsymbol{y}_{t-1};\tilde{\phi})$.
As $\log(\mathsf{n}_{p}(\boldsymbol{y}_{t-1};\tilde{\phi}))=-\frac{p}{2}\log(2\bar{\pi})-\frac{1}{2}\log(\det(\mathbf{\Gamma}_{1,p}))-\frac{1}{2}(\boldsymbol{y}_{t-1}-\mu_{1}\mathbf{1}_{p})'\mathbf{\Gamma}_{1,p}^{-1}(\boldsymbol{y}_{t-1}-\mu_{1}\mathbf{1}_{p})$
(see Section 2.2), we obtain, for each $i=1,\ldots,p+2$, (cf. Magnus
and Neudecker (1999, p. 325))\footnote{As before, $\nabla$ denotes differentiation with respect to $\tilde{\phi}=(\tilde{\phi}_{0},\tilde{\phi}_{1},\ldots,\tilde{\phi}_{p},\tilde{\sigma}_{1}^{2})$
and $\nabla_{i}$, $i=1,\ldots,p+2$, with respect to the $i$th component
of $\tilde{\phi}$.}
\[
\nabla_{i}\log(\mathsf{n}_{p}(\boldsymbol{y}_{t-1};\tilde{\phi}))=\frac{1}{2}tr\left(\frac{\partial\mathbf{\Gamma}_{1,p}^{-1}}{\partial\tilde{\phi}_{i}}\mathbf{\Gamma}_{1,p}\right)+(\boldsymbol{y}_{t-1}-\mu_{1}\mathbf{1}_{p})'\mathbf{\Gamma}_{1,p}^{-1}(\frac{\partial\mu_{1}}{\partial\tilde{\phi}_{i}}\mathbf{1}_{p})-\frac{1}{2}(\boldsymbol{y}_{t-1}-\mu_{1}\mathbf{1}_{p})'\frac{\partial\mathbf{\Gamma}_{1,p}^{-1}}{\partial\tilde{\phi}_{i}}(\boldsymbol{y}_{t-1}-\mu_{1}\mathbf{1}_{p}),
\]
where (see Section 2.2 for the notation) 
\[
\frac{\partial\mu_{1}}{\partial\tilde{\phi}}=\begin{bmatrix}(\tilde{\phi}(1))^{-1}\\
\tilde{\phi}_{0}(\tilde{\phi}(1))^{-2}\mathbf{1}_{p}\\
0
\end{bmatrix}.
\]
For the expression of $\frac{\partial\mathbf{\Gamma}_{1,p}^{-1}}{\partial\tilde{\phi}_{i}}$,
first note that $\mathbf{\Gamma}_{1,p}^{-1}$ can be expressed as
(see, e.g., Galbraith and Galbraith (1974)) $\mathbf{\Gamma}_{1,p}^{-1}=\frac{1}{\tilde{\sigma}_{1}^{2}}(U'U-V'V)$
with $U$ and $V$ being $p\times p$ Toeplitz matrices given by
\[
U=\begin{bmatrix}1\\
-\tilde{\phi}_{1} & \ddots\\
\vdots & \ddots & \ddots\\
-\tilde{\phi}_{p-1} & \cdots & -\tilde{\phi}_{1} & 1
\end{bmatrix},\,\,\,\,\,V=\begin{bmatrix}\tilde{\phi}_{p}\\
\tilde{\phi}_{p-1} & \ddots\\
\vdots & \ddots & \ddots\\
\tilde{\phi}_{1} & \cdots & \tilde{\phi}_{p-1} & \tilde{\phi}_{p}
\end{bmatrix}.
\]
Thus $\frac{\partial\mathbf{\Gamma}_{1,p}^{-1}}{\partial\tilde{\phi}_{i}}$
equals a zero matrix when differentiating with respect to $\tilde{\phi}_{0}$,
$-\frac{1}{\tilde{\sigma}_{1}^{2}}\mathbf{\Gamma}_{1,p}^{-1}$ when
differentiating with respect to $\tilde{\sigma}_{1}^{2}$, and 
\[
\frac{\partial\mathbf{\Gamma}_{1,p}^{-1}}{\partial\tilde{\phi}_{i}}=\frac{1}{\tilde{\sigma}_{1}^{2}}(\frac{\partial U'}{\partial\tilde{\phi}_{i}}U+U'\frac{\partial U}{\partial\tilde{\phi}_{i}}-\frac{\partial V'}{\partial\tilde{\phi}_{i}}V-V'\frac{\partial V}{\partial\tilde{\phi}_{i}})
\]
when differentiating with respect to to the autoregressive parameters.
To summarize, 
\begin{equation}
\frac{\nabla\mathsf{n}_{p}(\tilde{\phi})}{\mathsf{n}_{p}(\tilde{\phi})}=\nabla\log(\mathsf{n}_{p}(\boldsymbol{y}_{t-1};\tilde{\phi}))=\begin{bmatrix}d_{1}(\boldsymbol{y}_{t-1};\tilde{\phi})\\
d_{2}(\boldsymbol{y}_{t-1};\tilde{\phi})\\
d_{3}(\boldsymbol{y}_{t-1};\tilde{\phi})
\end{bmatrix},\label{eq:AppF3_nabla_np_np}
\end{equation}
where (note that $tr\left(\frac{\partial\mathbf{\Gamma}_{1,p}^{-1}}{\partial\tilde{\phi}_{i}}\mathbf{\Gamma}_{1,p}\right)=\frac{\partial vec(\mathbf{\Gamma}_{1,p}^{-1})'}{\partial\tilde{\phi}_{i}}vec(\mathbf{\Gamma}_{1,p})$)
\begin{align*}
d_{1}(\boldsymbol{y}_{t-1};\tilde{\phi}) & =(\tilde{\phi}(1))^{-1}(\boldsymbol{y}_{t-1}-\mu_{1}\mathbf{1}_{p})'\mathbf{\Gamma}_{1,p}^{-1}\mathbf{1}_{p},\\
d_{2}(\boldsymbol{y}_{t-1};\tilde{\phi}) & =\frac{1}{2}\frac{\partial vec(\mathbf{\Gamma}_{1,p}^{-1})'}{\partial(\tilde{\phi}_{1},\ldots,\tilde{\phi}_{p})}vec(\mathbf{\Gamma}_{1,p})+\tilde{\phi}_{0}(\tilde{\phi}(1))^{-2}\mathbf{1}_{p}(\mathbf{1}_{p}'\mathbf{\Gamma}_{1,p}^{-1}(\boldsymbol{y}_{t-1}-\mu_{1}\mathbf{1}_{p}))\\
 & -\frac{1}{2}\frac{\partial vec(\mathbf{\Gamma}_{1,p}^{-1})'}{\partial(\tilde{\phi}_{1},\ldots,\tilde{\phi}_{p})}((\boldsymbol{y}_{t-1}-\mu_{1}\mathbf{1}_{p})\otimes(\boldsymbol{y}_{t-1}-\mu_{1}\mathbf{1}_{p})),\\
d_{3}(\boldsymbol{y}_{t-1};\tilde{\phi}) & =-\frac{p}{2\tilde{\sigma}_{1}^{2}}+\frac{1}{2\tilde{\sigma}_{1}^{2}}(\boldsymbol{y}_{t-1}-\mu_{1}\mathbf{1}_{p})'\mathbf{\Gamma}_{1,p}^{-1}(\boldsymbol{y}_{t-1}-\mu_{1}\mathbf{1}_{p})
\end{align*}
(first and last scalars, middle one $p\times1$). Therefore 
\[
\frac{\nabla\mathsf{n}_{p}(\pi^{*})}{\mathsf{n}_{p}(\pi^{*})}=\nabla\log(\mathsf{n}_{p}(\boldsymbol{y}_{t-1};\pi^{*}))=\begin{bmatrix}d_{1}(\boldsymbol{y}_{t-1};\pi^{*})\\
d_{2}(\boldsymbol{y}_{t-1};\pi^{*})\\
d_{3}(\boldsymbol{y}_{t-1};\pi^{*})
\end{bmatrix}.
\]

Based on the preceding derivations, the derivatives appearing in (\ref{eq:AppBAss5Derivs1})\textendash (\ref{eq:AppBAss5Derivs3})
can now be expressed as

\begin{align*}
\nabla_{\beta}l_{t}^{\pi}(\alpha,\beta^{*},\pi^{*},0) & =\frac{1}{\tilde{\sigma}_{1}^{*}}\varepsilon_{t}\\
\nabla_{\pi}l_{t}^{\pi}(\alpha,\beta^{*},\pi^{*},0) & =\begin{bmatrix}\frac{1}{\tilde{\sigma}_{1}^{*}}\boldsymbol{y}_{t-1}\varepsilon_{t}\\
\frac{1}{2\tilde{\sigma}_{1}^{*2}}(\varepsilon_{t}^{2}-1)
\end{bmatrix}\,\,\,\,\,\,\,\,\,\,((p+1)\times1)\\
\nabla_{\varpi\varpi'}^{2}l_{t}^{\pi}(\alpha,\beta^{*},\pi^{*},0) & =\alpha(1-\alpha)\Biggl\{\begin{bmatrix}d_{2}(\boldsymbol{y}_{t-1};\pi^{*})\frac{1}{\tilde{\sigma}_{1}^{*}}\boldsymbol{y}_{t-1}^{\prime}\varepsilon_{t} & d_{2}(\boldsymbol{y}_{t-1};\pi^{*})\frac{1}{2\tilde{\sigma}_{1}^{*2}}(\varepsilon_{t}^{2}-1)\\
d_{3}(\boldsymbol{y}_{t-1};\pi^{*})\frac{1}{\tilde{\sigma}_{1}^{*}}\boldsymbol{y}_{t-1}^{\prime}\varepsilon_{t} & d_{3}(\boldsymbol{y}_{t-1};\pi^{*})\frac{1}{2\tilde{\sigma}_{1}^{*2}}(\varepsilon_{t}^{2}-1)
\end{bmatrix}\\
 & \qquad\qquad\qquad+\begin{bmatrix}\frac{1}{\tilde{\sigma}_{1}^{*}}\boldsymbol{y}_{t-1}\varepsilon_{t}d_{2}^{\prime}(\boldsymbol{y}_{t-1};\pi^{*}) & \frac{1}{\tilde{\sigma}_{1}^{*}}\boldsymbol{y}_{t-1}\varepsilon_{t}d_{3}(\boldsymbol{y}_{t-1};\pi^{*})\\
\frac{1}{2\tilde{\sigma}_{1}^{*2}}(\varepsilon_{t}^{2}-1)d_{2}^{\prime}(\boldsymbol{y}_{t-1};\pi^{*}) & \frac{1}{2\tilde{\sigma}_{1}^{*2}}(\varepsilon_{t}^{2}-1)d_{3}(\boldsymbol{y}_{t-1};\pi^{*})
\end{bmatrix}\\
 & \qquad\qquad\qquad+\begin{bmatrix}\frac{1}{\tilde{\sigma}_{1}^{*2}}\boldsymbol{y}_{t-1}\boldsymbol{y}_{t-1}'(\varepsilon_{t}^{2}-1) & \frac{1}{2\tilde{\sigma}_{1}^{*3}}\boldsymbol{y}_{t-1}(\varepsilon_{t}^{3}-3\varepsilon_{t})\\
\frac{1}{2\tilde{\sigma}_{1}^{*3}}\boldsymbol{y}_{t-1}'(\varepsilon_{t}^{3}-3\varepsilon_{t}) & \frac{1}{4\tilde{\sigma}_{1}^{*4}}(\varepsilon_{t}^{4}-6\varepsilon_{t}^{2}+3)
\end{bmatrix}\Biggr\}.
\end{align*}

\paragraph{Verification of Assumption \ref{assu:quadexp}(iii).}

Finiteness of $\mathcal{I}$ was already established in Supplementary
Appendix F.2. For positive definiteness, it suffices to show that
the components of the vector $s_{t}$ are linearly independent. Note
that for linear independence, it does not matter if the order of the
elements is changed or if some of the elements are multiplied by nonzero
constants. Therefore, making use of the explicit expressions given
above, it suffices to show that the components of the vector $\tilde{s}_{t}=(\tilde{s}_{t,1},\tilde{s}_{t,2},\tilde{s}_{t,3},\tilde{s}_{t,4},\tilde{s}_{t,5},\tilde{s}_{t,6})$
(where the dimensions of the six components are $1,p,1,p(p+1)/2,p,1$,
respectively) are linearly independent, where
\[
\begin{bmatrix}\tilde{s}_{t,1}\\
\tilde{s}_{t,2}\\
\tilde{s}_{t,3}\\
\tilde{s}_{t,4}\\
\tilde{s}_{t,5}\\
\tilde{s}_{t,6}
\end{bmatrix}=\begin{bmatrix}\frac{1}{\tilde{\sigma}_{1}^{*}}\varepsilon_{t}\\
\frac{1}{\tilde{\sigma}_{1}^{*}}\boldsymbol{y}_{t-1}\varepsilon_{t}\\
\frac{1}{2\tilde{\sigma}_{1}^{*2}}(\varepsilon_{t}^{2}-1)\\
vech[d_{2}(\boldsymbol{y}_{t-1};\pi^{*})\boldsymbol{y}_{t-1}^{\prime}+\boldsymbol{y}_{t-1}d_{2}^{\prime}(\boldsymbol{y}_{t-1};\pi^{*})]\frac{1}{\tilde{\sigma}_{1}^{*}}\varepsilon_{t}+vech[\boldsymbol{y}_{t-1}\boldsymbol{y}_{t-1}']\frac{1}{\tilde{\sigma}_{1}^{*2}}(\varepsilon_{t}^{2}-1)\\
d_{2}(\boldsymbol{y}_{t-1};\pi^{*})\frac{1}{2\tilde{\sigma}_{1}^{*2}}(\varepsilon_{t}^{2}-1)+\boldsymbol{y}_{t-1}d_{3}(\boldsymbol{y}_{t-1};\pi^{*})\frac{1}{\tilde{\sigma}_{1}^{*}}\varepsilon_{t}+\frac{1}{2\tilde{\sigma}_{1}^{*3}}\boldsymbol{y}_{t-1}(\varepsilon_{t}^{3}-3\varepsilon_{t})\\
d_{3}(\boldsymbol{y}_{t-1};\pi^{*})\frac{1}{\tilde{\sigma}_{1}^{*2}}(\varepsilon_{t}^{2}-1)+\frac{1}{4\tilde{\sigma}_{1}^{*4}}(\varepsilon_{t}^{4}-6\varepsilon_{t}^{2}+3)
\end{bmatrix}.
\]
To this end, suppose that $c'\tilde{s}_{t}=(c_{1},c_{2},c_{3},c_{4},c_{5},c_{6})'(\tilde{s}_{t,1},\tilde{s}_{t,2},\tilde{s}_{t,3},\tilde{s}_{t,4},\tilde{s}_{t,5},\tilde{s}_{t,6})=0$
(with the dimension of $c$ and its subvectors chosen conformably).
Note that the only random quantities $\tilde{s}_{t}$ depends on are
$\boldsymbol{y}_{t-1}$ and $\varepsilon_{t}$ which are independent.
First, as the term $\varepsilon_{t}^{4}$ only appears in $\tilde{s}_{t,6}$,
the equality $E[c'\tilde{s}_{t}\mid\varepsilon_{t}]=0$ can be expressed
as $c_{6}\varepsilon_{t}^{4}/(4\tilde{\sigma}_{1}^{*4})+P_{3}(\varepsilon_{t})=0$,
where $P_{3}(\varepsilon_{t})$ is a third-order polynomial in $\varepsilon_{t}$.
As the components of the vector $(\varepsilon_{t},\varepsilon_{t}^{2},\varepsilon_{t}^{3},\varepsilon_{t}^{4})$
are linearly independent (this clearly follows from normality), it
follows that $c_{6}=0$. Next, basic properties of the standard normal
distribution imply that $E[c'\tilde{s}_{t}(\varepsilon_{t}^{3}-3\varepsilon_{t})\mid\boldsymbol{y}_{t-1}]=c_{5}'\frac{1}{2\tilde{\sigma}_{1}^{*3}}\boldsymbol{y}_{t-1}E[(\varepsilon_{t}^{3}-3\varepsilon_{t})^{2}]=0$,
so that necessarily $c_{5}=0$ (as the components of $\boldsymbol{y}_{t-1}$
are linearly independent and $E[(\varepsilon_{t}^{3}-3\varepsilon_{t})^{2}]>0$).
Next note that (as $c_{5}=0$, $c_{6}=0$)
\[
0=E[c'\tilde{s}_{t}(\varepsilon_{t}^{2}-1)\mid\boldsymbol{y}_{t-1}]=c_{3}E[(\varepsilon_{t}^{2}-1)^{2}]/(2\tilde{\sigma}_{1}^{*2})+c_{4}^{\prime}vech[\boldsymbol{y}_{t-1}\boldsymbol{y}_{t-1}']E[(\varepsilon_{t}^{2}-1)^{2}]/\tilde{\sigma}_{1}^{*2}
\]
so that $c_{4}^{\prime}vech[\boldsymbol{y}_{t-1}\boldsymbol{y}_{t-1}']=-c_{3}/2$.
As the components of $vech[\boldsymbol{y}_{t-1}\boldsymbol{y}_{t-1}']$
are linearly independent (as $vech(\boldsymbol{y}_{t-1}\boldsymbol{y}_{t-1}')=D_{p}^{+}vec(\boldsymbol{y}_{t-1}\boldsymbol{y}_{t-1}')=D_{p}^{+}(\boldsymbol{y}_{t-1}\otimes\boldsymbol{y}_{t-1})$,
with $D_{p}^{+}$ denoting the Moore-Penrose inverse of the duplication
matrix $D_{p}$, $Cov[vech(\boldsymbol{y}_{t-1}\boldsymbol{y}_{t-1}')]=D_{p}^{+}Cov(\boldsymbol{y}_{t-1}\otimes\boldsymbol{y}_{t-1})D_{p}^{+\prime}$;
because $D_{p}^{+}$ is of full row rank and $Cov(\boldsymbol{y}_{t-1}\otimes\boldsymbol{y}_{t-1})$
has rank $p(p+1)/2$ (Magnus and Neudecker (1979, Thm 4.3(v)), $Cov[vech(\boldsymbol{y}_{t-1}\boldsymbol{y}_{t-1}')]$
is positive definite), it necessarily follows that $c_{4}=0$ and
$c_{3}=0$. Finally, as only $c_{1}$ and $c_{2}$ may be nonzero,
$E[c'\tilde{s}_{t}\varepsilon_{t}\mid\boldsymbol{y}_{t-1}]=c_{1}\frac{1}{\tilde{\sigma}_{1}^{*}}+c_{2}^{\prime}\frac{1}{\tilde{\sigma}_{1}^{*}}\boldsymbol{y}_{t-1}=0$,
from which $c_{2}=0$ and $c_{1}=0$ follow (as the components of
$\boldsymbol{y}_{t-1}$ are linearly independent).

\subsection{Verification of Assumption 5(iv)}

\noindent First consider $R_{T}^{(1)}(\alpha,\tilde{\pi},\varpi)$.
Of the quantities on the right hand side of (\ref{RemainderTerm}),
the first two are equal to zero because $\nabla_{\varpi}l_{t}^{\pi}(\alpha,\tilde{\pi}^{\ast},0)=0$
and $\nabla_{\tilde{\pi}\varpi^{\prime}}l_{t}^{\pi}(\alpha,\tilde{\pi}^{\ast},0)=0$;
for the other eight quantities, Lemma \ref{lem:XXX4} provides upper
bounds that aid in bounding them. Now, to verify Assumption \ref{assu:quadexp}(iv)
(for $R_{T}^{(1)}(\alpha,\tilde{\pi},\varpi)$), let $\{\gamma_{T},\ T\geq1\}$
be an arbitrary sequence of (non-random) positive scalars such that
$\gamma_{T}\rightarrow0$ as $T\rightarrow\infty$. Condition $\left\Vert (\beta,\pi,\varpi)-(\beta^{*},\pi^{\ast},0)\right\Vert \leq\gamma_{T}$
(appearing in Assumption \ref{assu:quadexp}(iv)), together with the
properties $\left\Vert \boldsymbol{\theta}(\alpha,\beta,\pi,\varpi)\right\Vert ^{2}=\left\Vert \tilde{\pi}-\tilde{\pi}^{\ast}\right\Vert ^{2}+\alpha_{1}^{2}\alpha_{2}^{2}\left\Vert v(\varpi)\right\Vert ^{2}$
and $\left\Vert v(\varpi)\right\Vert \leq\left\Vert \varpi\right\Vert ^{2}$,
implies that 
\[
\left\Vert \boldsymbol{\theta}(\alpha,\beta,\pi,\varpi)\right\Vert ^{1/2}\leq\left\Vert \tilde{\pi}-\tilde{\pi}^{\ast}\right\Vert ^{1/2}+\alpha_{1}^{1/2}\alpha_{2}^{1/2}\left\Vert v(\varpi)\right\Vert ^{1/2}\leq\gamma_{T}^{1/2}+\alpha_{1}^{1/2}\alpha_{2}^{1/2}\gamma_{T}.
\]
This, Lemma \ref{lem:XXX4}, and the fact that $\alpha$ is bounded
away from zero and one on $A$, imply that for some sequence $\{\tilde{\gamma}_{T},\ T\geq1\}$
of (non-random) positive scalars such that $\tilde{\gamma}_{T}\rightarrow0$
as $T\rightarrow\infty$ and for some finite $C$, 
\begin{align}
 & \sup_{\alpha\in A}\sup_{\substack{(\tilde{\pi},\varpi)\in B\times\Pi_{\alpha},\\
\left\Vert (\tilde{\pi},\varpi)-(\tilde{\pi}^{\ast},0)\right\Vert \leq\gamma_{T}
}
}\frac{\lvert R_{T}^{(1)}(\alpha,\tilde{\pi},\varpi)\rvert}{\left(1+\left\Vert T^{1/2}\boldsymbol{\theta}(\alpha,\tilde{\pi},\varpi)\right\Vert \right)^{2}}\nonumber \\
 & \leq\tilde{\gamma}_{T}\sum_{i,j,k}\sup_{\alpha\in A}\bigl[\lvert T^{-1}\nabla_{\tilde{\pi}_{i}\tilde{\pi}_{j}\tilde{\pi}_{k}}^{3}L_{T}^{\pi}(\alpha,\tilde{\pi}^{\ast},0)\rvert+\lvert T^{-1}\nabla_{\tilde{\pi}_{i}\tilde{\pi}_{j}\varpi_{k}}^{3}L_{T}^{\pi}(\alpha,\tilde{\pi}^{\ast},0)\rvert+\lvert T^{-1/2}\nabla_{\varpi_{i}\varpi_{j}\varpi_{k}}^{3}L_{T}^{\pi}(\alpha,\tilde{\pi}^{\ast},0)\rvert\bigr]\nonumber \\
 & +\tilde{\gamma}_{T}\sum_{i,j,k,l}\sup_{\alpha\in A}\sup_{\substack{(\tilde{\pi},\varpi)\in B\times\Pi_{\alpha},\\
\left\Vert (\tilde{\pi},\varpi)-(\tilde{\pi}^{\ast},0)\right\Vert \leq\gamma_{T}
}
}\bigl[\lvert T^{-1}\nabla_{\tilde{\pi}_{i}\tilde{\pi}_{j}\tilde{\pi}_{k}\tilde{\pi}_{l}}^{4}L_{T}^{\pi}(\alpha,\tilde{\pi},\varpi)\rvert+\lvert T^{-1}\nabla_{\tilde{\pi}_{i}\tilde{\pi}_{j}\tilde{\pi}_{k}\varpi_{l}}^{4}L_{T}^{\pi}(\alpha,\tilde{\pi},\varpi)\rvert\nonumber \\
 & \qquad\qquad\qquad\qquad\qquad\qquad\qquad+\lvert T^{-1}\nabla_{\tilde{\pi}_{i}\tilde{\pi}_{j}\varpi_{k}\varpi_{l}}^{4}L_{T}^{\pi}(\alpha,\tilde{\pi},\varpi)\rvert+\lvert T^{-1}\nabla_{\tilde{\pi}_{i}\varpi_{j}\varpi_{k}\varpi_{l}}^{4}L_{T}^{\pi}(\alpha,\tilde{\pi},\varpi)\rvert\bigr]\nonumber \\
 & +C\sum_{i,j,k,l}\sup_{\alpha\in A}\sup_{\substack{(\tilde{\pi},\varpi)\in B\times\Pi_{\alpha},\\
\left\Vert (\tilde{\pi},\varpi)-(\tilde{\pi}^{\ast},0)\right\Vert \leq\gamma_{T}
}
}\lvert T^{-1}(\nabla_{\varpi_{i}\varpi_{j}\varpi_{k}\varpi_{l}}^{4}L_{T}^{\pi}(\alpha,\tilde{\pi},\varpi)-\nabla_{\varpi_{i}\varpi_{j}\varpi_{k}\varpi_{l}}^{4}L_{T}^{\pi}(\alpha,\tilde{\pi}^{*},0))\rvert,\label{eq:R1T_Bound}
\end{align}
where the summations above are understood to contain counterparts
of each term in (\ref{RemainderTerm}). As the data is assumed to
be generated by a linear autoregression (Assumption \ref{assu:DGPParSpaceMixWeight}),
the $y_{t}$'s form a stationary and ergodic process. Moreover, as
the reparameterized log-likelihood of the GMAR model is four times
continuously differentiable (see Assumption \ref{assu:cont_differentiability}
and its verification), also the $\nabla_{\tilde{\pi}_{i}\tilde{\pi}_{j}\tilde{\pi}_{k}}^{3}l_{t}^{\pi}(\alpha,\tilde{\pi}^{\ast},0)$'s
form a stationary and ergodic process (for any $i$,$j$,$k$). An
analogous result holds for all the third and fourth partial derivatives
of $l_{t}^{\pi}(\alpha,\tilde{\pi},\varpi)$ appearing on the majorant
side of (\ref{eq:R1T_Bound}). 

Now, Lemma \ref{lem:XXX2}(iii) together with the ergodic theorem
implies that $\sup_{\alpha\in A}[\lvert T^{-1}\nabla_{\tilde{\pi}_{i}\tilde{\pi}_{j}\tilde{\pi}_{k}}^{3}L_{T}^{\pi}(\alpha,\tilde{\pi}^{\ast},0)\rvert]=O_{p}(1)$
(for any $i$,$j$,$k$). Similarly, $\sup_{\alpha\in A}[\lvert T^{-1}\nabla_{\tilde{\pi}_{i}\tilde{\pi}_{j}\varpi_{k}}^{3}L_{T}^{\pi}(\alpha,\tilde{\pi}^{\ast},0)\rvert]=O_{p}(1)$
(for any $i$,$j$,$k$). Expression of $\nabla_{\varpi_{i}\varpi_{j}\varpi_{k}}^{3}l_{t}^{\pi}(\alpha,\tilde{\pi}^{\ast},0)$
in Supplementary Appendix F.1, Lemmas \ref{lem:XXX1} and \ref{lem:XXX3},
and the compactness of $A$, imply that $\sup_{\alpha\in A}\lvert T^{-1/2}\nabla_{\varpi_{i}\varpi_{j}\varpi_{k}}^{3}L_{T}^{\pi}(\alpha,\tilde{\pi}^{\ast},0)\rvert\leq C\lvert T^{-1/2}\sum_{t=1}^{T}MDS_{t,i,j,k}(\pi^{*})\rvert$
for some finite $C$ and for some square integrable martingale difference
sequence $MDS_{t,i,j,k}(\pi^{*})$. Moreover, for any $i$,$j$,$k$,
the last upper bound is $O_{p}(1)$ by an appropriate central limit
theorem (\citet{billingsley1961lindeberg}). 

As for the fourth-order partial derivatives appearing on the majorant
side of (\ref{eq:R1T_Bound}), Lemma \ref{lem:XXX2}(iv) and a uniform
law of large numbers for stationary and ergodic processes (\citet{rangarao1962relations})
imply that $\sup_{\alpha\in A}\sup_{(\tilde{\pi},\varpi)\in B\times\Pi_{\alpha},\left\Vert (\tilde{\pi},\varpi)-(\tilde{\pi}^{\ast},0)\right\Vert \leq\gamma_{T}}\lvert T^{-1}\nabla_{\tilde{\pi}_{i}\tilde{\pi}_{j}\tilde{\pi}_{k}\tilde{\pi}_{l}}^{4}L_{T}^{\pi}(\alpha,\tilde{\pi},\varpi)\rvert=O_{p}(1)$
(for any $i$,$j$,$k$,$l$). The next three terms in (\ref{eq:R1T_Bound})
can be handled similarly. As for the last term on the majorant side
of (\ref{eq:R1T_Bound}), 
\begin{align*}
 & \sup_{\alpha\in A}\sup_{\substack{(\tilde{\pi},\varpi)\in B\times\Pi_{\alpha},\\
\left\Vert (\tilde{\pi},\varpi)-(\tilde{\pi}^{\ast},0)\right\Vert \leq\gamma_{T}
}
}\lvert T^{-1}(\nabla_{\varpi_{i}\varpi_{j}\varpi_{k}\varpi_{l}}^{4}L_{T}^{\pi}(\alpha,\tilde{\pi},\varpi)-\nabla_{\varpi_{i}\varpi_{j}\varpi_{k}\varpi_{l}}^{4}L_{T}^{\pi}(\alpha,\tilde{\pi}^{*},0))\rvert\\
 & \qquad\leq2\sup_{\alpha\in A}\sup_{\substack{(\tilde{\pi},\varpi)\in B\times\Pi_{\alpha},\\
\left\Vert (\tilde{\pi},\varpi)-(\tilde{\pi}^{\ast},0)\right\Vert \leq\gamma_{T}
}
}\lvert T^{-1}\nabla_{\varpi_{i}\varpi_{j}\varpi_{k}\varpi_{l}}^{4}L_{T}^{\pi}(\alpha,\tilde{\pi},\varpi)-E[\nabla_{\varpi_{i}\varpi_{j}\varpi_{k}\varpi_{l}}^{4}l_{t}^{\pi}(\alpha,\tilde{\pi},\varpi)]\rvert
\end{align*}
where the dominant side is $o_{p}(1)$ (again relying on Lemma \ref{lem:XXX2}(iv)
and a uniform LLN). To conclude, the upper bound in (\ref{eq:R1T_Bound})
is $\tilde{\gamma}_{T}O_{p}(1)+Co_{p}(1)=o_{p}(1)$. This completes
the verification of Assumption \ref{assu:quadexp}(iv) for the term
$R_{T}^{(1)}(\alpha,\tilde{\pi},\varpi)$.

Now consider $R_{T}^{(2)}(\alpha,\tilde{\pi},\varpi)=-\frac{1}{2}[T^{1/2}\boldsymbol{\theta}(\alpha,\tilde{\pi},\varpi)]^{\prime}(\mathcal{J}_{T}-\mathcal{I})[T^{1/2}\boldsymbol{\theta}(\alpha,\tilde{\pi},\varpi)]$.
We will below show that (a) $\mathcal{J}_{T}\overset{p}{\rightarrow}\mathcal{J}$
as $T\rightarrow\infty$, where the matrix $\mathcal{J}$ will be
specified below (and $\mathcal{J}_{T},\mathcal{J}$ do not depend
on $\alpha$). Write ($-2$ times) $R_{T}^{(2)}(\alpha,\tilde{\pi},\varpi)$
as 
\[
[T^{1/2}\boldsymbol{\theta}(\alpha,\tilde{\pi},\varpi)]^{\prime}(\mathcal{J}_{T}-\mathcal{J})[T^{1/2}\boldsymbol{\theta}(\alpha,\tilde{\pi},\varpi)]+[T^{1/2}\boldsymbol{\theta}(\alpha,\tilde{\pi},\varpi)]^{\prime}(\mathcal{J}-\mathcal{I})[T^{1/2}\boldsymbol{\theta}(\alpha,\tilde{\pi},\varpi)].
\]
We will below also show that (b) the latter term above equals zero.
The validity of Assumption \ref{assu:quadexp}(iv) for the term $R_{T}^{(2)}(\alpha,\tilde{\pi},\varpi)$
follows from results (a) and (b) (together with usual properties of
the Euclidean norm).

To prove claim (a), we first define the matrix $\mathcal{J}$ as 
\[
\mathcal{J}=\begin{bmatrix}\mathcal{J}_{\pi\pi} & \mathcal{J}_{\pi\varpi\varpi}^{\prime}\\
\mathcal{J}_{\pi\varpi\varpi} & \mathcal{J}_{\varpi\varpi\varpi\varpi}
\end{bmatrix}
\]
where the matrices $\mathcal{J}_{\tilde{\pi}\tilde{\pi}}$ ($(q_{1}+q_{2})\times(q_{1}+q_{2})$),
$\mathcal{J}_{\tilde{\pi}\varpi\varpi}^{\prime}$ ($(q_{1}+q_{2})\times q_{\vartheta}$),
and $\mathcal{J}_{\varpi\varpi\varpi\varpi}$ ($q_{\vartheta}\times q_{\vartheta}$)
are defined as 
\begin{align*}
\mathcal{J}_{\tilde{\pi}\tilde{\pi}} & =E\left[\frac{\nabla f_{t}^{*}}{f_{t}^{*}}\frac{\nabla'f_{t}^{*}}{f_{t}^{*}}\right]\\
\mathcal{J}_{\tilde{\pi}\varpi\varpi}^{\prime} & =E\left[\left[c_{ij}\frac{\nabla f_{t}^{*}}{f_{t}^{*}}X_{t,i,j}^{*}\right]_{(i,j)\in\mathfrak{I}}\right]\\
\mathcal{J}_{\varpi\varpi\varpi\varpi} & =\frac{1}{3}\left[c_{ij}c_{kl}\left(E[X_{t,i,j}^{*}X_{t,k,l}^{*}]+E[X_{t,i,k}^{*}X_{t,j,l}^{*}]+E[X_{t,i,l}^{*}X_{t,j,k}^{*}]\right)\right]_{(i,j,k,l)\in\mathfrak{I}\times\mathfrak{I}}
\end{align*}
where the $X_{t,i,j}^{*}$ ($i,j\in\{1,\ldots,q_{2}\}$) are as in
(\ref{eq:Notation_X*}). Finiteness of $\mathcal{J}$ follows from
Lemma \ref{lem:XXX1}. 

Now consider the convergence result $\mathcal{J}_{T}\overset{p}{\rightarrow}\mathcal{J}$
for each block at a time. For the top-left block, from Supplementary
Appendix F.1 we have $\nabla_{\tilde{\pi}\tilde{\pi}^{\prime}}l_{t}^{\pi*}=\frac{\nabla^{2}f_{t}^{*}}{f_{t}^{*}}-\frac{\nabla f_{t}^{*}}{f_{t}^{*}}\frac{\nabla'f_{t}^{*}}{f_{t}^{*}}$
so that ergodic theorem and Lemmas \ref{lem:XXX1} and \ref{lem:XXX3}
(latter ensuring the first term on the right-hand side of the previous
equation has zero expectation) imply that $\mathcal{J}_{T,\tilde{\pi}\tilde{\pi}}=-T^{-1}\nabla_{\tilde{\pi}\tilde{\pi}^{\prime}}^{2}L_{T}^{\pi*}\overset{p}{\rightarrow}\mathcal{J}_{\tilde{\pi}\tilde{\pi}}$.

For the off-diagonal block, consider the expression of $\nabla_{\tilde{\pi}_{i}\varpi_{j}\varpi_{k}}^{3}l_{t}^{\pi*}$
in Supplementary Appendix F.1. Lemma \ref{lem:XXX3} ensures that
of the ten summands in this expression, only the second, fourth, and
sixth ones have non-zero expectation. Therefore the ergodic theorem
and Lemma \ref{lem:XXX1} imply that 
\[
\mathcal{J}_{T,\tilde{\pi}\varpi\varpi}^{\prime}=-T^{-1}\frac{1}{\alpha_{1}\alpha_{2}}[c_{11}\nabla_{\tilde{\pi}\varpi_{1}\varpi_{1}}^{3}L_{T}^{\pi*}:\cdots:c_{q_{2}-1,q_{2}}\nabla_{\tilde{\pi}\varpi_{q_{2}-1}\varpi_{q_{2}}}^{3}L_{T}^{\pi*}]\overset{p}{\rightarrow}\mathcal{J}_{\tilde{\pi}\varpi\varpi}^{\prime}.
\]

Lastly, for the bottom-right block, consider the expression of $\nabla_{\varpi_{i}\varpi_{j}\varpi_{k}\varpi_{l}}^{4}l_{t}^{\pi*}$
in Supplementary Appendix F.1. Lemma \ref{lem:XXX3} reveals that
the terms in this expression that have non-zero expectation can be
expressed as
\[
-\alpha_{1}^{2}\alpha_{2}^{2}[X_{t,i,j}^{*}X_{t,k,l}^{*}+X_{t,i,k}^{*}X_{t,j,l}^{*}+X_{t,i,l}^{*}X_{t,j,k}^{*}].
\]
Therefore the ergodic theorem and Lemma \ref{lem:XXX1} imply that
\[
\mathcal{J}_{T,\varpi\varpi\varpi\varpi}=-T^{-1}\frac{8}{4!}\frac{1}{\alpha_{1}^{2}\alpha_{2}^{2}}\left[c_{ij}c_{kl}\nabla_{\varpi_{i}\varpi_{j}\varpi_{k}\varpi_{l}}^{4}L_{T}^{\pi*}\right]_{(i,j,k,l)\in\mathfrak{I}\times\mathfrak{I}}\overset{p}{\rightarrow}\mathcal{J}_{\varpi\varpi\varpi\varpi}.
\]
This completes the proof of claim (a).

To prove claim (b), first note that from the definitions of $\mathcal{J}$
and $\mathcal{I}$ (see (\ref{MatrixI})) it can be seen that only
the bottom-right blocks of $\mathcal{J}$ and $\mathcal{I}$ differ.
Therefore, as $T^{1/2}\boldsymbol{\theta}(\alpha,\tilde{\pi},\varpi)=(T^{1/2}(\tilde{\pi}-\tilde{\pi}^{\ast}),T^{1/2}(\alpha_{1}\alpha_{2}v(\varpi)))$,
claim (b) holds if $T(\alpha_{1}\alpha_{2})^{2}v(\varpi)^{\prime}(\mathcal{J}_{\varpi\varpi\varpi\varpi}-\mathcal{I}_{\varpi\varpi\varpi\varpi})v(\varpi)=0$
where 
\begin{align*}
\mathcal{J}_{\varpi\varpi\varpi\varpi} & =\frac{1}{3}\left[c_{ij}c_{kl}\left(E[X_{t,i,j}^{*}X_{t,k,l}^{*}]+E[X_{t,i,k}^{*}X_{t,j,l}^{*}]+E[X_{t,i,l}^{*}X_{t,j,k}^{*}]\right)\right]_{(i,j,k,l)\in\mathfrak{I}\times\mathfrak{I}},\\
\mathcal{I}_{\varpi\varpi\varpi\varpi} & =\left[c_{ij}c_{kl}E[X_{t,i,j}^{*}X_{t,k,l}^{*}]\right]_{(i,j,k,l)\in\mathfrak{I}\times\mathfrak{I}}.
\end{align*}
Note that the scalars $A_{ijkl}=E[X_{t,i,j}^{*}X_{t,k,l}^{*}]$ satisfy
$A_{ijkl}=A_{jikl}$ and $A_{ijkl}=A_{ijlk}$ for all $i,j,k,l$ so
that using property (\ref{PropertyIndexNotation}) we obtain 
\begin{align*}
v(\varpi)^{\prime}\mathcal{J}_{\varpi\varpi\varpi\varpi}v(\varpi) & =\frac{1}{3}v(\varpi)^{\prime}\left[c_{ij}c_{kl}\left(E[X_{t,i,j}^{*}X_{t,k,l}^{*}]+E[X_{t,i,k}^{*}X_{t,j,l}^{*}]+E[X_{t,i,l}^{*}X_{t,j,k}^{*}]\right)\right]_{(i,j,k,l)\in\mathfrak{I}\times\mathfrak{I}}v(\varpi)\\
 & =\frac{1}{3}\frac{1}{4}\sum_{i=1}^{q_{2}}\sum_{j=1}^{q_{2}}\sum_{k=1}^{q_{2}}\sum_{l=1}^{q_{2}}\left(E[X_{t,i,j}^{*}X_{t,k,l}^{*}]+E[X_{t,i,k}^{*}X_{t,j,l}^{*}]+E[X_{t,i,l}^{*}X_{t,j,k}^{*}]\right)\varpi_{i}\varpi_{j}\varpi_{k}\varpi_{l}\\
 & =\frac{1}{4}\sum_{i=1}^{q_{2}}\sum_{j=1}^{q_{2}}\sum_{k=1}^{q_{2}}\sum_{l=1}^{q_{2}}E[X_{t,i,j}^{*}X_{t,k,l}^{*}]\varpi_{i}\varpi_{j}\varpi_{k}\varpi_{l}\\
 & =v(\varpi)^{\prime}\mathcal{I}_{\varpi\varpi\varpi\varpi}v(\varpi).
\end{align*}
This completes the proof of claim (b). 

Therefore, the verification of Assumption 5(iv) is done. 

\subsection{Additional Lemmas}

The following four lemmas contain results needed in the proofs. Note
that the first and the third lemma are not specific to the examples
in this paper, whereas the second and fourth lemmas concern only the
GMAR example. In the first lemma, $\mathsf{n}_{p+1}(\tilde{\phi})=\mathsf{n}_{p+1}(y_{t},\boldsymbol{y}_{t-1};\tilde{\phi})$
denotes the $(p+1)$-dimensional density function of an AR($p$) process
based on parameter value $\tilde{\phi}$ evaluated at $(y_{t},\boldsymbol{y}_{t-1})$;
cf. equations (\ref{Mixing Weights})\textendash (\ref{Normal p})
for the $p$-dimensional counterpart $\mathsf{n}_{p}(\tilde{\phi})=\mathsf{n}_{p}(\boldsymbol{y}_{t-1};\tilde{\phi})$. 

\begin{lemma}\label{lem:XXX1}

For any $i,j,k,l\in\{1,\ldots,p+2\}$ and any positive $r$, the following
moments are all finite: 
\begin{align*}
\textrm{(i) } & E\bigl[\sup\nolimits _{\tilde{\phi}\in\tilde{\Phi}}\lvert\nabla_{i}f_{t}(\tilde{\phi})/f_{t}(\tilde{\phi})\rvert^{r}\bigr],\,E\bigl[\sup\nolimits _{\tilde{\phi}\in\tilde{\Phi}}\lvert\nabla_{ij}^{2}f_{t}(\tilde{\phi})/f_{t}(\tilde{\phi})\rvert^{r}\bigr],\ldots,\,E\bigl[\sup\nolimits _{\tilde{\phi}\in\tilde{\Phi}}\lvert\nabla_{ijkl}^{4}f_{t}(\tilde{\phi})/f_{t}(\tilde{\phi})\rvert^{r}\bigr],\\
\textrm{(ii) } & E\bigl[\sup\nolimits _{\tilde{\phi}\in\tilde{\Phi}}\lvert\nabla_{i}\mathsf{n}_{p}(\tilde{\phi})/\mathsf{n}_{p}(\tilde{\phi})\rvert^{r}\bigr],\,E\bigl[\sup\nolimits _{\tilde{\phi}\in\tilde{\Phi}}\lvert\nabla_{ij}^{2}\mathsf{n}_{p}(\tilde{\phi})/\mathsf{n}_{p}(\tilde{\phi})\rvert^{r}\bigr],\ldots,\,E\bigl[\sup\nolimits _{\tilde{\phi}\in\tilde{\Phi}}\lvert\nabla_{ijkl}^{4}\mathsf{n}_{p}(\tilde{\phi})/\mathsf{n}_{p}(\tilde{\phi})\rvert^{r}\bigr],\\
\textrm{(iii) } & E\bigl[\sup\nolimits _{\tilde{\phi}\in\tilde{\Phi}}\lvert\nabla_{i}\mathsf{n}_{p+1}(\tilde{\phi})\rvert^{r}\bigr],\,E\bigl[\sup\nolimits _{\tilde{\phi}\in\tilde{\Phi}}\lvert\nabla_{ij}^{2}\mathsf{n}_{p+1}(\tilde{\phi})\rvert^{r}\bigr],\ldots,\,E\bigl[\sup\nolimits _{\tilde{\phi}\in\tilde{\Phi}}\lvert\nabla_{ijkl}^{4}\mathsf{n}_{p+1}(\tilde{\phi})\rvert^{r}\bigr].
\end{align*}

\end{lemma}

\bigskip{}

\begin{lemma}\label{lem:XXX2}

In the GMAR example the following hold, where each of (the scalars)
$z_{1},z_{2},z_{3},z_{4}$ is a `placeholder' for any of $\tilde{\pi}_{i},\tilde{\pi}_{j},\tilde{\pi}_{k},\tilde{\pi}_{l}$
($i,j,k,l\in\{1,\ldots,q_{1}+q_{2}\}$) or $\varpi_{i},\varpi_{j},\varpi_{k},\varpi_{l}$
($i,j,k,l\in\{1,\ldots,q_{2}\}$):
\begin{align*}
\textrm{(i) } & E\left[\sup\nolimits _{\alpha\in A}\sup\nolimits _{(\tilde{\pi},\varpi)\in B\times\Pi_{\alpha}}\lvert\nabla_{z_{1}}l_{t}^{\pi}(\alpha,\tilde{\pi},\varpi)\rvert\right]<\infty,\\
\textrm{(ii) } & E\left[\sup\nolimits _{\alpha\in A}\sup\nolimits _{(\tilde{\pi},\varpi)\in B\times\Pi_{\alpha}}\lvert\nabla_{z_{1}z_{2}}^{2}l_{t}^{\pi}(\alpha,\tilde{\pi},\varpi)\rvert\right]<\infty,\\
\textrm{(iii) } & E\left[\sup\nolimits _{\alpha\in A}\sup\nolimits _{(\tilde{\pi},\varpi)\in B\times\Pi_{\alpha}}\lvert\nabla_{z_{1}z_{2}z_{3}}^{3}l_{t}^{\pi}(\alpha,\tilde{\pi},\varpi)\rvert\right]<\infty,\\
\textrm{(iv) } & E\left[\sup\nolimits _{\alpha\in A}\sup\nolimits _{(\tilde{\pi},\varpi)\in B\times\Pi_{\alpha}}\lvert\nabla_{z_{1}z_{2}z_{3}z_{4}}^{4}l_{t}^{\pi}(\alpha,\tilde{\pi},\varpi)\rvert\right]<\infty.
\end{align*}

\end{lemma}

\bigskip{}

\begin{lemma}\label{lem:XXX3}

For any $i,j,k,l\in\{1,\ldots,p+2\}$, 
\[
E\bigl[\begin{array}{c}
\frac{\nabla_{i}f_{t}^{*}}{f_{t}^{*}}\end{array}\mid\boldsymbol{y}_{t-1}\bigr]=E\bigl[\begin{array}{c}
\frac{\nabla_{ij}^{2}f_{t}^{*}}{f_{t}^{*}}\end{array}\mid\boldsymbol{y}_{t-1}\bigr]=E\bigl[\begin{array}{c}
\frac{\nabla_{ijk}^{3}f_{t}^{*}}{f_{t}^{*}}\end{array}\mid\boldsymbol{y}_{t-1}\bigr]=E\bigl[\begin{array}{c}
\frac{\nabla_{ijkl}^{4}f_{t}^{*}}{f_{t}^{*}}\end{array}\mid\boldsymbol{y}_{t-1}\bigr]=0.
\]

\end{lemma}

\bigskip{}

\begin{lemma}\label{lem:XXX4}

In the GMAR example the following hold for all $\alpha\in A$, $(\tilde{\pi},\varpi)\in B\times\Pi_{\alpha}$,
$T$, and $i,j,k,l\in\{1,\ldots,q_{1}+q_{2}\}$ (subindex in $\tilde{\pi}$)
or $i,j,k,l\in\{1,\ldots,q_{2}\}$ (subindex in $\varpi$): 
\begin{align*}
\text{(i) \ \ }T\left\vert \tilde{\pi}_{i}-\tilde{\pi}_{i}^{\ast}\right\vert |\tilde{\pi}_{j}-\tilde{\pi}_{j}^{\ast}||\tilde{\pi}_{k}-\tilde{\pi}_{k}^{\ast}| & \leq(1+\lVert T^{1/2}\boldsymbol{\theta}(\alpha,\tilde{\pi},\varpi)\rVert)^{2}\left\Vert \tilde{\pi}-\tilde{\pi}^{\ast}\right\Vert ,\\
\text{(ii) \ \ }T|\tilde{\pi}_{i}-\tilde{\pi}_{i}^{\ast}||\tilde{\pi}_{j}-\tilde{\pi}_{j}^{\ast}||\varpi_{k}| & \leq(1+\lVert T^{1/2}\boldsymbol{\theta}(\alpha,\tilde{\pi},\varpi)\rVert)^{2}\left\Vert \varpi\right\Vert ,\\
\text{(iii) \ \ }T^{1/2}|\varpi_{i}||\varpi_{j}||\varpi_{k}| & \leq(1+\lVert T^{1/2}\boldsymbol{\theta}(\alpha,\tilde{\pi},\varpi)\rVert)^{2}(\alpha_{1}\alpha_{2})^{-3/2}\left\Vert \boldsymbol{\theta}(\alpha,\tilde{\pi},\varpi)\right\Vert ^{1/2},\\
\text{(iv) \ \ }T(\tilde{\pi}_{i}-\tilde{\pi}_{i}^{\ast})(\tilde{\pi}_{j}-\tilde{\pi}_{j}^{\ast})(\tilde{\pi}_{k}-\tilde{\pi}_{k}^{\ast})(\tilde{\pi}_{l}-\tilde{\pi}_{l}^{\ast}) & \leq(1+\lVert T^{1/2}\boldsymbol{\theta}(\alpha,\tilde{\pi},\varpi)\rVert)^{2}\left\Vert \tilde{\pi}-\tilde{\pi}^{\ast}\right\Vert ^{2},\\
\text{(v) \ \ }T(\tilde{\pi}_{i}-\tilde{\pi}_{i}^{\ast})(\tilde{\pi}_{j}-\tilde{\pi}_{j}^{\ast})(\tilde{\pi}_{k}-\tilde{\pi}_{k}^{\ast})\varpi_{l} & \leq(1+\lVert T^{1/2}\boldsymbol{\theta}(\alpha,\tilde{\pi},\varpi)\rVert)^{2}\left\Vert \tilde{\pi}-\tilde{\pi}^{\ast}\right\Vert \left\Vert \varpi\right\Vert ,\\
\text{(vi) \ \ }T(\tilde{\pi}_{i}-\tilde{\pi}_{i}^{\ast})(\tilde{\pi}_{j}-\tilde{\pi}_{j}^{\ast})\varpi_{k}\varpi_{l} & \leq(1+\lVert T^{1/2}\boldsymbol{\theta}(\alpha,\tilde{\pi},\varpi)\rVert)^{2}\left\Vert \varpi\right\Vert ^{2},\\
\text{(vii) \ \ }T(\tilde{\pi}_{i}-\tilde{\pi}_{i}^{\ast})\varpi_{j}\varpi_{k}\varpi_{l} & \leq(1+\lVert T^{1/2}\boldsymbol{\theta}(\alpha,\tilde{\pi},\varpi)\rVert)^{2}(\alpha_{1}\alpha_{2})^{-3/2}\left\Vert \boldsymbol{\theta}(\alpha,\tilde{\pi},\varpi)\right\Vert ^{1/2},\\
\text{(viii) \ \ }T\varpi_{i}\varpi_{j}\varpi_{k}\varpi_{l} & \leq(1+\lVert T^{1/2}\boldsymbol{\theta}(\alpha,\tilde{\pi},\varpi)\rVert)^{2}(\alpha_{1}\alpha_{2})^{-2}.
\end{align*}

\end{lemma}

\subsection{Proofs of Lemmas \ref{lem:XXX1}\textendash \ref{lem:XXX4}}
\begin{proof}[\textbf{\emph{Proof of Lemma \ref{lem:XXX1}}}]
Writing $g_{t}(\tilde{\phi})=[y_{t}-(\tilde{\phi}_{0}+\tilde{\phi}_{1}y_{t-1}+\cdots+\tilde{\phi}_{p}y_{t-p})]/\tilde{\sigma}_{1}$
and recalling the definition of $f_{t}(\tilde{\phi})$ we can write
$f_{t}(\tilde{\phi})=\tilde{\sigma}_{1}^{-1}{\scriptstyle \mathscr{\mathfrak{N}}}(g_{t}(\tilde{\phi}))$
where ${\scriptstyle \mathscr{\mathfrak{N}}}(\cdot)$ denotes the
density function of a standard normal random variable. Recall also
that derivatives of ${\scriptstyle \mathscr{\mathfrak{N}}}(\cdot)$
can be expressed using (one version of) Hermite polynomials $H_{n}(x)$
as $\frac{d^{n}}{dx^{n}}{\scriptstyle \mathscr{\mathfrak{N}}}(x)=(-1)^{n}H_{n}(x){\scriptstyle \mathscr{\mathfrak{N}}}(x)$.
Using the chain rule for differentiation repeatedly, it can therefore
be seen that each of the functions $\nabla_{i}f_{t}(\tilde{\phi})/f_{t}(\tilde{\phi})$,
$\nabla_{ij}^{2}f_{t}(\tilde{\phi})/f_{t}(\tilde{\phi})$, $\nabla_{ijk}^{3}f_{t}(\tilde{\phi})/f_{t}(\tilde{\phi})$,
and $\nabla_{ijkl}^{4}f_{t}(\tilde{\phi})/f_{t}(\tilde{\phi})$ can
be expressed as a sum of terms each of which is a product involving
Hermite polynomials $H_{n}(g_{t}(\tilde{\phi}))$ and powers of derivatives
of $g_{t}(\tilde{\phi})$ (and functions of $\tilde{\phi}$). Thus,
each of these functions is a polynomial in terms of $y_{t},y_{t-1},\ldots,y_{t-p}$.
As the $y_{t}$'s are generated by a stationary linear Gaussian AR($p$)
model, they possess moments of all orders, implying (together with
the definition of $\tilde{\Phi}$, implying in particular that $\tilde{\sigma}_{1}$
is bounded away from zero on $\tilde{\Phi}$) the finiteness of the
moments listed in part (i) of the lemma. 

As for part (ii), note that $\mathsf{n}_{p}(\tilde{\phi})$ can be
expressed as $\mathsf{n}_{p}(\tilde{\phi})=g_{1}(\tilde{\phi}){\scriptstyle \mathscr{\mathfrak{N}}}(g_{2,t}(\tilde{\phi}))$
for some function $g_{1}(\tilde{\phi})$ not depending on the $y_{t}$'s
and $g_{2,t}(\tilde{\phi})$ the square root of a second-order polynomial
in $y_{t-1},\ldots,y_{t-p}$. Therefore the finiteness of the moments
listed in the part (ii) follows using similar arguments as above (noting
that the definition of $\tilde{\Phi}$ implies that the determinant
of the covariance matrix appearing in $\mathsf{n}_{p}(\tilde{\phi})$
is bounded away from zero on $\tilde{\Phi}$). 

Finally, for part (iii), similar arguments, together with the observation
that $\mathsf{n}_{p+1}(\tilde{\phi})$ is bounded on $\tilde{\Phi}$,
yield the desired result.
\end{proof}
\bigskip{}
\begin{proof}[\textbf{\emph{Proof of Lemma \ref{lem:XXX2}}}]
To prove (i), first consider the derivatives with respect to $\varpi$.
From the formulas in Supplementary Appendix F.7 we obtain
\begin{align*}
\nabla_{\varpi}l_{t}^{\pi}(\alpha,\tilde{\pi},\varpi) & =\alpha_{1,t}\alpha_{2,t}\left(\frac{D_{\tilde{\phi},\varpi}^{(1)\prime}\nabla\mathsf{n}_{p}(\tilde{\phi})}{\mathsf{n}_{p}(\tilde{\phi})}-\frac{D_{\tilde{\varphi},\varpi}^{(1)\prime}\nabla\mathsf{n}_{p}(\tilde{\varphi})}{\mathsf{n}_{p}(\tilde{\varphi})}\right)\frac{f_{t}(\tilde{\phi})}{f_{2,t}^{\pi}(\alpha,\tilde{\pi},\varpi)}+\frac{D_{\tilde{\phi},\varpi}^{(1)\prime}\nabla f_{t}(\tilde{\phi})}{f_{2,t}^{\pi}(\alpha,\tilde{\pi},\varpi)}\alpha_{1,t}\\
 & -\alpha_{1,t}\alpha_{2,t}\left(\frac{D_{\tilde{\phi},\varpi}^{(1)\prime}\nabla\mathsf{n}_{p}(\tilde{\phi})}{\mathsf{n}_{p}(\tilde{\phi})}-\frac{D_{\tilde{\varphi},\varpi}^{(1)\prime}\nabla\mathsf{n}_{p}(\tilde{\varphi})}{\mathsf{n}_{p}(\tilde{\varphi})}\right)\frac{f_{t}(\tilde{\varphi})}{f_{2,t}^{\pi}(\alpha,\tilde{\pi},\varpi)}+\frac{D_{\tilde{\varphi},\varpi}^{(1)\prime}\nabla f_{t}(\tilde{\varphi})}{f_{2,t}^{\pi}(\alpha,\tilde{\pi},\varpi)}(1-\alpha_{1,t})
\end{align*}
where $\tilde{\phi}$ and $\tilde{\varphi}$ are understood as functions
of $(\alpha,\tilde{\pi},\varpi)$ (i.e., $\tilde{\phi}=(\beta,\pi+\alpha_{2}\varpi)$
and $\tilde{\varphi}=(\beta,\pi-\alpha_{1}\varpi)$). Note that whenever
$\alpha\in A$ and $(\tilde{\pi},\varpi)\in B\times\Pi_{\alpha}$,
$\tilde{\phi}\in\tilde{\Phi}$ and $\tilde{\varphi}\in\tilde{\Phi}$.
Also note that over $\alpha\in A$ and $(\tilde{\pi},\varpi)\in B\times\Pi_{\alpha}$,
the quantities
\[
\lvert\alpha_{1,t}\rvert,\,\,\,\lvert\alpha_{2,t}\rvert,\,\,\,\lVert D_{\tilde{\phi},\varpi}^{(1)}\rVert,\,\,\,\lVert D_{\tilde{\varphi},\varpi}^{(1)}\rVert,\,\,\,\lvert f_{t}(\tilde{\phi})/f_{2,t}^{\pi}(\alpha,\tilde{\pi},\varpi)\rvert,\,\,\,\lvert f_{t}(\tilde{\varphi})/f_{2,t}^{\pi}(\alpha,\tilde{\pi},\varpi)\rvert
\]
are all bounded by finite constants. Therefore $E\bigl[\sup\nolimits _{\alpha\in A}\sup\nolimits _{(\tilde{\pi},\varpi)\in N(\tilde{\pi}^{\ast},0)}\lVert\nabla_{\varpi}l_{t}^{\pi}(\alpha,\tilde{\pi},\varpi)\rVert\bigr]<\infty$
as long as $E\bigl[\sup\nolimits _{\tilde{\phi}\in\tilde{\Phi}}\lVert\nabla f_{t}(\tilde{\phi})/f_{t}(\tilde{\phi})\rVert\bigr]<\infty$
and $E\bigl[\sup\nolimits _{\tilde{\phi}\in\tilde{\Phi}}\lVert\nabla\mathsf{n}_{p}(\tilde{\phi})/\mathsf{n}_{p}(\tilde{\phi})\rVert\bigr]<\infty$,
which is ensured by Lemma \ref{lem:XXX1}. The argument for $\nabla_{\tilde{\pi}}l_{t}^{\pi}(\alpha,\tilde{\pi},\varpi)$
is entirely similar and is omitted.

To prove (ii)\textendash (iv), entirely similar arguments can be used.
Tedious calculations (details omitted) show that the finiteness of
the required moments is ensured by the finiteness of the moments in
Lemma \ref{lem:XXX1}(i) and (ii). 
\end{proof}
\bigskip{}
\begin{proof}[\textbf{\emph{Proof of Lemma \ref{lem:XXX3}}}]
 For the first two derivatives, the stated result follows directly
from the expressions of $\nabla f_{t}^{*}/f_{t}^{*}$ and $\nabla^{2}f_{t}^{*}/f_{t}^{*}$
in (\ref{eq:AppF3_nabla2f_f}). The results for the third and fourth
derivatives can be obtained with straightforward calculation. 
\end{proof}
\bigskip{}
\begin{proof}[\textbf{\emph{Proof of Lemma \ref{lem:XXX4}}}]
First recall that $\left\Vert T^{1/2}\boldsymbol{\theta}(\alpha,\tilde{\pi},\varpi)\right\Vert ^{2}=T\left\Vert \tilde{\pi}-\tilde{\pi}^{\ast}\right\Vert ^{2}+T\alpha_{1}^{2}\alpha_{2}^{2}\left\Vert v(\varpi)\right\Vert ^{2}$.
(i) By an elementary inequality, $T\left\vert \tilde{\pi}_{i}-\tilde{\pi}_{i}^{\ast}\right\vert |\tilde{\pi}_{j}-\tilde{\pi}_{j}^{\ast}||\tilde{\pi}_{k}-\tilde{\pi}_{k}^{\ast}|\leq T\left\Vert \tilde{\pi}-\tilde{\pi}^{\ast}\right\Vert ^{3}$
and therefore the result follows by adding nonnegative terms on the
majorant side of this inequality. Parts (ii) and (iv)\textendash (vi)
are shown similarly. (vii) As $\boldsymbol{\theta}(\alpha,\tilde{\pi},\varpi)=(\tilde{\pi}-\tilde{\pi}^{\ast},\alpha_{1}\alpha_{2}v(\varpi))$,
each of the terms $|\tilde{\pi}_{i}-\tilde{\pi}_{i}^{\ast}|$, $\alpha_{1}\alpha_{2}\varpi_{j}^{2}$,
$\alpha_{1}\alpha_{2}\varpi_{k}^{2}$, and $\alpha_{1}\alpha_{2}\varpi_{l}^{2}$
are dominated by $\left\Vert \boldsymbol{\theta}(\alpha,\tilde{\pi},\varpi)\right\Vert $.
Therefore 
\[
T|\tilde{\pi}_{i}-\tilde{\pi}_{i}^{\ast}||\varpi_{j}||\varpi_{k}||\varpi_{l}|\leq T(\alpha_{1}\alpha_{2})^{-3/2}\left\Vert \boldsymbol{\theta}(\alpha,\tilde{\pi},\varpi)\right\Vert ^{5/2}\leq(\alpha_{1}\alpha_{2})^{-3/2}(1+\left\Vert T^{1/2}\boldsymbol{\theta}(\alpha,\tilde{\pi},\varpi)\right\Vert )^{2}\left\Vert \boldsymbol{\theta}(\alpha,\tilde{\pi},\varpi)\right\Vert ^{1/2}
\]
where the second inequality holds because nonnegative terms were added
to the majorant side. (viii) Similarly as in the previous part, 
\[
T|\varpi_{i}||\varpi_{j}||\varpi_{k}||\varpi_{l}|\leq(\alpha_{1}\alpha_{2})^{-2}\left\Vert T^{1/2}\boldsymbol{\theta}(\alpha,\tilde{\pi},\varpi)\right\Vert ^{2}\leq(\alpha_{1}\alpha_{2})^{-2}(1+\left\Vert T^{1/2}\boldsymbol{\theta}(\alpha,\tilde{\pi},\varpi)\right\Vert )^{2}.
\]
Finally, for (iii) we, similarly as above but scaling with $T^{1/2}$
instead of $T$, obtain 
\[
T^{1/2}|\varpi_{i}||\varpi_{j}||\varpi_{k}|\leq T^{1/2}(\alpha_{1}\alpha_{2})^{-3/2}\left\Vert \boldsymbol{\theta}(\alpha,\tilde{\pi},\varpi)\right\Vert ^{3/2}\leq(\alpha_{1}\alpha_{2})^{-3/2}(1+\left\Vert T^{1/2}\boldsymbol{\theta}(\alpha,\tilde{\pi},\varpi\alpha)\right\Vert )^{2}\left\Vert \boldsymbol{\theta}(\alpha,\tilde{\pi},\varpi)\right\Vert ^{1/2},
\]
which completes the proof.
\end{proof}

\subsection{Partial derivatives of the reparameterized log-likelihood function
(continued)}

Note that $l_{t}^{\pi}(\alpha,\beta,\pi,\varpi)=\log[f_{2,t}^{\pi}(\alpha,\beta,\pi,\varpi)]$
with
\begin{align*}
f_{2,t}^{\pi}(\alpha,\beta,\pi,\varpi) & =\alpha_{1,t}^{\pi}(\alpha,\beta,\pi,\varpi)f_{t}(\beta,\pi+\alpha_{2}\varpi)+(1-\alpha_{1,t}^{\pi}(\alpha,\beta,\pi,\varpi))f_{t}(\beta,\pi-\alpha_{1}\varpi),\\
\alpha_{1,t}^{\pi}(\alpha,\beta,\pi,\varpi) & =\alpha_{1,t}^{G}(\alpha,(\beta,\pi+\alpha_{2}\varpi),(\beta,\pi-\alpha_{1}\varpi)).
\end{align*}
For the sake of brevity, but with slight abuse of notation, we will
write these as
\begin{align*}
f_{2,t}^{\pi}(\alpha,\beta,\pi,\varpi) & =\alpha_{1,t}^{\pi}(\alpha,\beta,\pi,\varpi)f_{t}(\tilde{\phi})+(1-\alpha_{1,t}^{\pi}(\alpha,\beta,\pi,\varpi))f_{t}(\tilde{\varphi}),\\
\alpha_{1,t}^{\pi}(\alpha,\beta,\pi,\varpi) & =\frac{\alpha\mathsf{n}_{p}(\tilde{\phi})}{\alpha\mathsf{n}_{p}(\tilde{\phi})+(1-\alpha)\mathsf{n}_{p}(\tilde{\varphi})},
\end{align*}
where $\tilde{\phi}$ and $\tilde{\varphi}$ are understood as functions
of $(\alpha,\beta,\pi,\varpi)$, that is, $\tilde{\phi}=(\beta,\pi+\alpha_{2}\varpi)$
and $\tilde{\varphi}=(\beta,\pi-\alpha_{1}\varpi)$.

The following notation will be helpful: 
\begin{align*}
D_{\tilde{\phi},\tilde{\pi}}^{(1)} & =\frac{\partial(\beta,\pi+\alpha_{2}\varpi)}{\partial\tilde{\pi}'}=I_{1+q_{2}}\\
D_{\tilde{\phi},\varpi}^{(1)} & =\frac{\partial(\beta,\pi+\alpha_{2}\varpi)}{\partial\varpi'}=\begin{bmatrix}0\\
\alpha_{2}I_{q_{2}}
\end{bmatrix}\,\,\,\,\,((1+q_{2})\times q_{2})\\
D_{\tilde{\varphi},\tilde{\pi}}^{(1)} & =\frac{\partial(\beta,\pi-\alpha_{1}\varpi)}{\partial\tilde{\pi}'}=I_{1+q_{2}}\\
D_{\tilde{\varphi},\varpi}^{(1)} & =\frac{\partial(\beta,\pi-\alpha_{1}\varpi)}{\partial\varpi'}=\begin{bmatrix}0\\
-\alpha_{1}I_{q_{2}}
\end{bmatrix}\,\,\,\,\,((1+q_{2})\times q_{2})
\end{align*}

\paragraph{First-order partial derivatives. }

\noindent With straightforward differentation we obtain
\begin{align*}
\nabla_{\tilde{\pi}}l_{t}^{\pi}(\alpha,\tilde{\pi},\varpi) & =\frac{\nabla_{\tilde{\pi}}f_{2,t}^{\pi}(\alpha,\tilde{\pi},\varpi)}{f_{2,t}^{\pi}(\alpha,\tilde{\pi},\varpi)}\\
\nabla_{\varpi}l_{t}^{\pi}(\alpha,\tilde{\pi},\varpi) & =\frac{\nabla_{\varpi}f_{2,t}^{\pi}(\alpha,\tilde{\pi},\varpi)}{f_{2,t}^{\pi}(\alpha,\tilde{\pi},\varpi)}
\end{align*}
with
\begin{align*}
\nabla_{\tilde{\pi}}f_{2,t}^{\pi}(\alpha,\tilde{\pi},\varpi) & =\nabla_{\tilde{\pi}}\alpha_{1,t}f_{t}(\tilde{\phi})+\nabla f_{t}(\tilde{\phi})\alpha_{1,t}-\nabla_{\tilde{\pi}}\alpha_{1,t}f_{t}(\tilde{\varphi})+\nabla f_{t}(\tilde{\varphi})(1-\alpha_{1,t})\\
\nabla_{\varpi}f_{2,t}^{\pi}(\alpha,\tilde{\pi},\varpi) & =\nabla_{\varpi}\alpha_{1,t}f_{t}(\tilde{\phi})+D_{\tilde{\phi},\varpi}^{(1)\prime}\nabla f_{t}(\tilde{\phi})\alpha_{1,t}-\nabla_{\varpi}\alpha_{1,t}f_{t}(\tilde{\varphi})+D_{\tilde{\varphi},\varpi}^{(1)\prime}\nabla f_{t}(\tilde{\varphi})(1-\alpha_{1,t})
\end{align*}
and
\begin{align*}
\nabla_{\tilde{\pi}}\alpha_{1,t} & =\frac{\alpha_{1}\nabla\mathsf{n}_{p}(\tilde{\phi})}{\alpha_{1}\mathsf{n}_{p}(\tilde{\phi})+\alpha_{2}\mathsf{n}_{p}(\tilde{\varphi})}-\alpha_{1,t}\frac{\alpha_{1}\nabla\mathsf{n}_{p}(\tilde{\phi})+\alpha_{2}\nabla\mathsf{n}_{p}(\tilde{\varphi})}{\alpha_{1}\mathsf{n}_{p}(\tilde{\phi})+\alpha_{2}\mathsf{n}_{p}(\tilde{\varphi})}\\
\nabla_{\varpi}\alpha_{1,t} & =\frac{\alpha_{1}D_{\tilde{\phi},\varpi}^{(1)\prime}\nabla\mathsf{n}_{p}(\tilde{\phi})}{\alpha_{1}\mathsf{n}_{p}(\tilde{\phi})+\alpha_{2}\mathsf{n}_{p}(\tilde{\varphi})}-\alpha_{1,t}\frac{\alpha_{1}D_{\tilde{\phi},\varpi}^{(1)\prime}\nabla\mathsf{n}_{p}(\tilde{\phi})+\alpha_{2}D_{\tilde{\varphi},\varpi}^{(1)\prime}\nabla\mathsf{n}_{p}(\tilde{\varphi})}{\alpha_{1}\mathsf{n}_{p}(\tilde{\phi})+\alpha_{2}\mathsf{n}_{p}(\tilde{\varphi})}
\end{align*}
where simplification leads to 
\begin{align*}
\nabla_{\tilde{\pi}}\alpha_{1,t} & =\alpha_{1,t}\frac{\nabla\mathsf{n}_{p}(\tilde{\phi})}{\mathsf{n}_{p}(\tilde{\phi})}-\alpha_{1,t}\left(\alpha_{1,t}\frac{\nabla\mathsf{n}_{p}(\tilde{\phi})}{\mathsf{n}_{p}(\tilde{\phi})}+\alpha_{2,t}\frac{\nabla\mathsf{n}_{p}(\tilde{\varphi})}{\mathsf{n}_{p}(\tilde{\varphi})}\right)\\
 & =\alpha_{1,t}\alpha_{2,t}\left(\frac{\nabla\mathsf{n}_{p}(\tilde{\phi})}{\mathsf{n}_{p}(\tilde{\phi})}-\frac{\nabla\mathsf{n}_{p}(\tilde{\varphi})}{\mathsf{n}_{p}(\tilde{\varphi})}\right)\\
\nabla_{\varpi}\alpha_{1,t} & =\alpha_{1,t}\frac{D_{\tilde{\phi},\varpi}^{(1)\prime}\nabla\mathsf{n}_{p}(\tilde{\phi})}{\mathsf{n}_{p}(\tilde{\phi})}-\alpha_{1,t}\left(\alpha_{1,t}\frac{D_{\tilde{\phi},\varpi}^{(1)\prime}\nabla\mathsf{n}_{p}(\tilde{\phi})}{\mathsf{n}_{p}(\tilde{\phi})}+\alpha_{2,t}\frac{D_{\tilde{\varphi},\varpi}^{(1)\prime}\nabla\mathsf{n}_{p}(\tilde{\varphi})}{\mathsf{n}_{p}(\tilde{\varphi})}\right)\\
 & =\alpha_{1,t}\alpha_{2,t}\left(\frac{D_{\tilde{\phi},\varpi}^{(1)\prime}\nabla\mathsf{n}_{p}(\tilde{\phi})}{\mathsf{n}_{p}(\tilde{\phi})}-\frac{D_{\tilde{\varphi},\varpi}^{(1)\prime}\nabla\mathsf{n}_{p}(\tilde{\varphi})}{\mathsf{n}_{p}(\tilde{\varphi})}\right).
\end{align*}
Evaluated at $(\alpha,\tilde{\pi},\varpi)=(\alpha,\tilde{\pi}^{*},0)$
we get 
\[
\nabla_{\tilde{\pi}}\alpha_{1,t}^{\ast}=0,\,\,\,\,\,\,\,\,\,\,\nabla_{\varpi}\alpha_{1,t}^{\ast}=\alpha_{1}\alpha_{2}\frac{\nabla_{(2,\ldots,p+2)}\mathsf{n}_{p}(\tilde{\pi}^{\ast})}{\mathsf{n}_{p}(\tilde{\pi}^{\ast})},
\]
\[
f_{2,t}^{\pi}(\alpha,\tilde{\pi}^{*},0)=f_{t}(\tilde{\pi}^{*}),\ \ \ \ \ \ \ \ \ \ \nabla_{\tilde{\pi}}f_{2,t}^{\pi}(\alpha,\tilde{\pi}^{*},0)=\nabla f_{t}(\tilde{\pi}^{*}),\ \ \ \ \ \ \ \ \ \ \nabla_{\varpi}f_{2,t}^{\pi}(\alpha,\tilde{\pi}^{*},0)=0,
\]
so that 
\[
\nabla_{\tilde{\pi}}l_{t}^{\pi}(\alpha,\tilde{\pi}^{*},0)=\frac{\nabla f_{t}(\tilde{\pi}^{*})}{f_{t}(\tilde{\pi}^{*})},\ \ \ \ \ \ \ \ \ \ \nabla_{\varpi}l_{t}^{\pi}(\alpha,\tilde{\pi}^{*},0)=0.
\]

\paragraph{Second-order partial derivatives}

With straightforward differentation we obtain
\begin{align*}
\nabla_{\tilde{\pi}\tilde{\pi}^{\prime}}^{2}l_{t}^{\pi}(\alpha,\tilde{\pi},\varpi) & =\frac{\nabla_{\tilde{\pi}\tilde{\pi}^{\prime}}^{2}f_{2,t}^{\pi}(\alpha,\tilde{\pi},\varpi)}{f_{2,t}^{\pi}(\alpha,\tilde{\pi},\varpi)}-\frac{\nabla_{\tilde{\pi}}f_{2,t}^{\pi}(\alpha,\tilde{\pi},\varpi)}{f_{2,t}^{\pi}(\alpha,\tilde{\pi},\varpi)}\frac{\nabla_{\tilde{\pi}^{\prime}}f_{2,t}^{\pi}(\alpha,\tilde{\pi},\varpi)}{f_{2,t}^{\pi}(\alpha,\tilde{\pi},\varpi)}\\
\nabla_{\tilde{\pi}\varpi^{\prime}}^{2}l_{t}^{\pi}(\alpha,\tilde{\pi},\varpi) & =\frac{\nabla_{\tilde{\pi}\varpi^{\prime}}^{2}f_{2,t}^{\pi}(\alpha,\tilde{\pi},\varpi)}{f_{2,t}^{\pi}(\alpha,\tilde{\pi},\varpi)}-\frac{\nabla_{\tilde{\pi}}f_{2,t}^{\pi}(\alpha,\tilde{\pi},\varpi)}{f_{2,t}^{\pi}(\alpha,\tilde{\pi},\varpi)}\frac{\nabla_{\varpi^{\prime}}f_{2,t}^{\pi}(\alpha,\tilde{\pi},\varpi)}{f_{2,t}^{\pi}(\alpha,\tilde{\pi},\varpi)}\\
\nabla_{\varpi\varpi^{\prime}}^{2}l_{t}^{\pi}(\alpha,\tilde{\pi},\varpi) & =\frac{\nabla_{\varpi\varpi^{\prime}}^{2}f_{2,t}^{\pi}(\alpha,\tilde{\pi},\varpi)}{f_{2,t}^{\pi}(\alpha,\tilde{\pi},\varpi)}-\frac{\nabla_{\varpi}f_{2,t}^{\pi}(\alpha,\tilde{\pi},\varpi)}{f_{2,t}^{\pi}(\alpha,\tilde{\pi},\varpi)}\frac{\nabla_{\varpi^{\prime}}f_{2,t}^{\pi}(\alpha,\tilde{\pi},\varpi)}{f_{2,t}^{\pi}(\alpha,\tilde{\pi},\varpi)}
\end{align*}
 with 
\begin{align*}
\nabla_{\tilde{\pi}\tilde{\pi}^{\prime}}^{2}f_{2,t}^{\pi}(\alpha,\tilde{\pi},\varpi) & =\nabla_{\tilde{\pi}\tilde{\pi}^{\prime}}^{2}\alpha_{1,t}f_{t}(\tilde{\phi})+\nabla_{\tilde{\pi}}\alpha_{1,t}\nabla'f_{t}(\tilde{\phi})\\
 & +\nabla f_{t}(\tilde{\phi})\nabla_{\tilde{\pi}^{\prime}}\alpha_{1,t}+\alpha_{1,t}\nabla^{2}f_{t}(\tilde{\phi})\\
 & -\nabla_{\tilde{\pi}\tilde{\pi}^{\prime}}^{2}\alpha_{1,t}f_{t}(\tilde{\varphi})-\nabla_{\tilde{\pi}}\alpha_{1,t}\nabla'f_{t}(\tilde{\varphi})\\
 & -\nabla f_{t}(\tilde{\varphi})\nabla_{\tilde{\pi}^{\prime}}\alpha_{1,t}+(1-\alpha_{1,t})\nabla^{2}f_{t}(\tilde{\varphi})\\
\nabla_{\tilde{\pi}\varpi^{\prime}}^{2}f_{2,t}^{\pi}(\alpha,\tilde{\pi},\varpi) & =\nabla_{\tilde{\pi}\varpi^{\prime}}^{2}\alpha_{1,t}f_{t}(\tilde{\phi})+\nabla_{\tilde{\pi}}\alpha_{1,t}\nabla'f_{t}(\tilde{\phi})D_{\tilde{\phi},\varpi}^{(1)}\\
 & +\nabla f_{t}(\tilde{\phi})\nabla_{\varpi^{\prime}}\alpha_{1,t}+\alpha_{1,t}\nabla^{2}f_{t}(\tilde{\phi})D_{\tilde{\phi},\varpi}^{(1)}\\
 & -\nabla_{\tilde{\pi}\varpi^{\prime}}^{2}\alpha_{1,t}f_{t}(\tilde{\varphi})-\nabla_{\tilde{\pi}}\alpha_{1,t}\nabla'f_{t}(\tilde{\varphi})D_{\tilde{\varphi},\varpi}^{(1)}\\
 & -\nabla f_{t}(\tilde{\varphi})\nabla_{\varpi^{\prime}}\alpha_{1,t}+(1-\alpha_{1,t})\nabla^{2}f_{t}(\tilde{\varphi})D_{\tilde{\varphi},\varpi}^{(1)}\\
\nabla_{\varpi\varpi^{\prime}}^{2}f_{2,t}^{\pi}(\alpha,\tilde{\pi},\varpi) & =\nabla_{\varpi\varpi^{\prime}}^{2}\alpha_{1,t}f_{t}(\tilde{\phi})+\nabla_{\varpi}\alpha_{1,t}\nabla'f_{t}(\tilde{\phi})D_{\tilde{\phi},\varpi}^{(1)}\\
 & +\alpha_{1,t}D_{\tilde{\phi},\varpi}^{(1)\prime}\nabla^{2}f_{t}(\tilde{\phi})D_{\tilde{\phi},\varpi}^{(1)}+D_{\tilde{\phi},\varpi}^{(1)\prime}\nabla f_{t}(\tilde{\phi})\nabla_{\varpi^{\prime}}\alpha_{1,t}\\
 & -\nabla_{\varpi\varpi^{\prime}}^{2}\alpha_{1,t}f_{t}(\tilde{\varphi})-\nabla_{\varpi}\alpha_{1,t}\nabla'f_{t}(\tilde{\varphi})D_{\tilde{\varphi},\varpi}^{(1)}\\
 & +(1-\alpha_{1,t})D_{\tilde{\varphi},\varpi}^{(1)\prime}\nabla^{2}f_{t}(\tilde{\varphi})D_{\tilde{\varphi},\varpi}^{(1)}-D_{\tilde{\varphi},\varpi}^{(1)\prime}\nabla f_{t}(\tilde{\varphi})\nabla_{\varpi^{\prime}}\alpha_{1,t}
\end{align*}
For brevity, we omit the expressions of $\nabla_{\tilde{\pi}\tilde{\pi}^{\prime}}^{2}\alpha_{1,t}$,
$\nabla_{\tilde{\pi}\varpi^{\prime}}^{2}\alpha_{1,t}$, and $\nabla_{\varpi\varpi^{\prime}}^{2}\alpha_{1,t}$.
Evaluated at $(\alpha,\tilde{\pi},\varpi)=(\alpha,\tilde{\pi}^{*},0)$
we get 
\begin{align*}
\nabla_{\tilde{\pi}\tilde{\pi}^{\prime}}^{2}f_{2,t}^{\pi}(\alpha,\tilde{\pi}^{*},0) & =\nabla^{2}f_{t}(\tilde{\pi}^{*})\\
\nabla_{\tilde{\pi}\varpi^{\prime}}^{2}f_{2,t}^{\pi}(\alpha,\tilde{\pi}^{*},0) & =0\\
\nabla_{\varpi\varpi^{\prime}}^{2}f_{2,t}^{\pi}(\alpha,\tilde{\pi}^{*},0) & =\nabla_{\varpi}\alpha_{1,t}^{*}\nabla'_{(2,\ldots,p+2)}f_{t}(\tilde{\pi}^{*})+\nabla_{(2,\ldots,p+2)}f_{t}(\tilde{\pi}^{*})\nabla_{\varpi^{\prime}}\alpha_{1,t}^{*}+\alpha_{1}\alpha_{2}\nabla_{(2,\ldots,p+2)(2,\ldots,p+2)}^{2}f_{t}(\tilde{\pi}^{*})\\
 & =\nabla_{\varpi}\alpha_{1,t}^{*}\nabla'_{(2,\ldots,p+2)}f_{t}(\tilde{\pi}^{*})+\nabla_{(2,\ldots,p+2)}f_{t}(\tilde{\pi}^{*})\nabla_{\varpi^{\prime}}\alpha_{1,t}^{*}+\alpha_{1}\alpha_{2}\nabla_{(2,\ldots,p+2)(2,\ldots,p+2)}^{2}f_{t}(\tilde{\pi}^{*})
\end{align*}
so that
\begin{align*}
\nabla_{\tilde{\pi}\tilde{\pi}^{\prime}}^{2}l_{t}^{\pi}(\alpha,\tilde{\pi}^{*},0) & =\frac{\nabla^{2}f_{t}(\tilde{\pi}^{*})}{f_{t}(\tilde{\pi}^{*})}-\frac{\nabla f_{t}(\tilde{\pi}^{*})}{f_{t}(\tilde{\pi}^{*})}\frac{\nabla'f_{t}(\tilde{\pi}^{*})}{f_{t}(\tilde{\pi}^{*})}\\
\nabla_{\tilde{\pi}\varpi^{\prime}}^{2}l_{t}^{\pi}(\alpha,\tilde{\pi}^{*},0) & =0\\
\nabla_{\varpi\varpi^{\prime}}^{2}l_{t}^{\pi}(\alpha,\tilde{\pi}^{*},0) & =\frac{\nabla_{\varpi}\alpha_{1,t}^{\ast}\nabla_{(2,\ldots,p+2)}'f_{t}(\tilde{\pi}^{*})+\nabla_{(2,\ldots,p+2)}f_{t}(\tilde{\pi}^{*})\nabla_{\varpi^{\prime}}\alpha_{1,t}^{\ast}+\alpha_{1}\alpha_{2}\nabla_{(2,\ldots,p+2)(2,\ldots,p+2)}^{2}f_{t}(\tilde{\pi}^{*})}{f_{t}(\tilde{\pi}^{*})}\\
 & =\alpha_{1}\alpha_{2}\Biggl[\frac{\nabla_{(2,\ldots,p+2)}\mathsf{n}_{p}(\tilde{\pi}^{*})}{\mathsf{n}_{p}(\tilde{\pi}^{*})}\frac{\nabla_{(2,\ldots,p+2)}'f_{t}(\tilde{\pi}^{*})}{f_{t}(\tilde{\pi}^{*})}+\frac{\nabla_{(2,\ldots,p+2)}f_{t}(\tilde{\pi}^{*})}{f_{t}(\tilde{\pi}^{*})}\frac{\nabla_{(2,\ldots,p+2)}'\mathsf{n}_{p}(\tilde{\pi}^{*})}{\mathsf{n}_{p}(\tilde{\pi}^{*})}\\
 & \qquad\qquad\qquad+\frac{\nabla_{(2,\ldots,p+2)(2,\ldots,p+2)}^{2}f_{t}(\tilde{\pi}^{*})}{f_{t}(\tilde{\pi}^{*})}\Biggr].
\end{align*}

\bigskip{}

\section*{Additional References}

Aliprantis, C. D., Burkinshaw, O., 1998. Principles of Real Analysis.
Academic Press.\\
\\
Andrews, D. W., Ploberger, W., 1994. Optimal tests when a nuisance
parameter is present only under the alternative. Econometrica 62,
1383\textendash 1414.\\
\\
Galbraith, R., Galbraith, J., 1974. On the inverses of some patterned
matrices arising in the theory of stationary time series. Journal
of Applied Probability 11, 63\textendash 71.\\
\\
Magnus, J. R., Neudecker, H., 1979. The commutation matrix: some properties
and applications. Annals of Statistics 7, 381\textendash 394.\\
\\
Magnus, J. R., Neudecker, H., 1999. Matrix Differential Calculus with
Applications in Statistics and Econometrics, Revised Edition. Wiley.\\

\end{document}